\documentclass[a4paper]{article}
\usepackage{cite}
\usepackage{graphicx}
\usepackage{amsmath}
\usepackage{amssymb,bm}
\usepackage{array}
\usepackage{subfigure}
\usepackage{url}
\usepackage{algorithm}
\usepackage{algorithmic}
\usepackage{multirow}
\usepackage[margin=20truemm]{geometry}

\def\myinter#1{\boldsymbol{[}{#1}\boldsymbol{)}}
\def\myInter#1{\boldsymbol{\biggl [}{#1}\boldsymbol{\biggr )}}
\def\myINter#1{\boldsymbol{\Biggl [}{#1}\boldsymbol{\Biggr )}}

\makeatletter

\newcommand{\Rmnum}[1]{\expandafter\@slowromancap\romannumeral #1@}

\def\mytheory#1#2#3{\vspace{3mm} \hrule \vspace{1mm} \begin{#1}[#2] #3 \end{#1} \vspace{1mm} \hrule \vspace{3mm}}
\makeatother

\begin{document}
\title{Optimal Construction of N-bit-delay Almost Instantaneous Fixed-to-Variable-Length Codes}

\author{Ryosuke~Sugiura,
	Masaaki~Nishino,
	Norihito~Yasuda,
	Yutaka~Kamamoto,
	and~Takehiro~Moriya}
\date{}
\maketitle

\begin{abstract}
	This paper presents an optimal construction of $N$-bit-delay almost instantaneous fixed-to-variable-length (AIFV) codes, the general form of binary codes we can make when finite bits of decoding delay are allowed.
	The presented method enables us to optimize lossless codes among a broader class of codes compared to the conventional FV and AIFV codes.
	The paper first discusses the problem of code construction, which contains some essential partial problems, and defines three classes of optimality to clarify how far we can solve the problems.
	The properties of the optimal codes are analyzed theoretically, showing the sufficient conditions for achieving the optimum.
	Then, we propose an algorithm for constructing $N$-bit-delay AIFV codes for given stationary memory-less sources.
	The optimality of the constructed codes is discussed both theoretically and empirically.
	They showed shorter expected code lengths when $N\ge 3$ than the conventional AIFV-$m$ and extended Huffman codes.
	Moreover, in the random numbers simulation, they performed higher compression efficiency than the 32-bit-precision range codes under reasonable conditions.
\end{abstract}

\section{Introduction}
\label{sec:intro}
Lossless compression is one of the essential techniques for communication.
Especially, fixed-to-variable-length (FV) codes are useful for encoding a given size of data and are used in many coding applications like audio and video codecs \cite{ref:MSSN, ref:hbdc,ref:audio_coding, ref:speech_coding}.
In these situations, input signal distribution is often assumed based on some models.
We encode the signal values, or source symbols, into codes optimized for the given distribution.

When we have a distribution of a source symbol, we can construct Huffman codes \cite{ref:huffman, ref:candc}, which achieve the minimum expected code length among FV codes for 1-symbol-length inputs.
Huffman code can be represented by a code tree, requiring low computational complexity for encoding by using the tree as a coding table.
We can enhance the compression efficiency by constructing Huffman codes for Cartesian products of source symbols, namely, the extended Huffman codes \cite{ref:introDC}.
Here, the code trees of the extended Huffman codes become exponentially larger according to the input length.
There are trade-offs between the compression efficiency and the table size.

For longer source symbol sequences, the arithmetic coding \cite{ref:hbdc, ref:candc, ref:introDC} is also a well-known approach, which gives us some variable-to-variable-length (VV) codes without any table.
It does not achieve the minimum expected code length for a finite-size input, and thus,  Huffman codes perform better for sufficiently short sequences.
However, the arithmetic coding asymptotically achieves entropy rates when the input length is long enough, showing much higher efficiency than Huffman codes in many practical cases.

As a subclass of VV codes and an extension of FV codes, Yamamoto ${\it et}$ ${\it al}$.~have proposed the class of almost instantaneous FV (AIFV) codes \cite{ref:aifv1, ref:opt_aifv_dp, ref:opt_aifv_dp_fast}.
It loosens the constraint of FV codes that the decoder must be able to decode the fixed-length sequence instantaneously.
This relaxation enables us to achieve shorter expected code length than Huffman codes.

The coding rule of AIFV codes can be represented as a combination of multiple code trees.
AIFV-$m$ codes \cite{ref:aifv2, ref:opt_aifvm, ref:opt_aifvm_dp, ref:aifvm_red}, one of the conventional AIFV codes,
use sets of $m$ code trees to represent codes decodable with at most $m$ bits of decoding delay.
These code trees correspond to recursive structures of a single huge code tree like the one of extended Huffman codes.
Therefore, we can make more complex coding rules with smaller table sizes than the codes made by a single code tree.

In our previous work \cite{ref:n_aifv}, we have pointed out that the conventional AIFV codes can only represent a part of all codes decodable within a given decoding delay.
For example, AIFV-$m$ codes require a large difference in the code lengths of the source symbols when utilizing the permitted delay.
If we want to enhance the compression efficiency of AIFV-$m$ codes by using larger values for $m$,
we need to deal with heavily biased source distributions, such as sparse source symbol sequences containing many zeros.

Therefore, we have proposed $N$-bit-delay AIFV codes, which can represent every code decodable with at most $N$-bit delay.
It has been proven that any uniquely encodable and uniquely decodable VV codes can be represented by the code-tree sets of the proposed scheme when sufficient $N$ is given.
Owing to these facts, $N$-bit-delay AIFV codes are expected to outperform other codes by fully utilizing the permitted delay.

However, the construction for $N$-bit-delay AIFV codes has yet to be presented.
This paper aims to introduce an algorithm to construct the codes for given source distributions.
It is a complex problem to solve straightforwardly, containing combinatorial problems with huge freedom design.
So, we should understand the construction problem deeply, divide it into practically solvable ones and analyze in what range we can guarantee the optimality.

It is also important to know the sufficient constraints for constructing optimal codes and to limit the freedom:
Although $N$-bit-delay AIFV code is the necessary and sufficient representation of any uniquely encodable and uniquely decodable VV codes with $N$-bit decoding delay,
there are many codes achieving the same expected code length and being useless choices in code construction.
We discuss the structural properties of the optimal codes before introducing the construction algorithm.

The paper first reviews the idea of $N$-bit-delay AIFV codes in Section \ref{sec:prepare} for preparation.
Section \ref{sec:opt} discusses the problem of code construction.
We present its general form and break it down into partial problems. According to them, we introduce three classes of optimality.
We also show a decomposition that makes one of the problems tree-wise independent.
Then, Section \ref{sec:prop} focuses on the goal of the construction.
We analyze some properties of the optimal $N$-bit-delay AIFV codes, showing what condition can be sufficient for minimizing the expected code length.
In Section \ref{sec:design}, we present the code-construction algorithm.
Using the problem decomposition, we formulate an algorithm using tree-wise integer linear programming (ILP) problems,
which is guaranteed to give some class of optimal codes.
Finally, the proposed codes are evaluated experimentally in Section \ref{sec:eval}.
We compare the asymptotic expected code length and the average code length for finite-length sequences.
We also empirically check which optimality class is achieved by the constructed codes.

\section{Preliminaries}
The notations below are used for the following discussions.
\begin{itemize}
	\item $\mathbb{N}$: The set of all natural numbers.
	\item $\mathbb{R}$: The set of all real numbers.
	\item $\mathbb{Z}^+$: The set of all non-negative integers.
	\item $\mathbb{Z}^+_{<M}$: The set of all non-negative integers smaller than an integer $M$.
	\item $\mathbb{A}_M$: $\{a_m\mid m\in \mathbb{Z}^+_{<M}\}$, the source alphabet of size $M$.
	\item $\mathbb{S}_M$: the Kleene closure of $\mathbb{A}_M$, or the set of all $M$-ary source symbol sequences, including a zero-length sequence $\epsilon$.
	\item $\mathbb{W}$: The set of all binary strings, including a zero-length one `$\lambda$'. `$\lambda$' can be a prefix of any binary string.
	\item $\mathbb{W}_N$: The set of all binary strings of length $N$. Especially, $\mathbb{W}_0=\{\text{`$\lambda$'}\}$.
	\item $\mathbb{M}$: $\{\textsc{Words} \subseteq \mathbb{W}\mid \textsc{Words}\neq \emptyset\}$, the set of all non-empty subsets of $\mathbb{W}$.
	\item $\preceq$, $\npreceq$, $\prec$, $\nprec$: Dyadic relations defined in $\mathbb{W}$. $w\preceq w'\text{ (resp.~$w\npreceq w'$) }$ indicates that $w$ is (resp. is not) a prefix of $w'$. $\prec$ (resp. $\nprec$) excludes $=$ (resp. $\neq$) from $\preceq$ (resp. $\npreceq$).
	\item $\parallel$: A dyadic relation defined for $\mathbb{W}$. $w\parallel w'$ indicates that $w$ and $w'$ satisfy either $w\preceq w'$ or $w'\preceq w$.
	\item $\mathbb{PF}$: $\{\textsc{Words}\in \mathbb{M}\mid \forall w\neq w'\in \textsc{Words}: w\npreceq w'\}$, the set of all prefix-free binary string sets.
	\item $\|\cdot\|_{\rm len}$: The length of a string in $\mathbb{W}$.
	\item $\myinter{x_l, x_u}$: $\{x\in\mathbb{R}\mid x_l\le x<x_u\}$, an interval between $x_l$ and $x_u$ ($\in \mathbb{R}$).
	\item $\mathbb{PI}$: $\{R\subset \myinter{0, 1}\}$, the set of all probability intervals included between $0$ and $1$.
	\item $\oplus$: Appending operator. For $w, w'\in\mathbb{W}$, $w\oplus w'$ appends $w'$ to the right of $w$. It is defined for $\mathbb{S}_M$ similarly. For $w\in\mathbb{W}$ and $W'\in\mathbb{M}$, $w\oplus W'$ is defined to give $\{w\oplus w'\mid w'\in W'\}$ ($\in\mathbb{M}$).
	\item $\oslash$: Subtracting operator. For $w_{\rm pre}, w\in\mathbb{W}$, $w_{\rm pre}\oslash w$ subtracts the prefix $w_{\rm pre}$ from $w$. For $w_{\rm pre}\in\mathbb{W}$ and $W\in\mathbb{M}$, $w_{\rm pre}\oslash W$ is defined to give $\{w_{\rm pre}\oslash w\mid w\in W\}$ ($\in\mathbb{M}$).
	\item $\neg$: Bit-flipping operator. For $\mathbb{W}$, $\neg W$ gives a string by flipping every `0' and `1' in $W\in\mathbb{W}$. For $\mathbb{M}$, $\neg M$ gives a set by bit-flipping every string in $M\in\mathbb{M}$.
\end{itemize}
We also use the following terms related to some state set $S$ of a time-homogeneous Markov chain \cite{ref:markovchain,ref:markov_inv}.
\begin{itemize}
	\item A state $k'$ is {\it reachable from} $k$ ($k, k'\in S$), or $k$ {\it can reach} $k'$, when the state can transit to $k'$ within finite steps beginning from $k$.
	\item States $k$ and $k'$ ($\in S$) are {\it strongly connected} when they are reachable from each other.
	\item A subset $S'(\subset S)$ is {\it invariant} when it has no outgoing edge: If $k'$ in $S'$ reaches $k$ in $S$, $k$ must also be a member of $S'$.
	\item A non-empty subset $S'(\subset S)$ is a {\it closed set} when it is invariant and all the states in $S'$ are strongly connected to each other.
	\item A non-empty subset $S'(\subset S)$ is an {\it open set} when it contains no closed subset.
\end{itemize}

\section{$N$-bit-delay AIFV codes}
\label{sec:prepare}
\subsection{Linked code forest}
\begin{figure}[tb]
	\begin{center}
		\includegraphics[width=15cm,  bb=0 0 966 244]{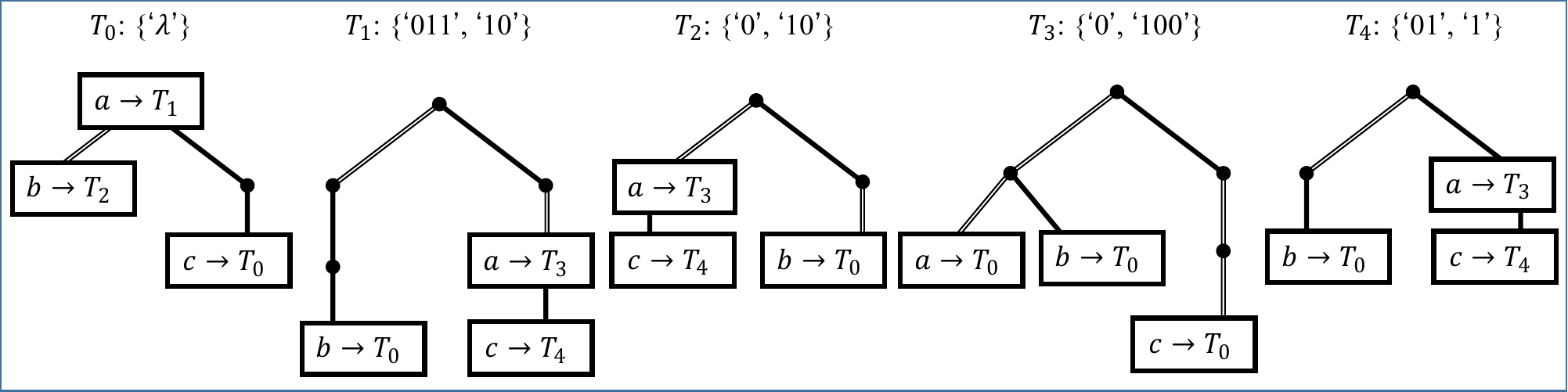}
	\end{center}
	\caption{An example of a linked code forest $\{T_k\mid 0\le k <5\}$ of the proposed code, with modes \{`$\lambda$'\}, \{`011', `10'\}, \{`0', `10'\}, \{`0', `100'\}, and \{`01', `1'\} for the respective code trees. }
	\label{fig:trees_ex}
\end{figure}
$N$-bit-delay AIFV codes are written as sets of code trees.
Unlike the conventional code trees, the ones used in the sets represent code-tree switching rules as well as codewords.
Additionally, each code tree is assigned a \emph{mode}, some binary-string set used to guarantee the decodability.
The code tree is written as follows.
\begin{equation}
	T_k = (\textrm{Cword}_k, \textrm{Link}_k, \textsc{Mode}_k),
\end{equation}
where $\textrm{Cword}_k:\mathbb{A}_M\to \mathbb{W}$, $\textrm{Link}_k: \mathbb{A}_{M}\to \mathbb{Z}^+$, and $\textsc{Mode}_k\in \mathbb{M}$. $k$ is an index of the code tree.
$\textrm{Cword}_k(a)$ is the codeword corresponding to the symbol $a$.
$\textrm{Link}_k(a)$ indicates the link, suggesting which code tree we should switch to after encoding/decoding the symbol $a$.
$\textsc{Mode}_k\in \mathbb{M}$ is the mode assigned to $T_k$.

Fig.~\ref{fig:trees_ex} shows an example of $N$-bit-delay AIFV codes.
Through the paper, we use squares and black dots for nodes and solid lines (resp. double lines) for edges representing `1' (resp. `0').
The source symbols and the links are assigned to the squared nodes of each code tree.
In general, multiple pairs of symbols and links can be assigned to a single node as long as they follow the rules shown in Section \ref{ssec:rule}.
The links combine the code trees and enable us to use multiple code trees to encode source symbols.
Based on them, each set $\{T_k\mid k\in \mathbb{Z}^+_{<K}\}$ forms a time-homogeneous Markov chain.
Here, we call such code-tree set a \emph{linked code forest}, or simply, a \emph{code forest}.
For simplicity, we write it as $\{T_k\}$.

\subsection{Coding processes}
The encoder and decoder using a linked code forest $\{T_k\}$ work as follows. 
\mytheory{prc}{Encoding a source symbol sequence into an $N$-bit-delay AIFV codeword sequence}
{
	\label{prc:enc}
	Follow the steps below with the $L$-length source symbol sequence $x_0 x_1\cdots x_{L-1}$ and code forest $\{T_k\}$ being the inputs of the encoder.
	\begin{enumerate}
		\item Start encoding from the initial $k=k_0$.
		\item For $i=0,1,\cdots, L-1$, output the codeword $\textrm{Cword}_k(x_i)$ in the current code tree $T_k$
		and switch the code tree by updating the index $k$ with $\textrm{Link}_k(x_i)$.
		\item Output the shortest binary string in the mode $\textsc{Mode}_k$ (here, we call it the termination codeword).
	\end{enumerate}
}
\mytheory{prc}{Decoding a source symbol sequence from an $N$-bit-delay AIFV codeword sequence}
{
	\label{prc:dec}
	Follow the steps below with the codeword sequence, code forest $\{T_k\}$, and output length $L$ being the inputs of the decoder.
	\begin{enumerate}
		\item Start decoding from the initial $k=k_0$.
		\item Compare the codeword sequence with the codewords in the current code tree $T_k$.
		If the codeword $\textrm{Cword}_k(a)$ matches the codeword sequence,
		and some codeword $\textrm{Query}\in \textsc{Mode}_{\textrm{Link}_k(a)}$ matches the codeword sequence after $\textrm{Cword}_k(a)$,
		output the source symbol $a$ and continue the process from the codeword sequence right after $\textrm{Cword}_k(a)$.
		\item Switch the code tree by updating the index $k$ with $\textrm{Link}_k(a)$.
		\item If the decoder has output less than $L$ symbols, return to b.
	\end{enumerate}
}
Any initial value $k_0$ will work as long as it is shared between the encoder and decoder, but the prefixes of the codeword sequence available for encoding and decoding will be restricted when $\textsc{Mode}_{k_0}\neq \{$`$\lambda$'$\}$. Unless otherwise noted, we use $k_0=0$, and thus $\textsc{Mode}_{k_0}= \{$`$\lambda$'$\}$, for simplicity. 

As an example of the coding processes, let us encode a source symbol sequence $acba$ using the code forest $\{T_k\mid 0\le k <5\}$ in Fig.~\ref{fig:trees_ex}.
We start encoding from $T_0$ and have a codeword `$\lambda$' because $a$ is assigned to the root.
Then, we switch the code tree to $T_1$, linked from the node of $a$.
$T_1$ gives a codeword `101' for $c$ and links to the next code tree $T_4$.
The source symbol $b$ is represented as `01' in $T_4$, and the code tree is switched to $T_0$ for encoding the remaining $a$.
The symbol $a$ is represented as `$\lambda$' in $T_0$, and links to $T_1$.
Since we have encoded all the source symbols, we get a termination codeword from $T_1$, i.e., `10'.
As a result, the codeword sequence for $acba$ becomes `$\lambda$10101$\lambda$10', i.e., `1010110'.

The decoder starts the decoding from $T_0$, checking at first whether the codeword `$\lambda$' of $a$ matches the codeword sequence `1010110'.
The codeword for $a$ is `$\lambda$' and thus matches the sequence.
Then, the decoder checks whether any codeword in the mode of $T_1$, linked from the node of $a$, matches the sequence.
Since `10' is included, it outputs $a$ and switches the code tree to $T_1$.
The next symbol is decoded from the codeword sequence following `$\lambda$', i.e., `1010110'.
Although `1010110' matches the codeword `10' of the symbol $a$ in $T_1$, no string in the mode of $T_3$ matches the following `10110'.
Thus, the decoder does not output $a$ but checks the codeword for $c$.
Since the codeword `101' of $c$ matches the sequence and the following `0110' matches `01' in the mode of $T_4$, the decoder can confirm $c$ as output and $T_4$ as the next tree.
The codeword `01' for $b$ in $T_4$ matches the `0110', and the following `10' obviously matches `$\lambda$' in the mode of $T_0$.
Therefore, $b$ is output, and the last symbol is decoded from `10' by $T_0$.
The codeword `$\lambda$' is for $a$ in $T_0$, and `10' is included in the mode of $T_1$.
So, the last symbol is determined as $a$, and we can get the correct source symbol sequence $acba$.

The termination codewords in step c of the encoding are necessary when the decoder only knows the total length $L$ of the source symbol sequence and cannot know the end of the codeword sequence.
In the above example, if we do not use the termination codeword `10', the decoder cannot stop the decoding process before confirming the last $a$ and starts reading the irrelevant binary strings following `10101'.
In this case, when some irrelevant strings such as `11' follow the codeword sequence, the decoder will fail to decode $a$ and output $c$ instead.

As we can see from the procedures, the modes $\{\textsc{Mode}_k\}$ are mainly used as queries for the decoder to determine the output symbol.
It reads the codeword sequence to check whether some binary string is included in the mode of the next tree.
When using code forests like Fig.~\ref{fig:trees_ex}, the encoder can uniquely know which code tree to switch according to the links.
However, the decoder cannot straightforwardly determine which one to switch.
For example, when the decoder has an input codeword `0' for $T_0$, it cannot determine whether it should output $a$ and switch to $T_1$ or output $b$ and switch to $T_2$.
Modes are set to help the decoder know what binary string it should read to determine the output.

\subsection{Decoding delay}
With the coding processes given, we can define the decoding delay of the codes.
For the encoder $V_{\rm enc}: \mathbb{S}_M\to\mathbb{W}$ and decoder $V_{\rm dec}: \mathbb{W}\to\mathbb{S}_M$, the decoding delay can be defined as follows.
\mytheory{dfn}{Decoding delay of a code for $s\in\mathbb{S}_M$}{
	$\max \|\text{Lookahd}\|_{\rm len}$ subject to
	\begin{equation}
		\exists \text{Prefix}\in\mathbb{W}:
		\left\{
		\begin{array}{rl}
			\forall \textrm{Tail}\in \mathbb{S}_M, \exists \textrm{Suffix}\in\mathbb{W}: & V_{\rm enc}(s\oplus\textrm{Tail})=\text{Prefix}\oplus\textrm{Suffix}                       \\
			\forall \textrm{Suffix}\in\mathbb{W}, \exists \textrm{Tail}\in \mathbb{S}_M: & V_{\rm dec}((\text{Prefix}\oplus\text{Lookahd})\oplus\textrm{Suffix})=s\oplus\textrm{Tail} \\
			\forall l\prec \text{Lookahd}, \exists \textrm{Suffix}\in\mathbb{W}, \forall \textrm{Tail}\in \mathbb{S}_M: & V_{\rm dec}((\text{Prefix}\oplus l)\oplus\textrm{Suffix})\neq s\oplus\textrm{Tail}
		\end{array}
		\right..
	\end{equation}
}
In other words, the decoding delay for $s$ is the maximum length of the binary string needed ($\text{Lookahd}$) for the decoder to determine $s$ as its output after reading the codeword ($\text{Prefix}$) that the encoder can immediately determine as its output when encoding $s$.
This delay can be defined for any code, including non-symbol-wise codes like the extended Huffman codes if we break them down into symbol-wise forms \cite{ref:n_aifv}.

In $N$-bit-delay AIFV codes, the decoding delay depends on the lengths of the binary strings in the modes.
For example, when decoding `10100' with the code tree $T_1$ in Fig.~\ref{fig:trees_ex}, there are two candidates for the output:
$\textrm{Cword}_1(a)$ $=$ `10' and $\textrm{Cword}_1(c)$ $=$ `101', the codewords that the encoder can immediately determine as its outputs when encoding $a$ and $c$, respectively.
The decoder can determine $a$ as its output by checking that the 3-bit string `100', following `10', belongs to the mode $\textsc{Mode}_{\textrm{Link}_1(a)}$.
In the case of Fig.~\ref{fig:trees_ex}, the decoder must check at most 3 bits after reading the codeword to confirm the output.
Therefore, the decoding delay of the code represented by this code forest is 3 bits.

\subsection{Rules of code forests}
\label{ssec:rule}
The combination of codewords, links, and modes must obey some rules to guarantee the decodability.
In defining the rule of $N$-bit-delay AIFV codes, it is useful to introduce the idea of expanded codewords, which are made of codewords and modes.
The sets of expanded codewords are written as
\begin{eqnarray}
	\textsc{Expand}_k(a)&=&\textrm{Cword}_k(a)\oplus\textsc{Mode}_{\textrm{Link}_k(a)},\\
	\textsc{Expands}_k&=&\bigcup_{a\in\mathbb{A}_M}\textsc{Expand}_k(a).
\end{eqnarray}
For example, $T_0$ in Fig.~\ref{fig:trees_ex} has codewords $\textrm{Cword}_0(a)=$ `$\lambda$', $\textrm{Cword}_0(b)=$ `0', and $\textrm{Cword}_0(c)=$ `11' with the corresponding modes $\textsc{Mode}_{\textrm{Link}_0(a)}=\{$`011', `10'$\}$, $\textsc{Mode}_{\textrm{Link}_0(b)}=\{$`0', `10'$\}$, and $\textsc{Mode}_{\textrm{Link}_0(c)}=\{$`$\lambda$'$\}$.
In this case, the sets of expanded codewords of $T_0$ for $a$, $b$, and $c$ are $\textsc{Expand}_0(a)=\{$`011', `10'$\}$, $\textsc{Expand}_0(b)=\{$`00', `010'$\}$, and $\textsc{Expand}_0(c)=\{$`11'$\}$, respectively.

Every code tree $T_k$ in the code forests representing $N$-bit-delay AIFV codes obeys the following rules.
\mytheory{rle}{Code trees of $N$-bit-delay AIFV code}
{
	\label{rle:decodable}
	\begin{enumerate}
		\item $\textsc{Expands}_k\in\mathbb{PF}$.
		\label{srle:prefix_free_expansion}
		\item $\forall \textrm{Expcw} \in \textsc{Expands}_k$, $\exists \textrm{Query} \in \textsc{Mode}_k$: $\textrm{Query} \preceq \textrm{Expcw}$.
		\label{srle:prefix_in_mode}
		\item $\textsc{Mode}_k\in\mathbb{BM}_N$.
		\label{srle:basic_mode}
	\end{enumerate}
}
In other words, $N$-bit-delay AIFV code is defined as a code given by {\it Procedure \ref{prc:enc}} with a linked code forest $\{T_k\}$ satisfying {\it Rule \ref{rle:decodable}}.
Unless otherwise specified, we use $\{$`$\lambda$'$\}$ for $\textsc{Mode}_0$.

The first two rules are necessary for the decodability.
{\it Rule \ref{rle:decodable}} \ref{srle:prefix_free_expansion} means that, for any code tree, the expanded codewords of a code tree must be prefix-free.
{\it Rule \ref{rle:decodable}} \ref{srle:prefix_in_mode} requires the prefix of every expanded codeword to be limited to its mode.
For example, $T_2$ in Fig.~\ref{fig:trees_ex} has expanded codewords as $\textsc{Expands}_2=\{$`00', `0100', `0101', `011', `10'$\}$, which are prefix-free and have some prefixes in $\textsc{Mode}_2=\{$`0', `10'$\}$.

\begin{figure}[t]
	\begin{center}
		\includegraphics[width=6cm,  bb=0 0 407 209]{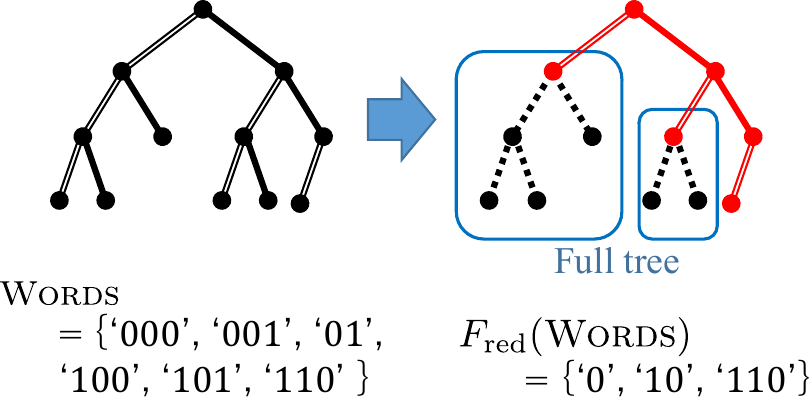}
	\end{center}
	\caption{Example of $F_{\rm red}(\textsc{Words})$. A tree representing $\textsc{Words}$ on the left, and the tree of $F_{\rm red}(\textsc{Words})$ on the right.}
	\label{fig:reduce_ex}
\end{figure}

\begin{figure}[t]
	\begin{center}
		\includegraphics[width=8cm,  bb=0 0 627 449]{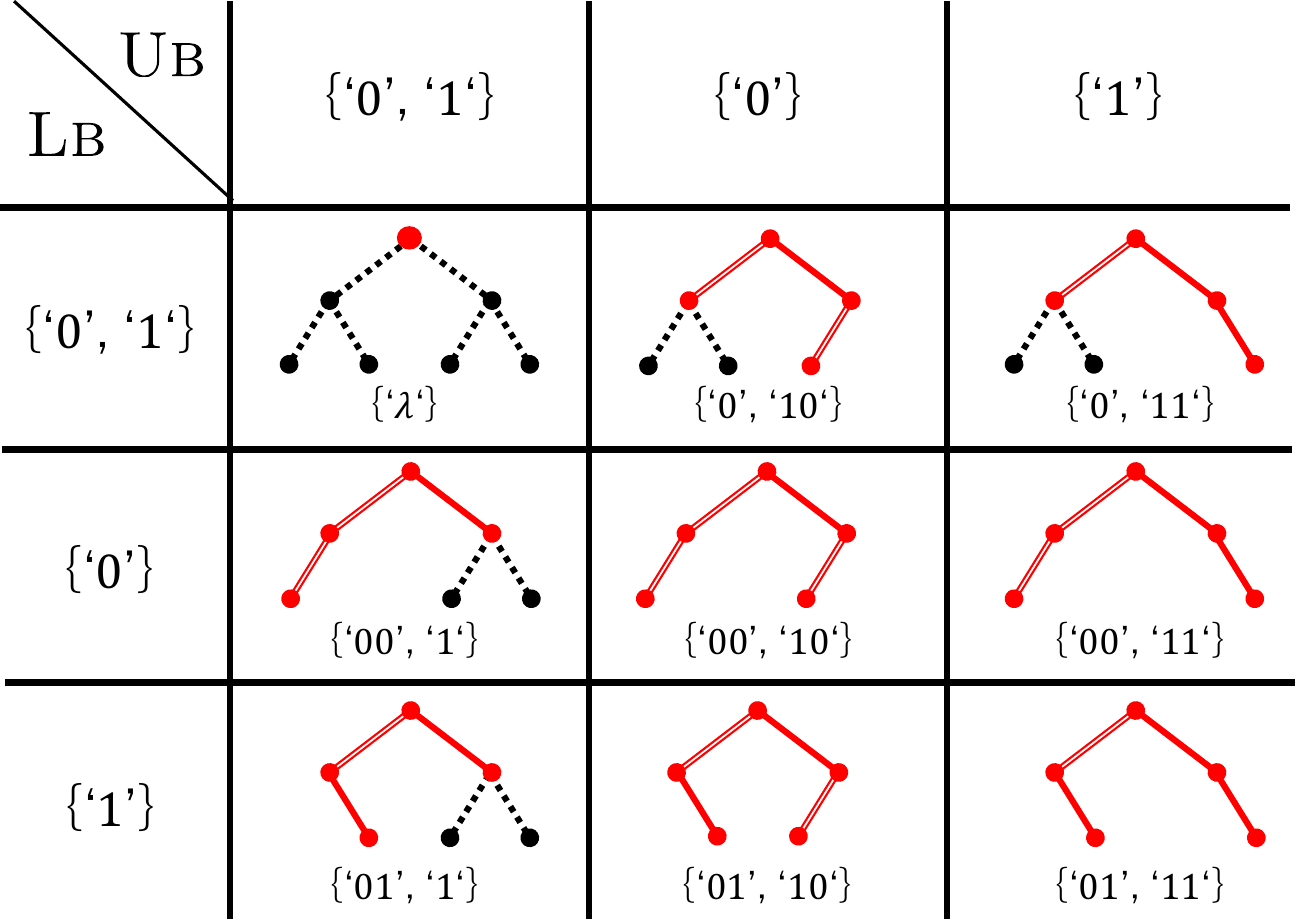}
	\end{center}
	\caption{2-bit binary-string sets available, based on {\it Rule \ref{rle:decodable}} \ref{srle:basic_mode}, for modes of 2-bit-delay AIFV codes.}
	\label{fig:2bit_mode_patt}
\end{figure}

{\it Rule \ref{rle:decodable}} \ref{srle:basic_mode} is for limiting the variation of modes without loss of generality. The notation $\mathbb{BM}_N$ is defined as
\begin{itemize}
	\item $\mathbb{BM}_N(\subset \mathbb{M})$: $\{ F_{\rm red}((\text{`0'}\oplus\textsc{Lb})\cup (\text{`1'}\oplus\textsc{Ub}))\mid \textsc{Lb},\textsc{Ub}\subset \mathbb{W}_{N-1}\}$, the set of all basic modes \cite{ref:n_aifv} with $N$-bit length.
\end{itemize}
It is known that every forest using modes other than the basic modes can be converted into one using only basic modes without increasing the decoding delay or expected code length.
The above definition uses the following functions.
\begin{itemize}
	\item $F_{\rm full}$: $\mathbb{M}\to\mathbb{M}$. $F_{\rm full}(\textsc{Words})=\{\textrm{Prefix}\in\mathbb{W}\mid \forall \textrm{Suffix}\in\mathbb{W}, \exists w\in \textsc{Words}: \textrm{Prefix}\oplus\textrm{Suffix}\parallel w\}$.
	\item $F_{\rm red}$: $\mathbb{M}\to\mathbb{PF}$. $F_{\rm red}(\textsc{Words})=\{\hat{w}\in F_{\rm full}(\textsc{Words})\mid \forall \textrm{Prefix}\in F_{\rm full}(\textsc{Words}): \textrm{Prefix}\nprec \hat{w}\}$.
\end{itemize}
$F_{\rm red}$ outputs a binary tree by cutting off all the full partial trees in the input, as illustrated in Fig.~\ref{fig:reduce_ex}.
Here, a full tree means a tree having no node with a single child.
According to {\it Rule \ref{rle:decodable}} \ref{srle:basic_mode}, for example, 2-bit-delay AIFV codes have 9 patterns of available modes depicted in Fig.~\ref{fig:2bit_mode_patt}.
Under {\it Rule \ref{rle:decodable}} \ref{srle:basic_mode}, every binary string is not longer than $N$ bits, and thus the decoding delay becomes at most $N$ bits.


We here use a notation to specify the code forests:
\begin{itemize}
	\item $\mathbb{LF}_{M,N}$: The set of all linked code forests representing $N$-bit-delay AIFV codes for $\mathbb{A}_M$.
\end{itemize}
For any code forest $\{T_k\}\in\mathbb{LF}_{M,N}$, every code tree $T_k$ satisfies {\it Rule \ref{rle:decodable}} and always links to a code tree belonging to $\{T_k\}$.

\subsection{Merits of code-forest representation}
\begin{figure}[t]
	\begin{center}
		\includegraphics[width=5cm,  bb=0 0 314 251]{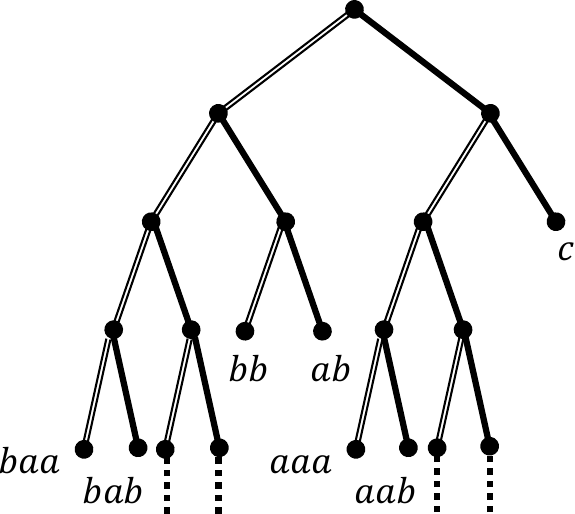}
	\end{center}
	\caption{A single code tree constructing the same code as the code forest in Fig.~\ref{fig:trees_ex}. }
	\label{fig:integ}
\end{figure}
$N$-bit-delay AIFV codes can represent a huge and complex code tree as a set of simple symbol-wise code trees.
Using the code forest in Fig.~\ref{fig:trees_ex}, we can encode a source symbol sequence $baa$ as `0000' with the code tree returning to $T_0$, the initial code tree.
Similarly, we have `0001', `010', and `011' respectively for $bab$, $bb$, and $ab$ before the code tree returns to $T_0$.
Fig.~\ref{fig:integ} represents this coding rule as a single tree.
It gives the same code as Fig.~\ref{fig:trees_ex}.
Note that the depth of the tree in Fig.~\ref{fig:integ} is infinite:
Some source symbol sequences never make the code tree return to $T_0$ since there is a self-transition in $T_4$.
As in this example, $N$-bit-delay AIFV codes can represent coding rules that a single tree cannot do in a finite way.

In the case of the extended Huffman codes, we have to fix the source symbol sequences to assign.
Meanwhile, optimizing the code forest of $N$-bit-delay AIFV codes is equivalent to optimizing simultaneously a huge code tree and a combination of symbol sequences assigned to it.
Therefore, higher compression efficiency is achievable compared to the extended Huffman codes.

The idea of modes allows us to assign the symbols more freely than the conventional AIFV codes.
In the case of AIFV-$m$ codes, for example, if we want to assign a symbol with a link to the $m$-th code tree, we have to assign another symbol to the node $m$ bits below.
This is because the switching rule of AIFV-$m$ codes depends on the symbol assignment, which makes the decodable condition simple but limits the freedom of the code-tree construction.
Introducing the idea of modes, $N$-bit-delay AIFV codes can use a combination of codewords and switching rules unavailable for AIFV-$m$ codes.
At the same time, we can guarantee the decodability in a code-tree-wise way by checking the expanded codewords.

Regardless of their representation ability, the encoding/decoding processes require low computational costs:
Compared with Huffman codes, $N$-bit-delay AIFV codes only add the symbol-wise switching of the code trees in the encoding process;
the decoding process requires only at most $M(=|\mathbb{A}_M|)$ times of additional check of codewords in the modes.
Indeed, the table size increases by $|\{T_k\}|$ times compared to a single code tree assigned with one symbol.
However, the table size of $N$-bit-delay AIFV code is much smaller than a single code tree assigned with a symbol sequence:
The table becomes much smaller when using the code forest in Fig.~\ref{fig:trees_ex} instead of the code tree in Fig.~\ref{fig:integ}, representing an equivalent coding rule.

\section{Code construction problem}
\label{sec:opt}
\subsection{Formal definition}
Let us discuss the problem of constructing $N$-bit-delay AIFV codes for given stationary memoryless sources of source symbols $\mathbb{A}_M$, achieving the minimum expected code length. 
Here, the expected code length is defined as an expectation of the code length per source symbol. 
For a source distribution $p_{\rm src}$, constructing a code forest $\{T_k\}$ can be written as the following optimization problem.
\mytheory{mip}{General $N$-bit-delay AIFV code construction}{
	\label{mip:totalcode}
	For given $N\in\mathbb{N}$ and $p_{\rm src}:\mathbb{A}_M\to \mathbb{R}$,
	\begin{equation}
		\mathcal{L}_{\rm general}(N, p_{\rm src})\equiv \min_{\{T_k\}\in\mathbb{LF}_{M,N}} \sum_{k'=0}^{|\{T_k\}|-1}\left(\sum_{m=0}^{M-1}\|\textrm{Cword}_{k'}(a_m)\|_{\rm len}\cdot p_{\rm src}(a_m)\right)\Pi_{k'}(\{T_k\}).
	\end{equation}
}
$\Pi_k(\{T_k\})$ is a stationary distribution for each code tree $T_k$, which is determined by the transition probabilities $\{P_{k', k''}\mid k',k''\in \mathbb{Z}^+_{<|\{T_k\}|}\}$ depending on the links $\{\textrm{Link}_{k'}\mid k'\in \mathbb{Z}^+_{<|\{T_k\}|}\}$:
\begin{equation}
	P_{k', k''} = \sum_{\{m\mid \textrm{Link}_{k'}(a_m)=k''\}} p_{\rm src}(a_m).
\end{equation}
When the stationary distribution is not unique, $\Pi_k(\{T_k\})$ gives the one corresponding to a closed subset in $\{T_k\}$ depending on the initial tree and source distribution.
This is because we never switch the code trees between the closed subsets during the encoding/decoding.

According to our previous work \cite{ref:n_aifv}, $N$-bit-delay AIFV codes can represent any VV codes decodable within $N$-bit decoding delay.
Owing to this fact, the global optimum of {\it Minimization problem \ref{mip:totalcode}} becomes the best code among such VV codes.
However, solving {\it Minimization problem \ref{mip:totalcode}} contains two complex problems simultaneously:
\begin{itemize}
	\item Finding a combination of the modes to be used.
	\item Constructing a code forest by assigning codewords and links to each tree.
\end{itemize}
Therefore, we next introduce some partial problems that are more reasonable to solve.
They are still meaningful problems, and defining the classes of optimality based on them helps us understand how far we can approach the optimal code construction.

\subsection{Partial problems and classes of optimality}
One of the partial problems is given by fixing the combination of modes:
\mytheory{mip}{Code construction under fixed modes}{
	\label{mip:modefixed}
	For given $N$, $p_{\rm src}$, $K$, and $\{\tilde{\textsc{Mode}}_k\mid k\in\mathbb{Z}^+_{<K}\}$,
	\begin{equation}
		\mathcal{L}_{\rm fixed}(N, p_{\rm src}, K,\{\tilde{\textsc{Mode}}_k\})\equiv \min_{\{T_k\}} \sum_{k'=0}^{K-1}\left(\sum_{m=0}^{M-1}\|\textrm{Cword}_{k'}(a_m)\|_{\rm len}\cdot p_{\rm src}(a_m)\right)\Pi_{k'}(\{T_k\})
		\label{eq:modefixed_obj}
	\end{equation}
	\begin{center}
		subject to: $\{T_k\}=\{(\textrm{Cword}_k, \textrm{Link}_k, \tilde{\textsc{Mode}}_k)\mid k\in \mathbb{Z}^+_{<K}\}\in\mathbb{LF}_{M,N}$.
	\end{center}
}
It means no other code forest having the modes $\{\tilde{\textsc{Mode}}_k\}$ achieves a shorter expected code length than the solution.
Of course, the achievable code length depends on the fixed modes and may be worse than the global optimum of {\it Minimization problem \ref{mip:totalcode}}.
However, this problem is important since the conventional AIFV-$m$ codes implicitly assume some fixed modes, and we may outperform them if we solve the problem with more modes allowed.

Especially when many modes are allowed, it is difficult to strictly guarantee that the constructed code forests achieve minimum expected code length among any other possible ones with the specified modes $\{\tilde{\textsc{Mode}}_k\}$.
Therefore, let us consider some further modification. We rewrite {\it Minimization problem \ref{mip:modefixed}} using an arbitrary subset of the specified modes:
\mytheory{smip}{Code construction under a subset of fixed modes}{
	\label{mip:modefixedsub}
	For given $N$, $p_{\rm src}$, $K$, $\{\tilde{\textsc{Mode}}_k\mid k\in\mathbb{Z}^+_{<K}\}$, and $A\subseteq \mathbb{Z}^+_{<K}$,
	\begin{equation}
		\mathcal{L}_{\rm subset}(N, p_{\rm src}, K,\{\tilde{\textsc{Mode}}_k\}, A)\equiv \min_{\{\tilde{T}_k\}} \sum_{k'\in A}\left(\sum_{m=0}^{M-1}\|\textrm{Cword}_{k'}(a_m)\|_{\rm len}\cdot p_{\rm src}(a_m)\right)\Pi_{k'}(\{\tilde{T}_k\})
		\label{eq:modefixedsub_obj}
	\end{equation}
	\begin{center}
		subject to: $\{\tilde{T}_k\}=\{(\textrm{Cword}_k, \textrm{Link}_k, \tilde{\textsc{Mode}}_k)\mid k\in A\}\in\mathbb{LF}_{M,N}$.
	\end{center}
}
The above problem is essentially the same as {\it Minimization problem \ref{mip:modefixed}} because it only replaces the modes $\{\tilde{\textsc{Mode}}_k\}$ with their subset.	However, the problem will be easier if we do not specify the subset $A$ in advance and focus on constructing a code forest that is a solution to {\it Minimization problem \ref{mip:modefixedsub}} for some $A$.
If we can guarantee that some subset $A$ exists and the constructed code forest becomes the solution to {\it Minimization problem \ref{mip:modefixedsub}} for such $A$, we can focus on the combination of the modes in investigating further improvement of the codes.
Therefore, even if we cannot control the subset $A$, solving this partial problem is more reasonable than completely heuristic approaches.

Note that
\begin{equation}
	\min_{K,\{\tilde{\textsc{Mode}}_k\}}\mathcal{L}_{\rm fixed}(N, p_{\rm src}, K,\{\tilde{\textsc{Mode}}_k\})=\mathcal{L}_{\rm general}(N, p_{\rm src})
\end{equation}
holds:
The solution to {\it Minimization problem \ref{mip:modefixed}} can be the optimum of {\it Minimization problem \ref{mip:totalcode}}, the main code-construction problem, when we choose the appropriate combination of modes.
It is also
\begin{equation}
	\min_{A\subseteq \mathbb{Z}^+_{<K}}\mathcal{L}_{\rm subset}(N, p_{\rm src}, K,\{\tilde{\textsc{Mode}}_k\}, A)=\mathcal{L}_{\rm fixed}(N, p_{\rm src}, K,\{\tilde{\textsc{Mode}}_k\}):
\end{equation}
The solution to {\it Minimization problem \ref{mip:modefixedsub}} can be the optimum of {\it Minimization problem \ref{mip:modefixed}} when we have the appropriate subset $A$.

Based on the above problems, we define some classes of optimality.
\mytheory{dfn}{Optimality classes of $N$-bit-delay AIFV code construction}{
	Say the decoding delay $N$ and source distribution $p_{\rm src}$ are given.
	\begin{itemize}
		\item \emph{G-optimality}: $\{T_k\}$ is \emph{G-optimal} when it is the minimum of {\it Minimization problem \ref{mip:totalcode}}.
		\item \emph{F-optimality}: For given $K$ and $\{\tilde{\textsc{Mode}}_k\}$, $\{T_k\}$ is \emph{F-optimal} when it is the minimum of {\it Minimization problem \ref{mip:modefixed}}.
		\item \emph{E-optimality}: For given $K$ and $\{\tilde{\textsc{Mode}}_k\}$, $\{T_k\}$ is \emph{E-optimal} when some $A$ exist such that $\{T_k\}$ becomes the minimum of  {\it Minimization problem \ref{mip:modefixedsub}}.
	\end{itemize}
}
\emph{G-optimality} refers to the global optimum of the main code-construction problem,
while \emph{F-optimality} corresponds to the optimum of the partial one.
Fixing the modes limits the freedom of code forests and makes it easier for us to guarantee the optimality.
We show in Section \ref{sec:prop} that, with appropriate modes specified, \emph{F-optimality} becomes sufficient for stationary memoryless sources.

\emph{E-optimality} is a relaxed form of \emph{F-optimality}.
Allowing an arbitrary subset $A$ of the specified modes $\{\tilde{\textsc{Mode}}_k\}$ makes it easier to guarantee. 
Of course, just finding any way to guarantee \emph{E-optimality} is trivial: 
Huffman codes are always \emph{E-optimal} because $\{\text{`$\lambda$'}\} \subseteq \{\tilde{\textsc{Mode}}_k\}$. 
However, if we have some code construction algorithms, such a class of optimality is worth discussing since we can be confident in the constructed codes for the mode set $A$. 
We show that we can guarantee \emph{E-optimality} for some non-trivial $A$ by decomposing the code-forest construction problem into code-tree-wise independent ones.

\subsection{Construction problem decomposition}
Even if the modes are fixed, it is hard to optimize code forests straightforwardly because the stationary distribution $\Pi_k$ depends on all code tree structures.
The links of each tree impact the construction of other trees, and we have to find the best combination of code trees at once.

So, based on the idea in the previous work \cite{ref:mcopt}, we propose a decomposition of the minimization problem.
By introducing a virtual cost $C_k\in\mathbb{R}$, we break down the code-forest construction into independent optimization problems for respective code trees:
\mytheory{mip}{Independent construction of $T_k$}{
	\label{mip:codetreewise}
	For given $K$, $\{\tilde{\textsc{Mode}}_k\mid k\in\mathbb{Z}^+_{<K}\}$, and $\{C_k\mid k\in\mathbb{Z}^+_{<K}\}$,
	\begin{equation}
		\min_{T_k} \sum_{m=0}^{M-1}\left(\|\textrm{Cword}_k(a_m)\|_{\rm len}+C_{\textrm{Link}_k(a_m)}\right)p_{\rm src}(a_m)
	\end{equation}
	\begin{center}
		subject to: $T_k=(\textrm{Cword}_k, \textrm{Link}_k, \tilde{\textsc{Mode}}_k)$ satisfying {\it Rule \ref{rle:decodable}}.
	\end{center}
}
The virtual cost $C_k$ represents the cost of linking to $T_k$.

If some algorithm is available for {\it Minimization problem \ref{mip:codetreewise}}, we can shorten the expected code length of an AIFV code by iteratively minimizing each decomposed objective function and updating the costs.
Therefore, we can focus on the code-tree-wise problems without dealing with the whole trees at once.

The cost updating goes as follows. It is equivalent to the previous work \cite{ref:mcopt} but is a much simpler formulation.
\mytheory{prc}{Cost updating for decomposed construction}{
	\label{prc:iterative}
	Follow the steps below with a given natural number $K$ and modes $\{\tilde{\textsc{Mode}}_k\mid k\in \mathbb{Z}^+_{<K}\}$.
	\begin{enumerate}
		\item Set an initial value ${\bm C}^{(0)}$ and $i=1$.
		\item Get code trees $\{T^{(i)}_k\}=\{(\textrm{Cword}^{(i)}_k, \textrm{Link}^{(i)}_k, \tilde{\textsc{Mode}}_k)\mid k\in\mathbb{Z}^+_{<K}\}$ by solving {\it Minimization problem \ref{mip:codetreewise}} for each tree $T^{(i)}_k$ with given $K$, $\tilde{\textsc{Mode}}_k$, and ${\bm C}^{(i-1)}$.
		\item Get the tree-wise expected code length ${\bm L}^{(i)}$ and the transition matrix $P_{\rm trans}^{(i)}$ respectively by
		\begin{eqnarray}
			L^{(i)}_k &=& \sum_m \left\|\textrm{Cword}^{(i)}_k(a_m)\right\|_{\rm len}\cdot p_{\rm src}(a_m), \\
			P^{(i)}_{k, k'} &=& \sum_{m : \textrm{Link}^{(i)}_k(a_m)=k'} p_{\rm src}(a_m).
		\end{eqnarray}
		\item Find a stationary distribution ${\bm \Pi}^{(i)}$ satisfying Eqs.~(\ref{eq:stabledist1}) and (\ref{eq:stabledist2}) and calculate the expected code length $\bar{L}^{(i)}={\bm \Pi}^{(i){\rm t}}{\bm L}^{(i)}$.
		\item Update the costs as
		\begin{eqnarray}
			C^{(i)}_0 &=& 0\nonumber\\
			{\bm C}^{(i)}_{\{0\}\emptyset} &=& \left(P_{{\rm trans}\{0\}\{0\}}^{(i)}-I_{K-1}\right)^{-1}\left({\bm 1}_{K-1}\bar{L}^{(i)}-{\bm L}^{(i)}_{\{0\}\emptyset}\right).
			\label{eq:cost_update}
		\end{eqnarray}
		\label{sprc:cost_update}
		\item If ${\bm C}^{(i)}_{A^{(i)}_z\emptyset}\neq {\bm C}^{(i-1)}_{A^{(i)}_z\emptyset}$ where $A^{(i)}_z=\{k\mid \Pi_k^{(i)}=0\}$, increment $i$ and return to step b.
		\item Output $\{T_k\}=\{T^{(i)}_k\mid k\in A^{(i)}\}$ where $A^{(i)}=\mathbb{Z}^+_{<K}\setminus A^{(i)}_z$.
	\end{enumerate}
}
Here, we use the following notations to represent the opertions on an arbitrary matrix (or vector) $P$.
Note that all the numbers of rows and columns here start with zero.
\begin{itemize}
	\item $P^{\rm t}$: Transpose of $P$.
	\item $P_{AB}$: Matrix (or vector) given by removing $i$-th row and $j$-th column from $P$ for all $i\in A$ and $j \in B$, where $A,B\subset \mathbb{Z}^+$.
\end{itemize}
The constants and variables are defined as below. 
\begin{itemize}
	\item $I_K$: $K$-by-$K$ identical matrix.
	\item ${\bm 0}_K$ (resp. ${\bm 1}_K$): $K$-th-order column vector with 0 (resp. 1) for every element.
	\item ${\bm L}^{(i)}$: $K$-th-order column vector containing the tree-wise expected code length $L_k^{(i)}$ of $T^{(i)}_k$ in the $k$-th element.
	\item ${\bm C}^{(i)}$: $K$-th-order column vector containing the cost $C_k^{(i)}$ in the $k$-th element.
	\item ${\bm \Pi}^{(i)}$: $K$-th-order column vector containing the stationary distribution $\Pi_k^{(i)}$ in the $k$-th element.
	\item $P_{\rm trans}^{(i)}$: $K$-by-$K$ transition matrix containing the probability $P_{k, k'}^{(i)}$ of the transition from $T^{(i)}_k$ to $T^{(i)}_{k'}$ in the $(k, k')$ element.
	\item $\bar{L}^{(i)}(={\bm \Pi}^{(i){\rm t}}{\bm L}^{(i)})$: Expected code length of the code using all the code trees.
\end{itemize}
The superscript $(i)$ are used for indicating the iteration number in {\it Procedure \ref{prc:iterative}}.

Note that $P_{\rm trans}^{(i)}$ and ${\bm \Pi}^{(i)}$ always satisfies
\begin{eqnarray}
	\label{eq:sumP}
	P_{\rm trans}^{(i)}{\bm 1}_K &=& {\bm 1}_K,\\
	\label{eq:stabledist1}
	{\bm \Pi}^{(i){\rm t}}{\bm 1}_K &=& 1,\\
	\label{eq:stabledist2}
	{\bm \Pi}^{(i){\rm t}}P_{\rm trans}^{(i)} &=& {\bm \Pi}^{(i){\rm t}},
\end{eqnarray}
and ${\bm C}^{(0)}$ can be an arbitrary value in theory. 
When there are several solutions for {\it Minimization problem \ref{mip:codetreewise}} in step b, we can use any of them. 

\subsection{Optimality guaranteed by cost updating}
Even if we focus on the code-tree-wise construction as {\it Minimization problem \ref{mip:codetreewise}},
we can still guarantee some optimality of its solution:
\mytheory{thm}{E-optimality of cost updating}{
	\label{thm:optimum}
	{\it Procedure \ref{prc:iterative}} gives an \emph{E-optimal} code forest $\{T_k\}$ for the given $K\in\mathbb{N}$ and $\{\tilde{\textsc{Mode}}_k\mid k\in\mathbb{Z}^+_{<K}\}$ at a finite iteration if every code tree can reach the code tree $T^{(i)}_0$ in every iteration.
}

Theorem 4.5. b) in Reference \cite{ref:markov_inv} helps us to justify {\it Procedure \ref{prc:iterative}}:
\mytheory{thm}{Sub-Markov matrix invertibility \cite{ref:markov_inv}}{
	\label{thm:detnzero}
	Suppose $S$ is a finite state set of a time-homogeneous Markov chain, $P_K$ is a $K$-by-$K$ matrix containing in the $(k, k')$ element the transition probability from $k$-th to $k'$-th states, and $S'$ is a non-empty subset of $S$.
	If $S\setminus S'$ is an open set, $(P_K-I_K)_{S'S'}$ is invertible.
}
The proof interprets the code forest as a time-homogeneous Markov chain whose states correspond to the respective code trees. \\\\
\underline{\it Proof of Theorem \ref{thm:optimum}}: Let us write the transition matrix as
\begin{equation}
	P_{\rm trans}^{(i)}=\left (
	\begin{array}{cc}
		P^{(i)}_{0,0}           & {\bm P}^{(i)\:{\rm t}}_{0, \ast} \\
		{\bm P}^{(i)}_{\ast, 0} & P_{\rm trans\{0\}\{0\}}^{(i)}
	\end{array}
	\right ).
\end{equation}
When every code tree can reach $T^{(i)}_0$, $\{T^{(i)}_k\mid k\neq 0\}$ is an open set, which makes the solution of Eqs.~(\ref{eq:stabledist1}) and (\ref{eq:stabledist2}) unique owing to Theorem \ref{thm:detnzero}.
Therefore, the stationary distributions are always uniquely determined in this algorithm as long as the assumption holds.

We take the following steps to prove the theorem.
\begin{enumerate}
	\item $\bar{L}^{(i-1)} \ge \bar{L}^{(i)}$ holds for all $i$.
	\item $\bar{L}^{(i-1)} > \bar{L}^{(i)}$ if ${\bm C}^{(i)}_{A^{(i)}_z\emptyset} \neq {\bm C}^{(i-1)}_{A^{(i)}_z\emptyset}$ where $A^{(i)}_z=\{k\mid \Pi_k^{(i)}=0\}$.
	\item When ${\bm C}^{(i)}_{A^{(i)}_z\emptyset}={\bm C}^{(i-1)}_{A^{(i)}_z\emptyset}$, $\bar{L}^{(i)}$ is the minimum value of {\it Minimization problem \ref{mip:modefixedsub}} for $A=A^{(i)}$.
\end{enumerate}
\vspace{5mm}

{\bf a}. The objective function of {\it Minimization problem \ref{mip:codetreewise}} for a code tree $T_k$ is written as
\begin{equation}
	L_k+\sum_{k'}P_{k, k'}\cdot C^{(i-1)}_{k'},
\end{equation}
where $L_k = \sum_m \|\textrm{Cword}_k(a_m)\|_{\rm len}\cdot p_{\rm src}(a_m)$ and $P_{k, k'} = \sum_{\{m\mid \textrm{Link}_k(a_m)=k'\}} p_{\rm src}(a_m)$.
Since $C^{(i-1)}_{k'}$ is a constant for the minimization problem,
\begin{equation}
	\label{eq:obj_single}
	L_k+\sum_{k'\neq k} P_{k,k'}\cdot C^{(i-1)}_{k'} + (P_{k,k}-1)\cdot C^{(i-1)}_{k}
\end{equation}
also takes the minimum at $L_k=L_k^{(i)}$ and $P_{k,k'}=P_{k,k'}^{(i)}$.

Placing Eq.~(\ref{eq:obj_single}) in a row, we can get a vector
\begin{equation}
	\label{eq:obj_vec}
	{\bm f}^{(i)}_{\rm obj}({\bm L}, P_{\rm trans}) \equiv {\bm L} + \left( P_{\rm trans} - I_K \right ){\bm C}^{(i-1)}.
\end{equation}
Every code tree can reach $T^{(i)}_0$ from the assumption, and thus $\{T^{(i)}_k\}\setminus \{T^{(i)}_0\}$ is an open set.
Therefore, {\it Theorem \ref{thm:detnzero}} guarantees $( P^{(i)}_{\rm trans\{0\}\{0\}} - I_{K-1})$ to be invertible for any $i$.
Since ${\bm \Pi}^{(i)}$ comprises only non-negative numbers and each element in ${\bm f}^{(i)}_{\rm obj}({\bm L}^{(i-1)}, P_{\rm trans}^{(i-1)})$ is not smaller than the corresponding element in ${\bm f}^{(i)}_{\rm obj}({\bm L}^{(i)}, P_{\rm trans}^{(i)})$, the inner products hold
\begin{equation}
	\label{eq:inp}
	{\bm \Pi}^{(i){\rm t}}{\bm f}^{(i)}_{\rm obj}({\bm L}^{(i-1)}, P_{\rm trans}^{(i-1)})\ge {\bm \Pi}^{(i){\rm t}} {\bm f}^{(i)}_{\rm obj}({\bm L}^{(i)}, P_{\rm trans}^{(i)}) .
\end{equation}

The right-hand side of Eq.~(\ref{eq:inp}) is written as
\begin{equation}
	\label{eq:Lb_i}
	{\bm \Pi}^{(i){\rm t}} {\bm f}^{(i)}_{\rm obj}({\bm L}^{(i)}, P_{\rm trans}^{(i)})
	= {\bm \Pi}^{(i){\rm t}} {\bm L}^{(i)} + {\bm \Pi}^{(i){\rm t}}\left ( P_{\rm trans}^{(i)} - I_K \right ) {\bm C}^{(i-1)}  = \bar{L}^{(i)}
\end{equation}
using Eq.~(\ref{eq:stabledist2}).
On the other hand, ${\bm f}^{(i)}_{\rm obj}({\bm L}^{(i-1)}, P_{\rm trans}^{(i-1)})$ is given as
\begin{eqnarray}
	{\bm f}^{(i)}_{\rm obj}({\bm L}^{(i-1)}, P_{\rm trans}^{(i-1)})
	&=& {\bm L}^{(i-1)} + \left (
	\begin{array}{c}
		{\bm P}^{(i-1)\:{\rm t}}_{0, \ast}{\bm C}^{(i-1)}_{\{0\}\emptyset} \\
		\left(P_{\rm trans\{0\}\{0\}}^{(i-1)}-I_{K-1}\right){\bm C}^{(i-1)}_{\{0\}\emptyset}
	\end{array}
	\right )\\
	&=& \left (
	\begin{array}{c}
		L^{(i-1)}_0 + {\bm P}^{(i-1)\:{\rm t}}_{0, \ast}\left(P_{{\rm trans}\{0\}\{0\}}^{(i-1)}-I_{K-1}\right)^{-1}\left({\bm 1}_{K-1}\bar{L}^{(i-1)}-{\bm L}^{(i-1)}_{\{0\}\emptyset}\right) \\
		{\bm 1}_{K-1}\bar{L}^{(i-1)}
	\end{array}
	\right ),
\end{eqnarray}
where the transformation in the second element comes from the update rule in {\it Procedure \ref{prc:iterative}} \ref{sprc:cost_update}.
We have from Eq.~(\ref{eq:stabledist2}) that
\begin{equation}
	\left (\Pi_{0}^{(i-1)}, {\bm \Pi}_{\{0\}\emptyset}^{(i-1)\:{\rm t}}\right )
	\left (
	\begin{array}{cc}
		P^{(i-1)}_{0,0}-1         & {\bm P}^{(i-1)\:{\rm t}}_{0, \ast}      \\
		{\bm P}^{(i-1)}_{\ast, 0} & P_{\rm trans\{0\}\{0\}}^{(i-1)}-I_{K-1}
	\end{array}
	\right )
	= {\bm 0}_{K}^{\rm t},
\end{equation}
which leads to
\begin{equation}
	{\bm P}^{(i-1)\:{\rm t}}_{0, \ast}\left ( P_{\rm trans\{0\}\{0\}}^{(i-1)}-I_{K-1}\right )^{-1}
	=-\frac{1}{\Pi_{0}^{(i-1)}}{\bm \Pi}_{\{0\}\emptyset}^{(i-1)\:{\rm t}},
\end{equation}
and thus
\begin{eqnarray}
	\label{eq:Lb_i-1}
	{\bm f}^{(i)}_{\rm obj}({\bm L}^{(i-1)}, P_{\rm trans}^{(i-1)}) &=&
	\left (
	\begin{array}{c}
		L^{(i-1)}_0-\frac{{\bm \Pi}_{\{0\}\emptyset}^{(i-1)\:{\rm t}}{\bm 1}_{K-1}}{\Pi_{0}^{(i-1)}}\bar{L}^{(i-1)}
		+ \frac{1}{\Pi_{0}^{(i-1)}}{\bm \Pi}_{\{0\}\emptyset}^{(i-1)\:{\rm t}}{\bm L}^{(i-1)}_{\{0\}\emptyset} \\
		{\bm 1}_{K-1}\bar{L}^{(i-1)}
	\end{array}
	\right )\nonumber\\
	&=& \left (
	\begin{array}{c}
		L^{(i-1)}_0 + \frac{\Pi_{0}^{(i-1)}-1}{\Pi_{0}^{(i-1)}}\bar{L}^{(i-1)}
		+ \frac{1}{\Pi_{0}^{(i-1)}}\left ( \bar{L}^{(i-1)} - \Pi_{0}^{(i-1)}L_{0}^{(i-1)} \right ) \\
		{\bm 1}_{K-1}\bar{L}^{(i-1)}
	\end{array}
	\right )\nonumber\\
	&=& {\bm 1}_{K}\bar{L}^{(i-1)}.
\end{eqnarray}
Note that $\Pi_{0}^{(i-1)}\neq 0$ because every code tree can reach $T^{(i-1)}_0$.
We can get the inner product as
\begin{equation}
	\label{eq:innerprod}
	{\bm \Pi}^{(i){\rm t}}{\bm f}^{(i)}_{\rm obj}({\bm L}^{(i-1)}, P_{\rm trans}^{(i-1)})
	= \bar{L}^{(i-1)},
\end{equation}
resulting in
\begin{equation}
	\bar{L}^{(i-1)} \ge \bar{L}^{(i)}.
\end{equation}
\\

{\bf b}. Let us assume ${\bm C}^{(i)}_{A^{(i)}_z\emptyset}\neq {\bm C}^{(i-1)}_{A^{(i)}_z\emptyset}$ when $\bar{L}^{(i)}=\bar{L}^{(i-1)}$.
From Eq.~(\ref{eq:Lb_i-1}), we have
\begin{equation}
	\label{eq:if_Li_eq_Li-1}
	{\bm f}^{(i)}_{\rm obj}({\bm L}^{(i-1)}, P_{\rm trans}^{(i-1)})= {\bm 1}_{K} \bar{L}^{(i-1)} = {\bm 1}_{K} \bar{L}^{(i)}= {\bm f}^{(i+1)}_{\rm obj}({\bm L}^{(i)}, P_{\rm trans}^{(i)}).
\end{equation}
$\det(P_{\rm trans\{0\}\{0\}}^{(i)}-I_{K-1})\neq 0$ and $C_0^{(i)}=C_0^{(i-1)}$ gives
\begin{equation}
	{\bm C}^{(i)}\neq {\bm C}^{(i-1)} \Longrightarrow {\bm f}^{(i+1)}_{\rm obj}({\bm L}^{(i)}, P_{\rm trans}^{(i)})\neq {\bm f}^{(i)}_{\rm obj}({\bm L}^{(i)}, P_{\rm trans}^{(i)}).
\end{equation}
It is well-known that a probability matrix of a Markov chain is called irreducible when the states are strongly connected to each other, and there is always a unique stationary distribution that is non-zero for any state \cite{ref:probmat}.
Since every code tree can reach the code tree $T^{(i)}_0$, the ones in $\{T^{(i)}_k\mid k\in A^{(i)}_z\}$ cannot be reached from $T^{(i)}_0$: If $T^{(i)}_0$ can reach them, the transition matrix becomes irreducible, and every $\Pi^{(i)}_k$ would be non-zero.
Due to this fact,
\begin{equation}
	{\bm C}^{(i)}_{A^{(i)}_z\emptyset}\neq {\bm C}^{(i-1)}_{A^{(i)}_z\emptyset} \Longrightarrow {\bm f}^{(i+1)}_{\rm obj}({\bm L}^{(i)}, P_{\rm trans}^{(i)})_{A^{(i)}_z\emptyset}\neq {\bm f}^{(i)}_{\rm obj}({\bm L}^{(i)}, P_{\rm trans}^{(i)})_{A^{(i)}_z\emptyset}
	\label{eq:f_is_neq}
\end{equation}
is given.
Note that this holds because ${\bm C}$ is postmultiplied to $P_{\rm trans}$.

Under the assumption, Eqs.~(\ref{eq:if_Li_eq_Li-1}) and (\ref{eq:f_is_neq}) give
\begin{equation}
	{\bm f}^{(i)}_{\rm obj}({\bm L}^{(i-1)}, P_{\rm trans}^{(i-1)})_{A^{(i)}_z\emptyset} \neq {\bm f}^{(i)}_{\rm obj}({\bm L}^{(i)}, P_{\rm trans}^{(i)})_{A^{(i)}_z\emptyset},
\end{equation}
which means at least one element in ${\bm f}^{(i)}_{\rm obj}({\bm L}^{(i)}, P_{\rm trans}^{(i)})_{A^{(i)}_z\emptyset}$ is smaller than the counterpart of ${\bm f}^{(i)}_{\rm obj}({\bm L}^{(i-1)}, P_{\rm trans}^{(i-1)})_{A^{(i)}_z\emptyset}$ because of Eq.~(\ref{eq:inp}).
The other elements in ${\bm f}^{(i)}_{\rm obj}({\bm L}^{(i)}, P_{\rm trans}^{(i)})_{A^{(i)}_z\emptyset}$ are smaller than or equal to the counterparts of ${\bm f}^{(i)}_{\rm obj}({\bm L}^{(i-1)}, P_{\rm trans}^{(i-1)})_{A^{(i)}_z\emptyset}$,
and all of the elements in ${\bm \Pi}^{(i)}_{A^{(i)}_z\emptyset}$ are larger than 0.
Thus, taking inner products gives
\begin{eqnarray}
	{\bm \Pi}^{(i){\rm t}}_{A^{(i)}_z\emptyset}{\bm f}^{(i)}_{\rm obj}({\bm L}^{(i-1)}, P_{\rm trans}^{(i-1)})_{A^{(i)}_z\emptyset} &>& {\bm \Pi}^{(i){\rm t}}_{A^{(i)}_z\emptyset}{\bm f}^{(i)}_{\rm obj}({\bm L}^{(i)}, P_{\rm trans}^{(i)})_{A^{(i)}_z\emptyset}\nonumber\\
	\iff \bar{L}^{(i-1)} &>& \bar{L}^{(i)},
\end{eqnarray}
which conflicts with $\bar{L}^{(i)}=\bar{L}^{(i-1)}$.
Combining the above fact with step a of this proof, we have
\begin{equation}
	{\bm C}^{(i)}_{A^{(i)}_z\emptyset} \neq {\bm C}^{(i-1)}_{A^{(i)}_z\emptyset}\Longrightarrow \bar{L}^{(i-1)} > \bar{L}^{(i)}.
\end{equation}
\\

{\bf c}. Let us assume $\bar{L}^{(i)}>\mathcal{L}_{\rm subset}(N, p_{\rm src}, K,\{\tilde{\textsc{Mode}}_k\}, A^{(i)})$ when ${\bm C}^{(i)}_{A^{(i)}_z\emptyset}={\bm C}^{(i-1)}_{A^{(i)}_z\emptyset}$.
From this assumption, we can set some feasible code forest $\{T^*_k\}=\{(\textrm{Cword}^*_k, \textrm{Link}^*_k, \tilde{\textsc{Mode}}_k)\mid k\in\mathbb{Z}^+_{<K}\}$ whose subset $\{T^*_k\mid k\in A^{(i)}\}$ belongs to $\mathbb{LF}_{M,N}$ and achieves the minimum expected code length $\bar{L}^{\ast}=\mathcal{L}_{\rm subset}(N, p_{\rm src}, K,\{\tilde{\textsc{Mode}}_k\}, A^{(i)})$ ($<\bar{L}^{(i)}$).

We can define the code-tree-wise expected code lengths, transition matrix, and stable distribution of $\{T^*_k\}$ respectively as ${\bm L}^{\ast}$, $P_{\rm trans}^{\ast}$, and ${\bm \Pi}^{\ast}$.
Since $\{T^*_k\}$ is defined to make its subset $\{T^*_k\mid k\in A^{(i)}\}$ achieve the minimum $\bar{L}^{\ast}$, we can assume without loss of generality that ${\bm \Pi}^{\ast}_{A^{(i)}_z\emptyset}$ adds up to 1.

According to step a of this proof, we can formulate as
\begin{eqnarray}
	\label{eq:obj_vec_l1}
	{\bm 1}_{|A^{(i)}|} \bar{L}^{(i)}&=&\left[{\bm L}^{(i)} +  \left( P_{{\rm trans}}^{(i)} - I_K \right ){\bm C}^{(i)}\right ]_{A^{(i)}_z\emptyset}\\
	\label{eq:obj_vec_l2}
	&=&{\bm L}^{(i)}_{A^{(i)}_z\emptyset}+ \left( P_{{\rm trans}A^{(i)}_zA^{(i)}_z}^{(i)} - I_{|A^{(i)}|} \right ){\bm C}^{(i)}_{A^{(i)}_z\emptyset}\\
	\label{eq:obj_vec_l3}
	&=&{\bm L}^{(i)}_{A^{(i)}_z\emptyset}+\left( P_{{\rm trans}A^{(i)}_zA^{(i)}_z}^{(i)} - I_{|A^{(i)}|} \right ){\bm C}^{(i-1)}_{A^{(i)}_z\emptyset}\\
	\label{eq:obj_vec_l4}
	&=&\left[{\bm L}^{(i)}+  \left( P_{{\rm trans}}^{(i)} - I_K \right ){\bm C}^{(i-1)}\right ]_{A^{(i)}_z\emptyset}.
\end{eqnarray}
The transformations from Eq.~(\ref{eq:obj_vec_l1}) to Eq.~(\ref{eq:obj_vec_l2}) and from Eq.~(\ref{eq:obj_vec_l3}) to Eq.~(\ref{eq:obj_vec_l4}) come from the fact that the code trees in $\{T^{(i)}_k\mid k\in A^{(i)}\}$, including $T^{(i)}_0$, cannot reach the ones in $\{T^{(i)}_k\mid k\in A^{(i)}_z\}$.
Since solving {\it Minimization problem \ref{mip:codetreewise}} in the $i$-th iteration minimizes Eq.~(\ref{eq:obj_vec_l4}), each element does not become smaller when replacing ${\bm L}^{(i)}$ and $P_{\rm trans}^{(i)}$ with ${\bm L}^{\ast}$ and $P_{\rm trans}^{\ast}$.
Therefore, taking an inner product with ${\bm \Pi}^{\ast}_{A^{(i)}_z\emptyset}$, which has only non-negative values and takes a value of 1 in total, leads to an inequality
\begin{eqnarray}
	{\bm \Pi}^{\ast\:{\rm t}}_{A^{(i)}_z\emptyset}{\bm 1}_{|A^{(i)}|} \bar{L}^{(i)} &\le&
	{\bm \Pi}^{\ast\:{\rm t}}_{A^{(i)}_z\emptyset}\left [{\bm L}^{\ast}+\left( P_{\rm trans}^* - I_K \right ){\bm C}^{(i-1)} \right ]_{A^{(i)}_z\emptyset}\nonumber\\
	\iff \bar{L}^{(i)} &\le& \bar{L}^{\ast},
\end{eqnarray}
which conflicts with the assumption $\bar{L}^{\ast}<\bar{L}^{(i)}$.
Thus, $\bar{L}^{(i)}=\mathcal{L}_{\rm subset}(N, p_{\rm src}, K,\{\tilde{\textsc{Mode}}_k\}, A^{(i)})$ when ${\bm C}^{(i)}_{A^{(i)}_z\emptyset}={\bm C}^{(i-1)}_{A^{(i)}_z\emptyset}$. \\

From step b, the cost ${\bm C}$ does not oscillate unless ${\bm C}^{(i)}_{A^{(i)}_z\emptyset}={\bm C}^{(i-1)}_{A^{(i)}_z\emptyset}$.
When we ignore trivial cases, the pattern of the code forest is finite, and thus ${\bm C}^{(i)}_{A^{(i)}_z\emptyset}={\bm C}^{(i-1)}_{A^{(i)}_z\emptyset}$ must happen within a finite iteration.
Therefore, from step c, {\it Procedure \ref{prc:iterative}} always gives an \emph{E-optimal} code forest for given $K$ and $\{\tilde{\textsc{Mode}}_k\}$ within a finite iteration.
$\qquad \blacksquare$\\\\

If we want to claim that it is always ${\bm C}^{(i)} \neq {\bm C}^{(i-1)}\Longrightarrow \bar{L}^{(i-1)} > \bar{L}^{(i)}$,
$\Pi^{(i)}_k$ must be non-zero for every $k$:
If some $\Pi^{(i)}_k$ is zero, the update in $C^{(i-1)}_k\to C^{(i)}_k$ does not influence the value in $\bar{L}^{(i)}$ since the code trees with non-zero stationary probabilities do not use the cost $C^{(i)}_k$.
Due to this fact, when $P_{\rm trans}$ is not irreducible, it can be $\bar{L}^{(i-1)} = \bar{L}^{(i)}$ even if ${\bm C}^{(i)} \neq {\bm C}^{(i-1)}$, resulting in an oscillation of the solution.

In some cases, we can also achieve the further optimality.
The proposed cost updating also allows us to easily check whether it is such a case:
\mytheory{thm}{F-optimality check}{
	\label{thm:optimum_check}
	If we have ${\bm C}^{(i)}={\bm C}^{(i-1)}$ in {\it Procedure \ref{prc:iterative}},
	the output $\{T_k\}$ is \emph{F-optimal} for the given $K$ and $\{\tilde{\textsc{Mode}}_k\mid k\in\mathbb{Z}^+_{<K}\}$.
}
\underline{\it Proof}: If ${\bm C}^{(i)}={\bm C}^{(i-1)}$, for any code forest $\{T^*_k\}=\{(\textrm{Cword}^*_k, \textrm{Link}^*_k, \tilde{\textsc{Mode}}_k)\mid k\in\mathbb{Z}^+_{<K}\}$ with the tree-wise expected code length ${\bm L}^{\ast}$, transition matrix $P_{\rm trans}^{\ast}$, stationary distribution ${\bm \Pi}^{\ast}$, and expected code length $\bar{L}^{\ast}$, we have
\begin{eqnarray}
	\label{eq:fopt}
	\bar{L}^{(i)}&=& {\bm \Pi}^{\ast{\rm t}}{\bm f}^{(i+1)}_{\rm obj}({\bm L}^{(i)}, P_{\rm trans}^{(i)})\\
	&=& {\bm \Pi}^{\ast{\rm t}}{\bm f}^{(i)}_{\rm obj}({\bm L}^{(i)}, P_{\rm trans}^{(i)})
	\le {\bm \Pi}^{\ast{\rm t}}{\bm f}^{(i)}_{\rm obj}({\bm L}^{\ast}, P_{\rm trans}^{\ast})=\bar{L}^{\ast}.
\end{eqnarray}
Therefore, $\bar{L}^{(i)}=\mathcal{L}_{\rm fixed}(N, p_{\rm src}, K,\{\tilde{\textsc{Mode}}_k\})$.
Note that Eq.~(\ref{eq:fopt}) comes from Eq.~(\ref{eq:innerprod}), independent of $A^{(i)}_z$.
$\qquad \blacksquare$\\\\

The cost updating proposed here requires all of the code trees to reach $T^{(i)}_0$.
Otherwise, it stops with an error at the update of ${\bm C}$ because the inverse does not exist.
This assumption may seem strict, and actually, the algorithm fails when the value of ${\bm C}$ is too far from the optimum.
However, if we use practical initial costs shown later, code trees rarely occur which cannot reach $T_0$.
We cannot guarantee they will never happen, but in the experiments shown in Section \ref{sec:eval}, such cases did not occur in any condition.

In fact, we can extend {\it Procedure \ref{prc:iterative}} to deal with code forests forming general Markov chains if we analyze the graphical features of the transition matrix.
The extended cost updating guarantees the \emph{E-optimality} without the assumption of reachability as in {\it Theorem \ref{thm:optimum}} and will never stop even if some code trees cannot reach $T^{(i)}_0$, whose proof is shown in Appendix \ref{app:gen_opt}.
Therefore, we can decompose, in general, the construction problem of $N$-bit-delay AIFV codes into {\it Minimization problem \ref{mip:codetreewise}} if we focus on \emph{E-optimality}.

\section{General properties of optimal $N$-bit-delay AIFV codes}
\label{sec:prop}
\subsection{Mode-wise single tree}
\begin{figure*}[t]
	\begin{center}
		\includegraphics[width=12cm,  bb=0 0 875 476]{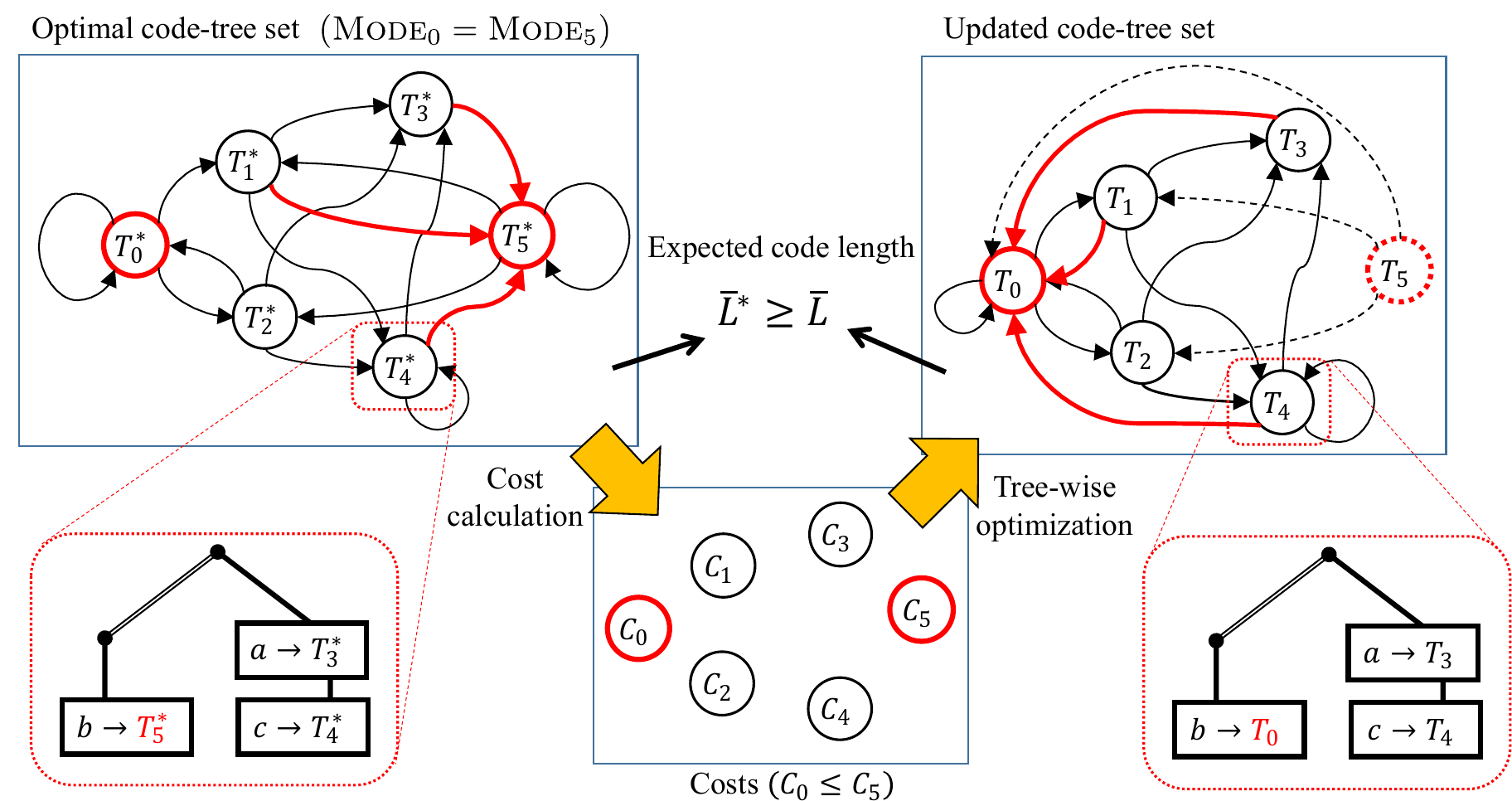}
	\end{center}
	\caption{Outline of {\it Proof of Theorem \ref{thm:mos}}. }
	\label{fig:mos}
\end{figure*}
Understanding what kind of structure the optimal codes should have helps design a reasonable construction algorithm.
It clarifies how much we can reduce the complexity without losing the optimality.
Section \ref{sec:opt} has revealed that considering the code-tree-wise independent problems as {\it Minimization problem \ref{mip:codetreewise}} is sufficient for the \emph{E-optimality} and enables us to check the \emph{F-optimality} empirically.
Here, we show that the gap from the \emph{G-optimality} can be filled.
\mytheory{thm}{Generality of F-optimality}{
	\label{thm:Gopt_if_Fopt}
	For any decoding delay $N$ and stationary memoryless source distribution $p_{\rm src}$
	\begin{equation}
		\text{$\{T^*_k\}$ is \emph{F-optimal} for $K=|\mathbb{BM}_N|$ and $\{\tilde{\textsc{Mode}}_k\}=\mathbb{BM}_N$}\Longrightarrow \text{$\{T^*_k\}$ is \emph{G-optimal}}.
	\end{equation}
}
It means that if we have a sufficient variety of modes, {\it Minimization problem \ref{mip:modefixed}} is enough to get \emph{G-optimal} code forests.
This fact is consistent with the recent results focusing on $N=1,2$ \cite{ref:aifv1_opt_proof,ref:aifv2_opt_proof}.
Additionally, since $|\mathbb{BM}_N|=(2^{2^{N-1}}-1)^2$, it gives a little stricter guarantee than the recent work \cite{ref:kdelay_prop} suggesting that we need at least $2^{2^N}$ coding tables to represent an optimal $N$-bit-delay decodable code.

The above theorem is based on the following fact. 
\mytheory{thm}{Mode-wise single tree}{
	\label{thm:mos}
	Say $\{T^*_k\mid k\in\mathbb{Z}^+_{<K^*}\}$, where $T^*_k=(\textrm{Cword}^*_k, \textrm{Link}^*_k, \textrm{Mode}^*_k)$ and $K^*\ge 2$, is \emph{G-optimal}. If $\textrm{Mode}^*_\kappa=\textrm{Mode}^*_{\kappa'}$ for some $\kappa\neq\kappa'\in\mathbb{Z}^+_{<K^*}$, the solution to {\it Minimization problem \ref{mip:modefixed}} with given $K=K^*-1$ and $\{\textrm{Mode}_k\mid k\in\mathbb{Z}^+_{<K}\}=\{\textrm{Mode}^*_k\mid k\neq \kappa'\}$ can also be \emph{G-optimal}.
}
It implies that a single tree for each mode is enough to make an optimal code forest, which is consistent with the results of some recent work \cite{ref:kdelay_prop}.
The decomposition trick introduced in Section \ref{sec:opt} helps us to prove it, owing to the non-increasing property of the cost updating.
\\\\
\underline{\it Proof of Theorem \ref{thm:mos}}:
Fig.~\ref{fig:mos} describes the outline with an example.
We can assume $\{T^*_k\}$ to be a closed set without loss of generality:
The objective of {\it Minimization problem \ref{mip:totalcode}} depends on the stationary distribution of $\{T^*_k\}$, and we only have to consider the code trees that have non-zero stationary probability;
in cases where the stationary distribution is not unique, there are some closed subsets in $\{T^*_k\}$ unreachable from each other so that we only have to pick up one of them as $\{T^*_k\}$.

Since $\{T^*_k\}$ is a closed set, we can calculate the cost ${\bm C}$ as Eq.~(\ref{eq:cost_update}) in step e of {\it Procedure \ref{prc:iterative}}:
\begin{eqnarray}
	C_0 &=& 0\nonumber\\
	{\bm C}_{\{0\}\emptyset} &=& \left(P^*_{{\rm trans}\{0\}\{0\}}-I_{K^*-1}\right)^{-1}\left({\bm 1}_{K^*-1}\bar{L}^*-{\bm L}^*_{\{0\}\emptyset}\right)
\end{eqnarray}
where the $k$-th element of ${\bm L}^*$ and $(k,k')$ element of $P^*_{{\rm trans}}$ are
\begin{eqnarray}
	L^*_k &=& \sum_m \left\|\textrm{Cword}^*_k(a_m)\right\|_{\rm len}\cdot p_{\rm src}(a_m), \\
	P^*_{k, k'} &=& \sum_{m : \textrm{Link}^*_k(a_m)=k'} p_{\rm src}(a_m),
\end{eqnarray}
and $\bar{L}^*=\mathcal{L}_{\rm general}(N, p_{\rm src})$.

Say $C_\kappa\le C_{\kappa'}$, and
think of optimizing {\it Minimization problem \ref{mip:codetreewise}} for each code tree using $K^*$, $\{\textsc{Mode}^*_k\mid k\in \mathbb{Z}^+_{<K^*}\}$, and the above cost ${\bm C}$.
Since $\textrm{Mode}^*_\kappa=\textrm{Mode}^*_{\kappa'}$, replacing $\textrm{Link}_{k}(a)=\kappa'$ with $\textrm{Link}_{k}(a)=\kappa$, for any $k$ and $a$, always keeps {\it Rule \ref{rle:decodable}} \ref{srle:prefix_free_expansion} and \ref{srle:prefix_in_mode} and never increases the objective function.
Therefore, the code tree $T_k$ with $\textrm{Link}_k\neq \kappa'$ can be the optimum of {\it Minimization problem \ref{mip:codetreewise}} for any $k$.
For the code forest $\{T_k\}$ with such trees, at least one of the possible stationary probabilities of $T_{\kappa'}$ becomes zero.

As in step a of {\it Proof of Theorem \ref{thm:optimum}}, the expected code length $\bar{L}$ of $\{T_k\mid k\neq \kappa'\}$ given by the above tree-wise optimization is guaranteed to be
\begin{equation}
	\bar{L}^*\ge \bar{L}.
\end{equation}
Since $\{T^*_k\}$ is \emph{G-optimal}, $\bar{L}=\bar{L}^*=\mathcal{L}_{\rm general}(N, p_{\rm src})$, and thus $\{T_k\}$ is also \emph{G-optimal}.
Note that step a of {\it Proof of Theorem \ref{thm:optimum}} holds for any stationary distribution of $P^{(i)}_{\rm trans}$.
So, this theory holds even if the stationary distribution of $\{T_k\}$ is not unique.
$\qquad \blacksquare$\\\\
\underline{\it Proof of Theorem \ref{thm:Gopt_if_Fopt}}:
From {\it Rule \ref{rle:decodable}} \ref{srle:basic_mode}, every mode $\textrm{Mode}^*_k$ of an arbitrary \emph{G-optimal} code forest should be a member of $\mathbb{BM}_N$.
If $\{T^*_k\}$ is \emph{F-optimal} for $K=|\mathbb{BM}_N|$ and $\{\tilde{\textsc{Mode}}_k\}=\mathbb{BM}_N$ but is not \emph{G-optimal}, the \emph{G-optimal} code forest must contain two or more code trees with the same mode.
However, if so, we can get the \emph{G-optimal} one from {\it Minimization problem \ref{mip:modefixed}} with $K=|\mathbb{BM}_N|$ and $\{\tilde{\textsc{Mode}}_k\}=\mathbb{BM}_N$ due to {\it Theorem \ref{thm:mos}}, which conflicts with the assumption.
Therefore, $\{T^*_k\}$ is \emph{G-optimal}.
$\qquad \blacksquare$\\\\

In particular, {\it Theorem \ref{thm:mos}} is helpful to analyze useless modes theoretically.
For example, the binary AIFV, or AIFV-2, codes \cite{ref:aifv1, ref:aifv2} belong to a subclass of 2-bit-delay AIFV codes.
They use only two code trees whose modes are respectively $\{$`$\lambda$'$\}$ and $\{$`01', `1'$\}$ while general 2-bit-delay AIFV codes can use 9 patterns of modes as in Fig.~\ref{fig:2bit_mode_patt}.
However, it is recently reported \cite{ref:aifv2_opt_proof} that AIFV-2 codes can achieve the optimal expected code length among the codes decodable with 2 bits of decoding delay.
This fact means that, for stationary memoryless sources, the modes except $\{$`$\lambda$'$\}$ and $\{$`01', `1'$\}$ are useless in minimizing the code length under the condition of 2-bit decoding delay.
We can explain why using {\it Theorem \ref{thm:mos}}, which we leave to Appendix \ref{app:gen_aifv2}.

\subsection{Code-tree symmetry}
Furthermore, we can derive a useful constraint which reduces the freedom of code forests while still keeping the generality. 
\mytheory{thm}{Code-tree symmetry}{
	\label{thm:cos}
	For any decoding delay $N$ and stationary memoryless source distribution $p_{\rm src}$, some \emph{G-optimal} code forest $\{T_k\}$ always exists that satisfies
	\begin{equation}
		\textsc{Mode}_\kappa=\neg \textsc{Mode}_{\kappa'}\Longrightarrow \forall a\in\mathbb{A}_M:(\textrm{Cword}_\kappa(a)=\neg \textrm{Cword}_{\kappa'}(a),  \textsc{Mode}_{\textrm{Link}_{\kappa}(a)}=\neg \textsc{Mode}_{\textrm{Link}_{\kappa'}(a)})
		\label{eq:cos}
	\end{equation}
	for any $\kappa\neq\kappa'$.
}
Some work \cite{ref:red_aifvk} has recently reported on code-table reduction based on the same idea. 
Owing to this theorem, we can omit the optimization processes for code trees having symmetric modes:
We can just reuse the results of {\it Minimization problem \ref{mip:codetreewise}} in step b of {\it Procedure \ref{prc:iterative}} for symmetric modes by bit-flipping the code trees. \\\\
\underline{\it Proof of Theorem \ref{thm:cos}}:
Say $\{T^*_k\mid k\in\mathbb{Z}^+_{<K^*}\}$, where $T^*_k=(\textrm{Cword}^*_k, \textrm{Link}^*_k, \textrm{Mode}^*_k)$ and $K^*\ge 1$, is \emph{G-optimal}.
Similar to {\it Proof of Theorem \ref{thm:mos}}, we assume $\{T^*_k\}$ to be a closed set without loss of generality.
Note that assuming $\textsc{Mode}^*_0=\{$`$\lambda$'$\}$ also does not make any loss of generality:
$\textsc{Mode}^*_0\neq\{$`$\lambda$'$\}$ means there are some codewords unused for representing symbols, which never be the optimum.
$\{T^*_k\}$ may include pairs of code trees not satisfying Eq.~(\ref{eq:cos}) and ones satisfying it.

\begin{figure}[!tb]
	\begin{center}
		\subfigure[Bit-{flipped} copies of Fig.~\ref{fig:trees_ex}. ]{
			\includegraphics[width=10cm,  bb=0 0 805 242]{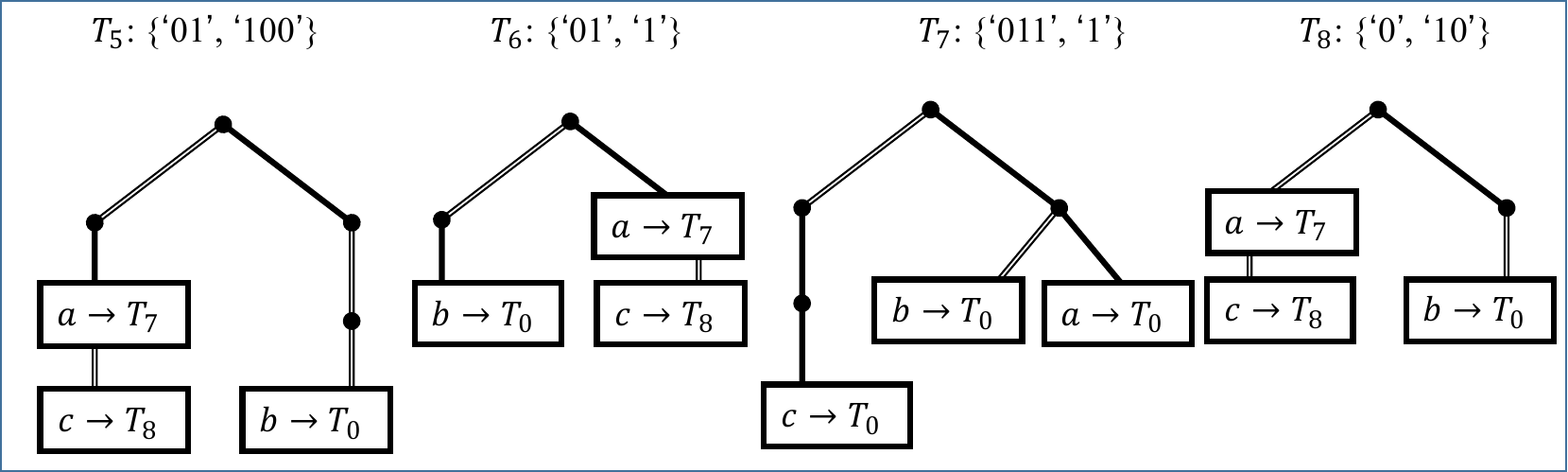}}
		\subfigure[Transition paths of the code trees in Fig.~\ref{fig:trees_ex} and their copies. ]{
			\includegraphics[width=6cm,  bb=0 0 385 227]{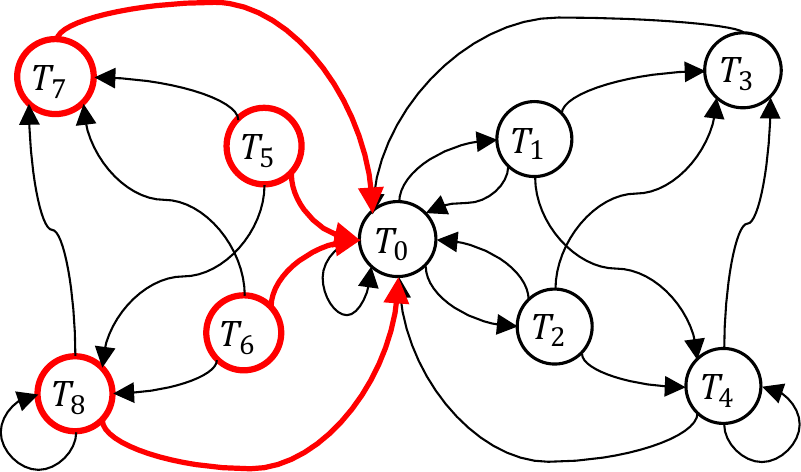}}
	\end{center}
	\caption{Example of the copying operation used in {\it Proof of Theorem \ref{thm:cos}}. }
	\label{fig:cos}
\end{figure}

Let us make some bit-flipped copies of the code trees as
\begin{equation}
	T^*_{K^*+k-1}\equiv (\neg\textrm{Cword}^*_k, \textrm{Link}^*_{K^*+k-1}, \neg\textrm{Mode}^*_k),
\end{equation}
for $1\le k<K^*$, where
\begin{equation}
	\textrm{Link}^*_{K^*+k-1}(a) =
	\left\{
	\begin{array}{cl}
		0                            & \text{(if $\textrm{Link}^*_{k}(a)=0$)} \\
		K^*+\textrm{Link}^*_{k}(a)-1 & \text{(otherwise)}
	\end{array}
	\right.
\end{equation}
for every $a$.
Since all the codewords and modes are bit-{flipped} from $\{T^*_k\}$ and $\textrm{Mode}^*_0=\{\text{`$\lambda$'}\}$,
every copied tree in $\{T^*_{K^*+k-1}\mid k\in\mathbb{Z}^+_{<K^*}\}$ also meets {\it Rule \ref{rle:decodable}} \ref{srle:prefix_free_expansion} and \ref{srle:prefix_in_mode}.
Fig.~\ref{fig:cos} shows an example of this copying by using the code trees in Fig.~\ref{fig:trees_ex}.

Think of the code forest $\{T^*_k\mid k\in\mathbb{Z}^+_{<2K^*-1}\}$.
The code trees in $\{T^*_k\mid k\in\mathbb{Z}^+_{<K^*}\}$ never reach the ones in $\{T^*_{K^*+k-1}\mid k\in\mathbb{Z}^+_{<K^*}\}$.
Besides, every tree in $\{T^*_{K^*+k-1}\mid k\in\mathbb{Z}^+_{<K^*}\}$ {can} reach $T_0$ because the original code forest forms a closed set.
Therefore, the transition matrix of $\{T^*_k\mid k\in\mathbb{Z}^+_{<2K^*-1}\}$ can be written as
\begin{equation}
	\left (
	\begin{array}{ccc}
		P^*_{0,0}             & {\bm P}^{*\:{\rm t}}_{0, \ast} &                             \\
		{\bm P}^{*}_{\ast, 0} & P_{\rm trans\{0\}\{0\}}^{*}    &                             \\
		{\bm P}^{*}_{\ast, 0} &                                & P_{\rm trans\{0\}\{0\}}^{*}
	\end{array}
	\right )
\end{equation}
with the transition matrix $P_{\rm trans}^{*}$ of the original code forest {$\{T^*_k\mid k\in\mathbb{Z}^+_{<K^*}\}$}.
Bit {flipping} never affects the length, and thus the tree-wise expected code lengths of $\{T^*_k\mid k\in\mathbb{Z}^+_{<2K^*-1}\}$ can be written as
\begin{equation}
	\left (
	\begin{array}{c}
		L^*_0                          \\
		{\bm L}^{*}_{\{0\}, \emptyset} \\
		{\bm L}^{*}_{\{0\}, \emptyset}
	\end{array}
	\right )
\end{equation}
using the tree-wise expected code lengths ${\bm L}^{*}$ of the original set.
Additionally, the stationary probabilities of the copied trees $\{T^*_{K^*+k-1}\mid k\in\mathbb{Z}^+_{<K^*}\}$ are zero, and the expected code length of the total set $\{T^*_k\mid k\in\mathbb{Z}^+_{<2K^*-1}\}$ is the same as the global minimum value $\bar{L}^*$.

Since every code tree in $\{T^*_k\mid k\in\mathbb{Z}^+_{<2K^*-1}\}$ {can} reach $T_0$, we can calculate the cost ${\bm C}$ as Eq.~(\ref{eq:cost_update}) in step e of {\it Procedure \ref{prc:iterative}}:
\begin{eqnarray}
	C_0 &=& 0\nonumber\\
	{\bm C}_{\{0\}\emptyset} &=& \left (
	\begin{array}{cc}
		P_{\rm trans\{0\}\{0\}}^{*}-I_{K^*-1} &                                       \\
		& P_{\rm trans\{0\}\{0\}}^{*}-I_{K^*-1}
	\end{array}
	\right )^{-1}\left (
	\begin{array}{c}
		{\bm 1}_{K^*-1}\bar{L}^*-{\bm L}^{*}_{\{0\}, \emptyset} \\
		{\bm 1}_{K^*-1}\bar{L}^*-{\bm L}^{*}_{\{0\}, \emptyset}
	\end{array}
	\right ).
\end{eqnarray}
As a result, the cost given by $\{T^*_k\mid k\in\mathbb{Z}^+_{<2K^*-1}\}$ satisfies
\begin{equation}
	C_k=C_{K^*+k-1}
	\label{eq:cost_equivalence}
\end{equation}
for $1\le k<K^*$.

Think of optimizing {\it Minimization problem \ref{mip:codetreewise}} for each code tree using $K^*$, $\{\textsc{Mode}^*_k\mid k\in\mathbb{Z}^+_{<K^*}\}$, and the above cost ${\bm C}$.
If there are some $1\le \kappa, \kappa'<K^*$ ($\kappa\neq\kappa'$) that have symmetry modes $\textrm{Mode}^*_{\kappa}=\neg\textrm{Mode}^*_{\kappa'}$ {with costs $C_{\kappa}\le C_{\kappa'}$}, there are always $T_{K^*+\kappa-1}$ and $T_{K^*+\kappa'-1}$ that satisfy
\begin{eqnarray}
	\left \{
	\begin{array}{rcl}
		\textrm{Mode}^*_{K^*+\kappa-1} & =                    & \textrm{Mode}^*_{\kappa'}, \\
		C_{K^*+\kappa-1}               & {\le} & C_{\kappa'}
	\end{array}
	\right.,
	\label{eq:cost_mode_symmetry1}\\
	\left \{
	\begin{array}{rcl}
		\textrm{Mode}^*_{\kappa} & =                    & \textrm{Mode}^*_{K^*+\kappa'-1}, \\
		C_{\kappa}               & {\le} & C_{K^*+\kappa'-1}
	\end{array}
	\right.,
	\label{eq:cost_mode_symmetry2}
\end{eqnarray}
owing to Eq.~(\ref{eq:cost_equivalence}).
Since Eqs.~(\ref{eq:cost_mode_symmetry1}) and (\ref{eq:cost_mode_symmetry2}), replacing $\textrm{Link}_{k}(a)=\kappa'$ (resp. $\textrm{Link}_{k}(a)=K^*+\kappa'-1$ ) with $\textrm{Link}_{k}(a)=K^*+\kappa-1$ (resp. $\textrm{Link}_{k}(a)=\kappa$), for any $k$ and $a$, always keeps {\it Rule \ref{rle:decodable}} \ref{srle:prefix_free_expansion} and \ref{srle:prefix_in_mode} and never increases the objective function.

Therefore, the code-tree $T_k$ with $\textrm{Link}_k\neq \kappa', K^*+\kappa'-1$ can be the optimum of {\it Minimization problem \ref{mip:codetreewise}} for any $k$.
Additionally, from Eq.~(\ref{eq:cost_equivalence}), we can say that when $(\textrm{Cword}_{\kappa}, \:\textrm{Link}_{\kappa},\textrm{Mode}^*_\kappa)$ is optimal for $T_\kappa$, $(\neg\textrm{Cword}_{\kappa}, K^*+\textrm{Link}_{\kappa}-1,\textrm{Mode}^*_{K^*+\kappa-1})$ is optimal for $T_{K^*+\kappa-1}$.

So, every code tree in a code forest $\{T_k\}$ satisfying Eq.~(\ref{eq:cos}) can be the optimum of {\it Minimization problem \ref{mip:codetreewise}}.
For the same reason as {\it Proof of Theorem \ref{thm:mos}}, such $\{T_k\}$ is also \emph{G-optimal}.
$\qquad \blacksquare$\\\\

\section{Code-forest construction algorithm}
\label{sec:design}
\subsection{Outline of construction}
\begin{figure}[!tb]
	\begin{center}
		\includegraphics[width=5cm,  bb=0 0 345 363]{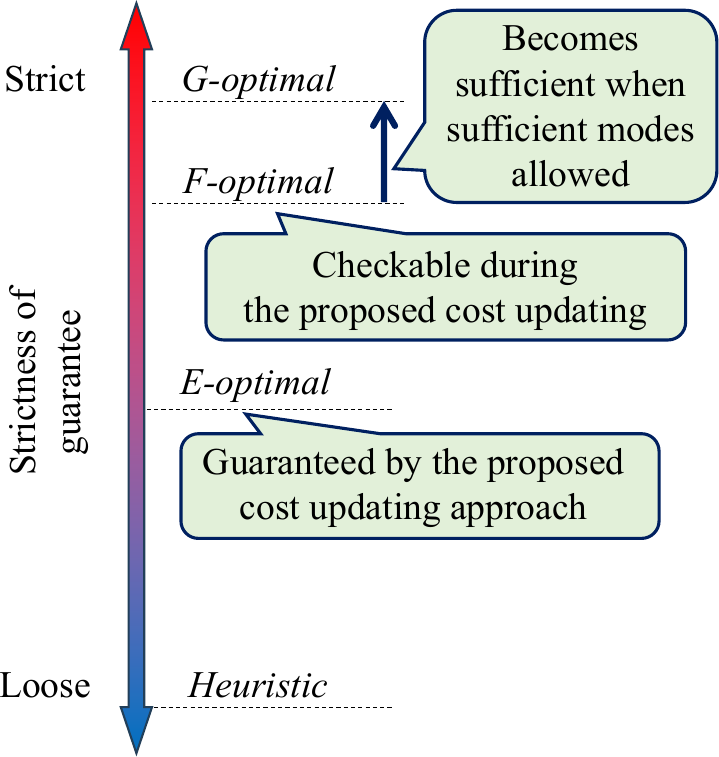}
	\end{center}
	\caption{Mapping of the optimality for code construction. }
	\label{fig:optimality}
\end{figure}
According to the previous discussions, the cost updating for the decomposed construction problem guarantees the output code forest to be \emph{E-optimal}, as illustrated in Fig.~\ref{fig:optimality}.
Additionally, the constructed code forest may be \emph{F-optimal}, and we can empirically check whether it is by {\it Theorem \ref{thm:optimum_check}}.
In case we use every mode in $\mathbb{BM}_N$ as fixed, \emph{F-optimality} can be sufficient for \emph{G-optimality}.

Without disturbing these facts, we can set the following constraints on the code trees to reduce their freedom of design.
\mytheory{rle}{Full code forest condition \cite{ref:n_aifv}}{
	\label{rle:full}
	Code forest $\{T_k\}$ with modes $\{\textsc{Mode}_k\}\subseteq \mathbb{BM}_N$ follows the conditions below.
	\begin{equation}
		\textsc{Mode}_0=\{\text{`$\lambda$'}\} \text{ and }
		\forall k:\textsc{Mode}_k=F_{\rm red}(\textsc{Expands}_k).
		\label{eq:full_tree_cond}
	\end{equation}
}
Non-full code forests have some codewords unused for representing source symbols and thus cannot be optimal.
\mytheory{rle}{CoSMoS condition}{
	\label{rle:cosmos}
	Code forest follows the conditions below.
	\begin{itemize}
		\item Code-tree symmetry (CoS) condition: Every code tree holds Eq.~(\ref{eq:cos}).
		\item Mode-wise single-tree (MoS) condition: Each code tree in the set has a different mode.
	\end{itemize}
}
These conditions are justified by {\it Theorems \ref{thm:mos}} and {\it \ref{thm:cos}} for stationary memoryless sources.
Note that under CoS condition, the encoder and decoder only need to memorize one codebook for each pair of code trees with symmetric modes.

\begin{figure}[!tb]
	\begin{center}
		\includegraphics[width=3cm,  bb=0 0 235 388]{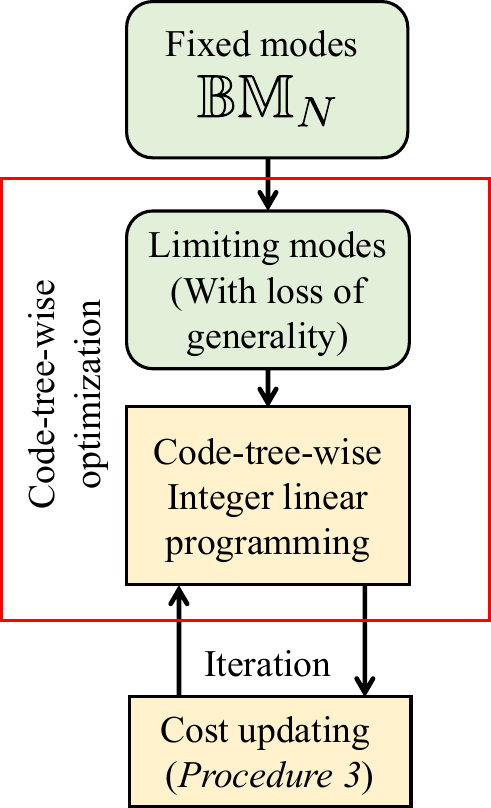}
	\end{center}
	\caption{Outline of the proposed code construction algorithm. }
	\label{fig:alg_outline}
\end{figure}
Based on the above conditions, we introduce a code-forest construction algorithm as in Fig.~\ref{fig:alg_outline}.
Although we can decompose the problem as {\it Minimization problem \ref{mip:codetreewise}},
it is still very complex to solve in general because we need to find for each symbol the combination of codeword and link.
Therefore, we formulate {\it Minimization problem \ref{mip:codetreewise}} as an ILP and use a common integer programming solver based on a similar idea to the previous works \cite{ref:aifv1,ref:ilpcode}.

Here, dealing with $\mathbb{BM}_N$ is too complex to make an ILP.
$N$-bit-delay AIFV codes allow $|\mathbb{BM}_N|=(2^{2^{N-1}}-1)^2$ patterns of modes, which explode by $N$ too rapidly.
Therefore, to make a practically-solvable problem, we should limit the variety of modes.
Indeed, the limitation of modes makes \emph{F-optimality} insufficient for guaranteeing \emph{G-optimality}.
However, it still allows a much broader class of codes than the conventional codes, such as AIFV-$m$ codes, whose advantage will be experimentally shown in Section \ref{sec:eval}.

Note that the proposed construction guarantees not to be worse than Huffman codes.
This is because as long as the assumption in {\it Theorem \ref{thm:optimum}} holds, $\{\tilde{\textsc{Mode}}_k\mid k\in A^{(i)}\}$ includes $\tilde{\textsc{Mode}}_0=\{$`$\lambda$'$\}$, and Huffman codes are code forests of size 1 with a fixed mode $\{$`$\lambda$'$\}$.

\subsection{Code-tree-wise optimization}
\subsubsection{Additional constraint for limiting complexity}
\begin{figure}[!tb]
	\begin{center}
		\includegraphics[width=10cm,  bb=0 0 656 235]{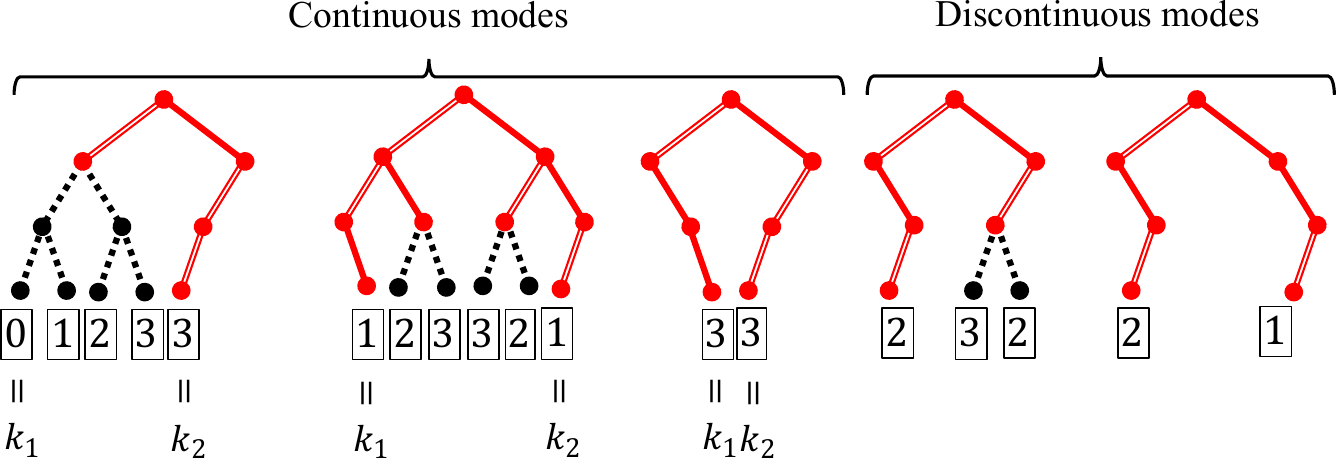}
	\end{center}
	\caption{Examples of continuous and discontinuous modes with numbers identifying their leaves. }
	\label{fig:modes_ex}
\end{figure}
To introduce the additional constraint for modes, we rewrite the rules using intervals in the real-number line, which is a similar approach to the range coding \cite{ref:hbdc}.
{The constraints of decodability can be written in a numerical form \cite{ref:n_aifv} as}
\begin{equation}
	\bigcup_{\textrm{Expcw}\neq\textrm{Expcw}'\in \textsc{Expands}_k} (F_{\rm prob}(\textrm{Expcw})\cap F_{\rm prob}(\textrm{Expcw}'))=\emptyset,
	\label{eq:pi_rule3_a}
\end{equation}
for {\it Rule \ref{rle:decodable}} \ref{srle:prefix_free_expansion}, and
\begin{equation}
	\bigcup_{\textrm{Expcw} \in \textsc{Expands}_k} F_{\rm prob}(\textrm{Expcw}) {\subseteq} \bigcup_{\textrm{Query} \in \textsc{Mode}_k} F_{\rm prob}(\textrm{Query}),
	\label{eq:pi_rule3_b}
\end{equation}
for {\it Rule \ref{rle:decodable}} \ref{srle:prefix_in_mode}.
{Especially, the equality holds for full code forests.}
$F_{\rm prob}$: $\mathbb{W}\to \mathbb{PI}$ is a function outputting a probability interval as
\begin{equation}
	F_{\rm prob}(y_0 y_1 \cdots  y_{L-1})=\myINter{\sum_{i=0}^{L-1}\frac{y_i}{2^{i+1}}, \sum_{i=0}^{L-1}\frac{y_i}{2^{i+1}}+2^{-L}}
\end{equation}
for $y_i \in \{0, 1\}$.
Owing to this formulation, we can check the decodability of the code forests by comparing the intervals in the real-number line.

Based on the interval representation, we define a class of modes:
\mytheory{dfn}{Continuous mode}{
	$\textsc{Mode}\in\mathbb{BM}_N$ is continuous if its probability interval
	\begin{equation}
		\bigcup_{\textrm{Query}\in \textsc{Mode}} F_{\rm prob}(\textrm{Query})
	\end{equation}
	is continuous.
}
Fig.~\ref{fig:modes_ex} shows some examples.
Since every mode in $\mathbb{BM}_N$ can be written as $F_{\rm red}((\text{`0'}\oplus\textsc{Lb})\cup (\text{`1'}\oplus\textsc{Ub}))$ with some $\textsc{Lb},\textsc{Ub}\subset \mathbb{W}_{N-1}$, the modes can be identified using the $N$-bit strings $(\text{`0'}\oplus\textsc{Lb})\cup (\text{`1'}\oplus\textsc{Ub})$.
We number them for convenience:
The string $y_0y_1\cdots y_{N-1}\in\mathbb{W}_N$ ($y_0, y_1,\cdots\in \{0,1\}$) is numbered as
\begin{eqnarray}
	\sum_{i=1}^{N-1}y_i2^{N-1-i}&\text{(if $y_0=0$)}&
	\label{eq:leaf_number_1}\\
	\sum_{i=1}^{N-1}(1-y_i)2^{N-1-i}&\text{(if $y_0=1$)}&
	\label{eq:leaf_number_2}
\end{eqnarray}
Every continuous mode is represented by two groups of sequential numbers both containing $(2^{N-1}-1)$ and thus can be identified by using two numbers $k_1$ and $k_2$:
$k_1$ (resp. $k_2$) is the minimum of Eq.~(\ref{eq:leaf_number_1}) (resp. Eq.~(\ref{eq:leaf_number_2})) among $y_0y_1\cdots y_{N-1}$ included in each mode.
In the following discussions, we write $(k_1,k_2)$ to represent the continuous modes.
Note that the mode $(0,0)$ is $\{$`$\lambda$'$\}$ regardless of $N$.

Using the above definition, we introduce the following condition.
\mytheory{rle}{Continuous mode (CoM) condition}{
	\label{rle:cos}
	Every mode in the code forest is continuous.
}
Continuous modes have $2^{2(N-1)}$ patterns, which is much more limited than the discontinuous ones.
Therefore, CoM condition greatly reduces the complexity of the mode selection.
Moreover, the intervals corresponding to $\textsc{Expand}_k(a)$ also become continuous.
So, when we write as
\begin{eqnarray}
	\textsc{Inter}_k(a) &\equiv& \bigcup_{\textrm{Expcw} \in \textsc{Expand}_k(a)} F_{\rm prob}(\textrm{Expcw}),\\
	\textsc{Inter}_k &\equiv& \bigcup_{\textrm{Query}\in \textsc{Mode}_k} F_{\rm prob}(\textrm{Query}),
\end{eqnarray}
the rule in Eqs.~(\ref{eq:pi_rule3_a}) and (\ref{eq:pi_rule3_b}) can be simplified as
\begin{equation}
	\bigcup_{a\neq a'} (\textsc{Inter}_k(a)\cap \textsc{Inter}_k(a'))=\emptyset,
	\label{eq:pi_rule3_a_alt}
\end{equation}
\begin{equation}
	{\bigcup_{a\in\mathbb{A}_M} \textsc{Inter}_k(a) \subseteq \textsc{Inter}_k},
	\label{eq:pi_rule3_b_alt}
\end{equation}
which can be checked only by comparing the lower and upper bounds of $\textsc{Inter}_k(a)$ and $\textsc{Inter}_k$.
The lower (resp. upper) bound of $\textsc{Inter}_k(a)$ depends only on the codeword $\textrm{Cword}_k(a)$ and the mode number $k_1$ (resp. $k_2$) of the linked mode $\textrm{Mode}_{\textrm{Link}_k(a)}$.
It should also be noted that code trees can be uniquely determined by their modes under MoS condition:
Using CoM condition, we can uniquely write every code tree as $T_{k_1}^{k_2}$ where $(k_1,k_2)$ is its mode.

\subsubsection{Integer programming formulation}
Under CoM condition, the code-tree-wise {\it Minimization problem \ref{mip:codetreewise}} can be rewritten into an ILP problem as follows.
\mytheory{ipp}{Minimization for a code tree of mode $(K_1, K_2)$}
{
	\label{ipp:core}
	\leavevmode\\Variables:
	\begin{equation}
		\begin{array}{c}
			\displaystyle m\in\mathbb{Z}^+_{<M}, \:\:\:\:
			d\in \mathbb{Z}^+_{<D+1}, \:\:\:\:
			k_1,k_2\in \mathbb{Z}^+_{<2^{N-1}}, \:\:\:\:
			i\in \mathbb{Z}^+_{<D}, \:\:\:\:
			j\in\{1,2\} \\
			\displaystyle t_{m,d},\:\:u_{m,k_1,k_2},\:\:v_{m,m'},\:\:{v_{{\rm L},m}},\:\:v_{m,{\rm R}},\:\:
			w_{m,i},\:\:\bar{w}_{m,i}\:\:\in \{0,1\}, \:\:\:\:
			k_{j,m,d}\in \mathbb{Z}^+_{<2^{N-1}}
		\end{array}
	\end{equation}
	Objective function:
	\begin{equation}
		\sum_{m,d} p_{\rm src}(a_m)\cdot d\cdot t_{m,d}
		+ \sum_{m,k_1,k_2} p_{\rm src}(a_m)\cdot C_{k_1,k_2}\cdot u_{m,k_1,k_2}
	\end{equation}
	Subject to
	\begin{eqnarray}
		\label{eq:cw_consis1}
		w_{m,i}+\bar{w}_{m,i}&\le& \displaystyle 1\:\:\:\text{for all $(m,i)$} \\
		\label{eq:cw_consis2}
		w_{m,i+1} + \bar{w}_{m,i+1} - w_{m,i} - \bar{w}_{m,i} &\le&  0 \:\:\:\text{for all $m$ and $i<D-1$}
	\end{eqnarray}
	\begin{eqnarray}
		\label{eq:pick_ou1}
		\sum_d t_{m,d} = 1, \:\:\:\:\sum_{k_1,k_2} u_{m,k_1,k_2} = 1, \:\:\:\:
		\sum_m {v_{{\rm L},m}} = 1, \:\:\:\:\sum_m v_{m,{\rm R}} = 1\\
		\label{eq:pick_ou2}
		\sum_{m'} v_{m',m} + {v_{{\rm L},m}} = 1, \:\:\:\:\sum_{m'} v_{m,m'} + v_{m,{\rm R}} = 1, \:\:\:\:
		v_{m,m} = 0\:\:\:\text{for all $m$}
	\end{eqnarray}
	\begin{eqnarray}
		\label{eq:d_consis1}
		\sum_{i}w_{m,i}+\sum_i \bar{w}_{m,i} - \sum_{d} d\cdot t_{m,d} &=& \displaystyle 0\:\:\:\text{for all $m$} \\
		\label{eq:d_consis2}
		k_{j,m,d} - (2^{N-1}-1)\cdot t_{m,d} &\le& \displaystyle 0\:\:\:\text{for all $(m,d,j)$}\\
		\label{eq:d_consis3}
		\sum_{k_1,k_2} k_j\cdot u_{m,k_1,k_2} - \sum_d k_{j,m,d} &=& \displaystyle 0\:\:\:\text{for all $(m,j)$}
	\end{eqnarray}
	{
		\begin{equation}
			\label{eq:color1}
			-\sum_i \frac{1}{2^{i+1}}\cdot \bar{w}_{m,i} - \sum_{i}\frac{1}{2^{i+1}}\cdot w_{m',i}
			-\sum_d \frac{1}{2^{d+N}}\cdot k_{2,m,d} -\sum_d \frac{1}{2^{d+N}}\cdot k_{1,m',d}
			+ v_{m,m'}  \le   0\:\:\:\text{for all $m\neq m'$} 
		\end{equation}
		\begin{equation}
			\label{eq:color2L}
			-\sum_i \frac{1}{2^{i+1}}\cdot w_{m,i}-\sum_d \frac{1}{2^{d+N}}\cdot k_{1,m,d} + v_{{\rm L},m}
			\le   1-\frac{K_1}{2^N}\:\:\:\text{for all $m$} 
		\end{equation}
		\begin{equation}
			\label{eq:color2R}
			-\sum_i \frac{1}{2^{i+1}}\cdot \bar{w}_{m,i}-\sum_d \frac{1}{2^{d+N}}\cdot k_{2,m,d} + v_{m,{\rm R}}
			\le  1-\frac{K_2}{2^N}\:\:\:\text{for all $m$} 
		\end{equation}
	}
}
$C_{k_1,k_2}$ is a cost for selecting $T_{k_1}^{k_2}$ as the next code tree.
$t_{m,d}$ takes a value of 1 only if the node assigned with the source symbol $a_m$ is located at depth $d$ of the code tree.
$u_{m,k_1,k_2}$ takes a value of 1 only if the code tree switches to $T_{k_1}^{k_2}$ after encoding $a_m$.
$v_{m, m'}$ takes a value of 1 only if {there is no probability interval between} the upper bound of the probability interval corresponding to $a_m$ and the lower bound of the one corresponding to $a_{m'}$.
$v_{{\rm L}, m}$ (resp. $v_{m, {\rm R}}$) takes a value of 1 only if {there is no probability interval between} the lower (resp. upper) bounds of the probability intervals corresponding to $a_m$ and the mode of the code tree.
$w_{m,i}$ (resp. $\bar{w}_{m,i}$) is the $i$-th code symbol (resp. bit-{flipped} code symbol) in the codeword for $a_m$.
When the codeword is shorter than $i$, both $w_{m,i}$ and $\bar{w}_{m,i}$ take a value of 0.
$k_{j, m,d}$ takes the value of $k_j$ of the mode of the code tree to which the encoder switches after encoding $a_m$ if $a_m$ is located at depth $d$ of the code tree and 0 otherwise.
$D$ is a predetermined maximum depth of the code tree, depending on the distribution of the source:
It is about several times as large as $\log_2 M$.

Eq.~(\ref{eq:cw_consis1}) forces $\bar{w}_{m,i}$ to be the bit-{flipped} code symbol of $w_{m,i}$.
It uses an inequality to allow $w_{m,i}=\bar{w}_{m,i}=0$, when the depth $i$ is deeper than the node assigned with the source symbol $a_m$:
When the codeword for $a_m$ is shorter than $i$, $w_{m,i}+\bar{w}_{m,i}=0$, otherwise $w_{m,i}+\bar{w}_{m,i}=1$.
Eq.~(\ref{eq:cw_consis2}) forces $w_{m,i+1}+\bar{w}_{m,i+1}$ to be 0 when $w_{m,i}+\bar{w}_{m,i}$ is 0, which guarantees that $w_{m,i}+\bar{w}_{m,i}$ for $i$ larger than the codeword length always becomes 0.
Eqs.~(\ref{eq:pick_ou1}) and (\ref{eq:pick_ou2}) force the variable sets $\{t_{m,d}\mid d\}$, $\{u_{m,k_1,k_2}\mid k_1,k_2\}$, $\{v_{{\rm L}, m}\mid m\}$, $\{v_{m',m}\mid m'\}\cup \{v_{{\rm L}, m}\}$, $\{v_{m, {\rm R}}\mid m\}$, and $\{v_{m,m'}\mid m'\}\cup \{v_{m, {\rm R}}\}$ to have only one non-zero member, respectively.
The summation of $w_{m,i}+\bar{w}_{m,i}$ in Eq.~(\ref{eq:d_consis1}) becomes equivalent to the node assigned with the source symbol $a_m$.
Eq.~(\ref{eq:d_consis1}) makes $t_{m,d}$ to be 1 at $d$ representing the depth of $a_m$'s node.
Eq.~(\ref{eq:d_consis2}) ensures $k_{j, m,d}$ becomes 0 for the depth $d$ that does not represent the one of $a_m$'s node.
Eq.~(\ref{eq:d_consis3}) ensures $u_{m,k_1,k_2}$ becomes 1 for the mode $(k_1,k_2)$ only if the code tree switches to $T_{k_1}^{k_2}$ after encoding $a_m$.

Eq.~(\ref{eq:color1}) corresponds to Eq.~(\ref{eq:pi_rule3_a_alt}) based on {\it Rule \ref{rle:decodable}} \ref{srle:prefix_free_expansion}.
If the codeword assigned to $a_m$ is `$w_{m,0}w_{m,1}\cdots w_{m, d'-1}$' and the code tree of mode $(k_{1,m,d'},k_{2,m,d'})$ is selected as the next code tree, the union of the probability intervals of the respective expanded codeword becomes
\begin{equation}
	\textsc{Inter}_{k_{1,m,d'},k_{2,m,d'}}(a_m)=\myInter{\sum_i \frac{w_{m,i}}{2^{i+1}}+\sum_d \frac{k_{1,m,d}}{2^{d+N}},\:\:1-\sum_i \frac{\bar{w}_{m,i}}{2^{i+1}}-\sum_d \frac{k_{2,m,d}}{2^{d+N}}}.
\end{equation}
This is because $w_{m,i}$, $\bar{w}_{m,i}$, $k_{1,m,d}$, and $k_{2,m,d}$ are 0 at $i\ge d'$ and $d\neq d'$.
{The inequality in Eq.~(\ref{eq:color1}) is made from}
\begin{equation}
	\left(1-\sum_i \frac{\bar{w}_{m,i}}{2^{i+1}}-\sum_d \frac{k_{2,m,d}}{2^{d+N}}\right) - \left(\sum_i \frac{w_{m',i}}{2^{i+1}}+\sum_d \frac{k_{1,m',d}}{2^{d+N}}\right) \le 1- v_{m,m'},
\end{equation}
{which forces the upper bound for $a_m$ not to be higher than the lower bound for $a_{m'}$ only when $v_{m,m'}=1$ and otherwise become trivial constraints.
	Eqs.~(\ref{eq:color2L}) and (\ref{eq:color2R}) are for Eq.~(\ref{eq:pi_rule3_b_alt}) based on {\it Rule \ref{rle:decodable}} \ref{srle:prefix_in_mode}, derived in a similar way to Eq.~(\ref{eq:color1}).}

{Although the constraints in the above ILP formulation are enough to follow {\it Rule \ref{rle:decodable}}, we know that the code forests should be full, as stated above. Therefore, forcing the code forests to be full practically hastens the convergence and enhances the numerical stability. Such conditions can be realized by making the upper bound for $a_m$ and the lower bound for $a_{m'}$ equivalent when $v_{m,m'}=1$:}
\begin{equation}
	\left(1-\sum_i \frac{\bar{w}_{m,i}}{2^{i+1}}-\sum_d \frac{k_{2,m,d}}{2^{d+N}}\right) - \left(\sum_i \frac{w_{m',i}}{2^{i+1}}+\sum_d \frac{k_{1,m',d}}{2^{d+N}}\right) \ge -1+ v_{m,m'},
\end{equation}
{giving the constraint 
	\begin{equation}
		\label{eq:color1_sub}
		\sum_i \frac{1}{2^{i+1}}\cdot \bar{w}_{m,i} + \sum_{i}\frac{1}{2^{i+1}}\cdot w_{m',i}
		+\sum_d \frac{1}{2^{d+N}}\cdot k_{2,m,d} +\sum_d \frac{1}{2^{d+N}}\cdot k_{1,m',d}
		+ v_{m,m'} \le   2\:\:\:\text{for all $m\neq m'$}.
	\end{equation}
	Similarly, we have the additional constraints 
	\begin{equation}
		\label{eq:color2L_sub}
		\sum_i \frac{1}{2^{i+1}}\cdot w_{m,i}+\sum_d \frac{1}{2^{d+N}}\cdot k_{1,m,d} + v_{{\rm L},m}
		\le   1+\frac{K_1}{2^N}\:\:\:\text{for all $m$},
	\end{equation}
	\begin{equation}
		\label{eq:color2R_sub}
		\sum_i \frac{1}{2^{i+1}}\cdot \bar{w}_{m,i}+\sum_d \frac{1}{2^{d+N}}\cdot k_{2,m,d} + v_{m,{\rm R}}
		\le   1+\frac{K_2}{2^N}\:\:\:\text{for all $m$},
	\end{equation}
	corresponding to Eqs.~(\ref{eq:color2L}) and (\ref{eq:color2R}).
	In the evaluations in Section \ref{sec:eval}, we added Eqs.~(\ref{eq:color1_sub}), (\ref{eq:color2L_sub}), and (\ref{eq:color2R_sub}) to {\it ILP problem \ref{ipp:core}} in constructing the code forests. 
} 

It is well-known that ILP problems can be solved with a finite number of steps, for example, by the cutting-plane algorithm \cite{ref:ilp_converge1, ref:ilp_converge2}.
Therefore, by combining the ILP problem above with the cost updating presented previously, we can get an \emph{E-optimal} code forest within a finite number of steps.

The costs $C_{k_1,k_2}$ are updated in each iteration.
Although their initial values can be set to arbitrary numbers, it is preferable to use the value near the optimum.
The union of the probability interval allowed for a code tree gets limited when $k_1$ and $k_2$ become larger.
The interval on the right-hand side of the constraint in Eq.~(\ref{eq:pi_rule3_b_alt}) is $\myinter{K_1/2^N,1-K_2/2^N}$, and thus the codewords given by $T_{K_1}^{K_2}$ will be about $-\log_2 (1-K_1/2^N-K_2/2^N)$ bits longer compared to those given by $T_0^0$.
According to this fact, it is reasonable to set the initial values as
\begin{equation}
	\label{eq:init_C}
	C_{k_1,k_2}= N-\log_2(2^N - k_1- k_2).
\end{equation}
However, the above values assume all nodes have the same weight, which does not hold for general switching rules.
This is why we need the iterative update of the costs to get the optimum.

Indeed, the above ILP problem still requires very high computational complexity compared to the prior works.
Since it is a combinatorial optimization, in essence, it becomes rapidly complex, especially when $M$ increases.
However, once we can construct the code tree, the encoding and decoding can be realized with simple procedures.
Moreover, as we show in Section \ref{sec:eval}, the proposed code has the potential for high compression efficiency.

\subsubsection{Relationship with conventional codes}
To understand whether CoM condition is reasonable, we here compare the constraint with the conventional codes by interpreting them as AIFV ones by breaking down them into symbol-wise coding rules \cite{ref:n_aifv}.
Under such interpretation, CoM condition does not necessarily hold for the extended Huffman codes but for the arithmetic codes.
Note that the arithmetic codes also satisfy a condition that their codewords preserve the lexicographic order of source sequences. 
The conventional AIFV-$m$ codes implicitly use it, too. The relationship between the arithmetic and AIFV codes has been reported in the previous work \cite{ref:aifv_arith1}. 

Of course, {\it Theorem \ref{thm:Gopt_if_Fopt}} does not hold under CoM condition, and we cannot guarantee the \emph{G-optimality} even if we can get \emph{F-optimal} code forests.
However, CoM condition is expected to be reasonable enough because it is implicitly used in the practical method as the arithmetic codes.

\subsubsection{Construction of AIFV-$m$ codes}
\label{ssec:aifvm_constraint}
The conventional AIFV-$m$ codes is a special case of $N$-bit-delay AIFV codes,
the case where $N=m$ with their code trees limited to modes $(0,0)$ and $(2^n,0)$ for $n=0,1,\cdots,N-2$.
Therefore, the proposed algorithm also enables us to construct optimal AIFV-$m$ codes:
Only adding the following constraint to {\it ILP problem \ref{ipp:core}} will do.
\begin{equation}
	u_{m,0,k_2}+\sum_{n=0}^{N-2} u_{m,2^n,k_2} = 1\:\:\:\text{for all $m$}.
	\label{eq:aifvm_constraint}
\end{equation}
Note that there are methods for constructing AIFV-$m$ codes with lower computational costs: We can use a dynamic programming approach for AIFV-$m$ codes\cite{ref:opt_aifvm_dp, ref:aifvm_dp_speed}, and especially AIFV-$2$ codes can be constructed by a polynomial time algorithm \cite{ref:poly_aifv2}. 
Speeding up the proposed code construction is one of the challenges remaining. 
The recent work on code table reduction \cite{ref:red_aifvk} may be useful for further improvement.

\section{Evaluations}
\label{sec:eval}
\subsection{Asymptotic expected code length}

\begin{figure}[!tb]
	\begin{center}
		\subfigure[AIFV-$m$ codes for $m=2,3,4,5,6$.]{
			\includegraphics[width=8.5cm,  bb=0 0 824 519]{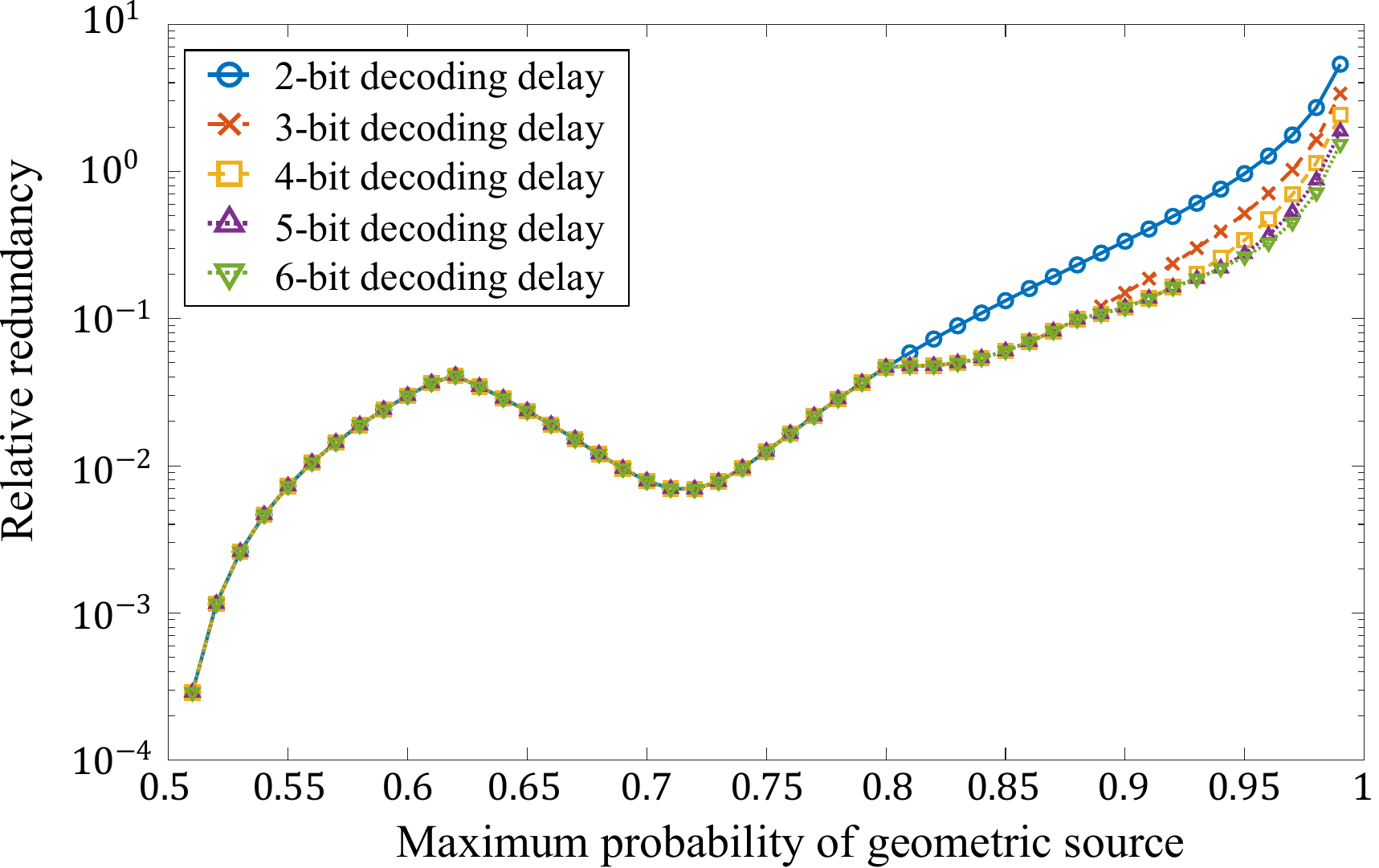}}
		\subfigure[Proposed $N$-bit-delay AIFV codes for $N=2,3,4,5,6$.]{
			\includegraphics[width=8.5cm,  bb=0 0 824 519]{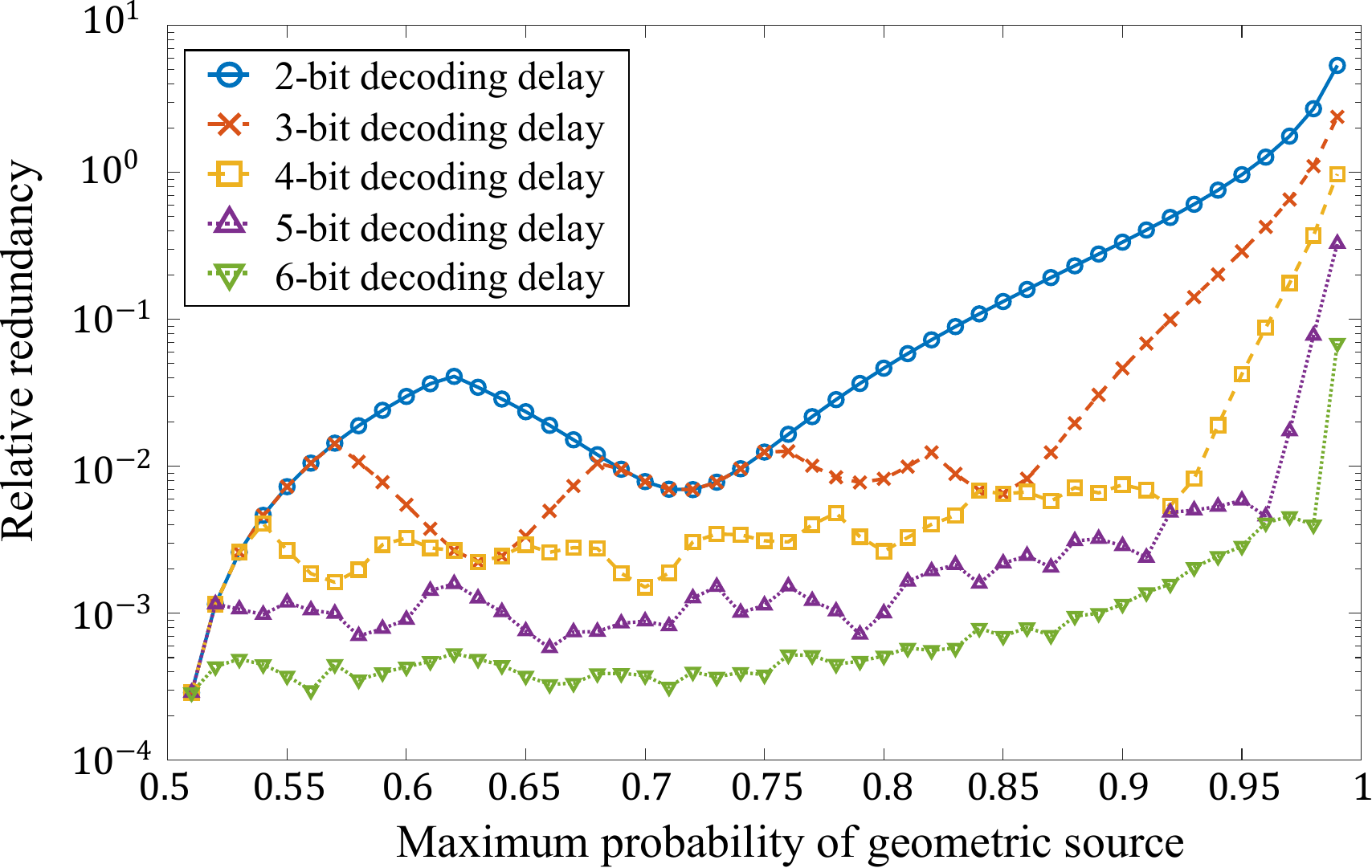}}
		\subfigure[Comparison with the extended Huffman codes.]{
			\includegraphics[width=8.5cm,  bb=0 0 824 519]{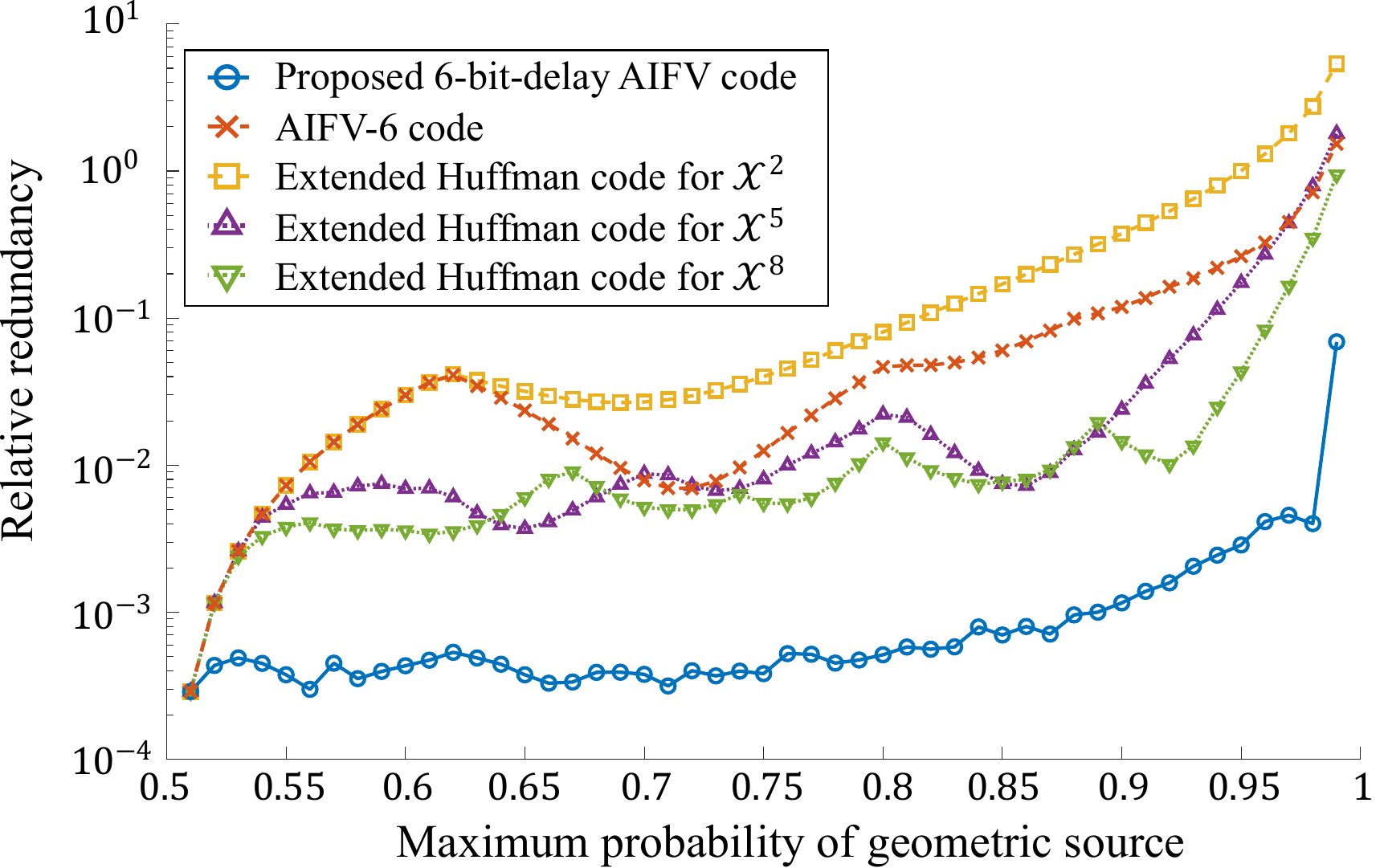}}
	\end{center}
	\caption{Theoretical relative redundancy of the constructed codebooks in binary-source cases.}
	\label{fig:comp_dict_geo}
\end{figure}

\subsubsection{Comparison of codebooks}
To evaluate the compression performance, we first compared codebooks for binary source symbols $\mathcal{X}=\mathbb{A}_2$.
It is a very simple case, but the constructed codebooks can be used for non-binary exponential sources, which play essential roles in practical use.
If we want to apply the codebooks to such cases, we only need to represent the input source symbol by unary before encoding.
Each bit in the unary representation of exponential sources behaves as an independent Bernoulli trial, which is optimally encoded by the codebook for $\mathbb{A}_2$.

For a variety of binary random sources, where $p_{\rm src}(a_0)=0.51,0.52,\cdots,0.99$ ($p_{\rm src}(a_1)=1-p_{\rm src}(a_0)$), the codebooks of AIFV-$m$, $N$-bit-delay AIFV, and the extended Huffman codes are respectively constructed.
Their theoretical relative redundancy is calculated as $(L_{\rm code}/H_{\rm src}-1)$, where $L_{\rm code}$ and $H_{\rm src}$ are the expected code length and the source entropy \cite{ref:hamming}, respectively.
The codebooks of AIFV-$m$ and $N$-bit-delay AIFV were given by the proposed construction algorithm using the constraint in Eq.~(\ref{eq:aifvm_constraint}) for AIFV-$m$ codes.
We here used the relative redundancy to make the results easy to compare, but there were similar relative merits even if we used the absolute redundancy.

Fig.~\ref{fig:comp_dict_geo} (a) plots the results for the conventional AIFV-$m$ codes of each $m$.
Note that $m$ corresponds to the decoding delay for AIFV-$m$ codes.
Even if we permit a longer decoding delay, the theoretical performance increases only for the higher values of $p_{\rm src}(a_0)$.
AIFV-$m$ codes must assign $a_0$ to the root to take advantage of the allowed delay.
In this case, the codeword for $a_1$ must be $m$-bit length, which does not fit the lower value cases well.

Besides, the proposed $N$-bit-delay AIFV codes in (b) show much higher performance by permitting a longer decoding delay.
Note that $2$-bit-delay AIFV codes showed exactly the same performance as AIFV-$2$ codes in any case.
This is because, as explained previously, AIFV-$2$ codes are sufficient to achieve \emph{G-optimality} for the 2-bit-delay condition.

A comparison among the codes is presented in (c).
Since the extended Huffman codes are not designed to limit the decoding delay, we compared them based on their codebook size.
Here, to make a roughly fair comparison, we defined the codebook size as the number of pairs of source symbols and codewords to be memorized.
Basically, the codebook size was calculated by multiplying the number of code trees in the constructed code forest and the number of source symbols.
However, owing to {\it Theorem \ref{thm:cos}}, the code trees with symmetric modes can be represented by bit-flipping, and thus we counted as one code tree for such a pair.

In this comparison, the number of code trees for each source was at most $90$, and thus, the maximum codebook size was $180$.
So, we compared them with the extended Huffman codes for $\mathcal{X}^2$, $\mathcal{X}^5$, and $\mathcal{X}^8$, whose maximum codebook size is $256$.
The proposed $6$-bit-delay AIFV codes outperformed the conventional AIFV-$6$ and extended Huffman codes at $p_{\rm src}(a_0) \ge 0.52$.
We can also compare the results with Golomb-Rice codes \cite{ref:grc1,ref:grc2}, well-known codes effective for exponential sources,
by checking the results in the previous works \cite{ref:xdgr,ref:xdg,ref:grc3}.
The proposed $6$-bit-delay AIFV codes are much more efficient than Golomb-Rice codes except for the sparse sources with high $p_{\rm src}(a_0)$.
AIFV codes need more decoding delay to compress such sources efficiently.

\begin{figure}[!tb]
	\begin{center}
		\subfigure[Comparison for $P_\mathcal{X}^{(0)}$.]{
			\includegraphics[width=8.5cm,  bb=0 0 896 519]{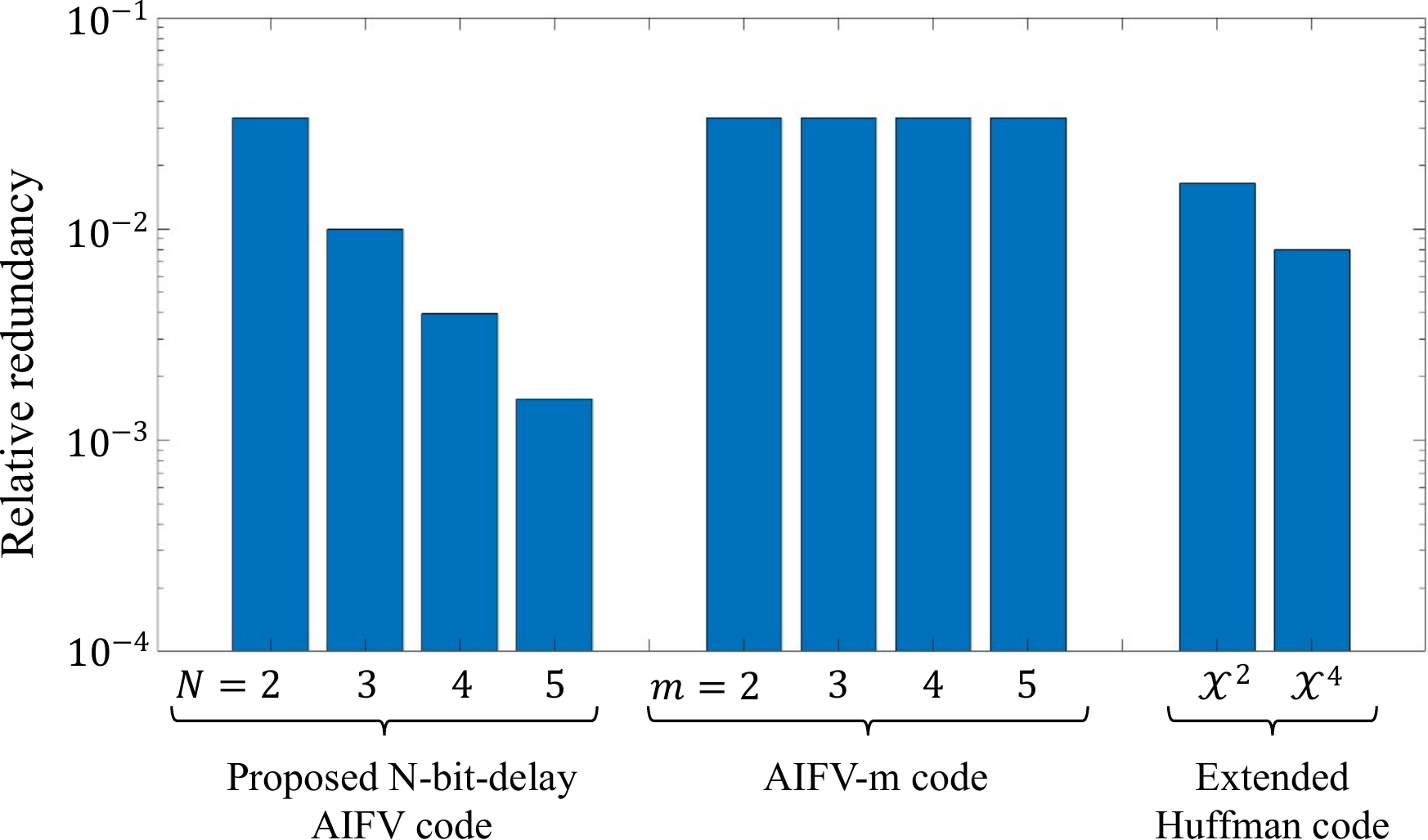}}
		\subfigure[Comparison for $P_\mathcal{X}^{(1)}$. ]{
			\includegraphics[width=8.5cm,  bb=0 0 896 519]{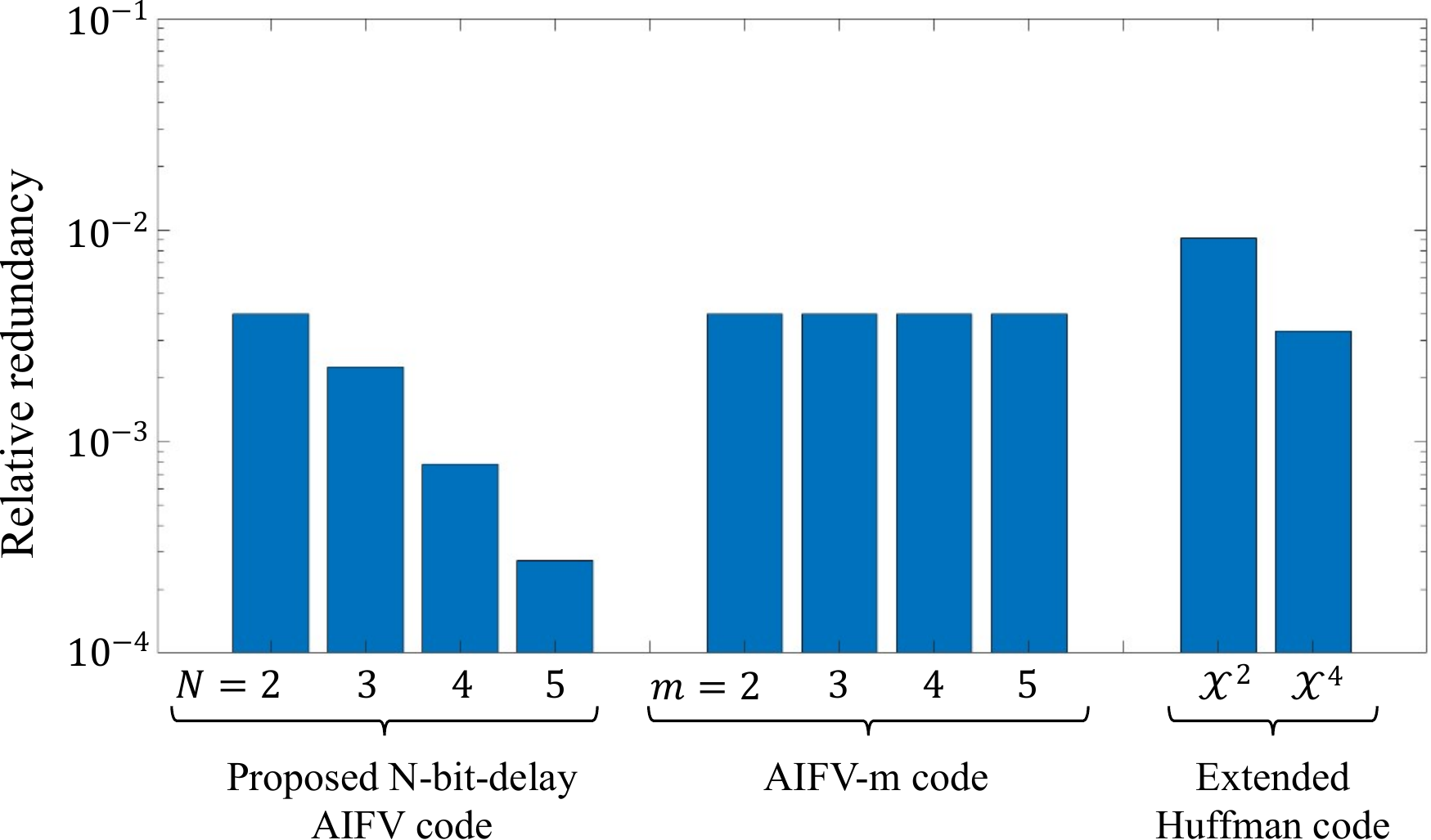}}
		\subfigure[Comparison for $P_\mathcal{X}^{(2)}$.]{
			\includegraphics[width=8.5cm,  bb=0 0 896 519]{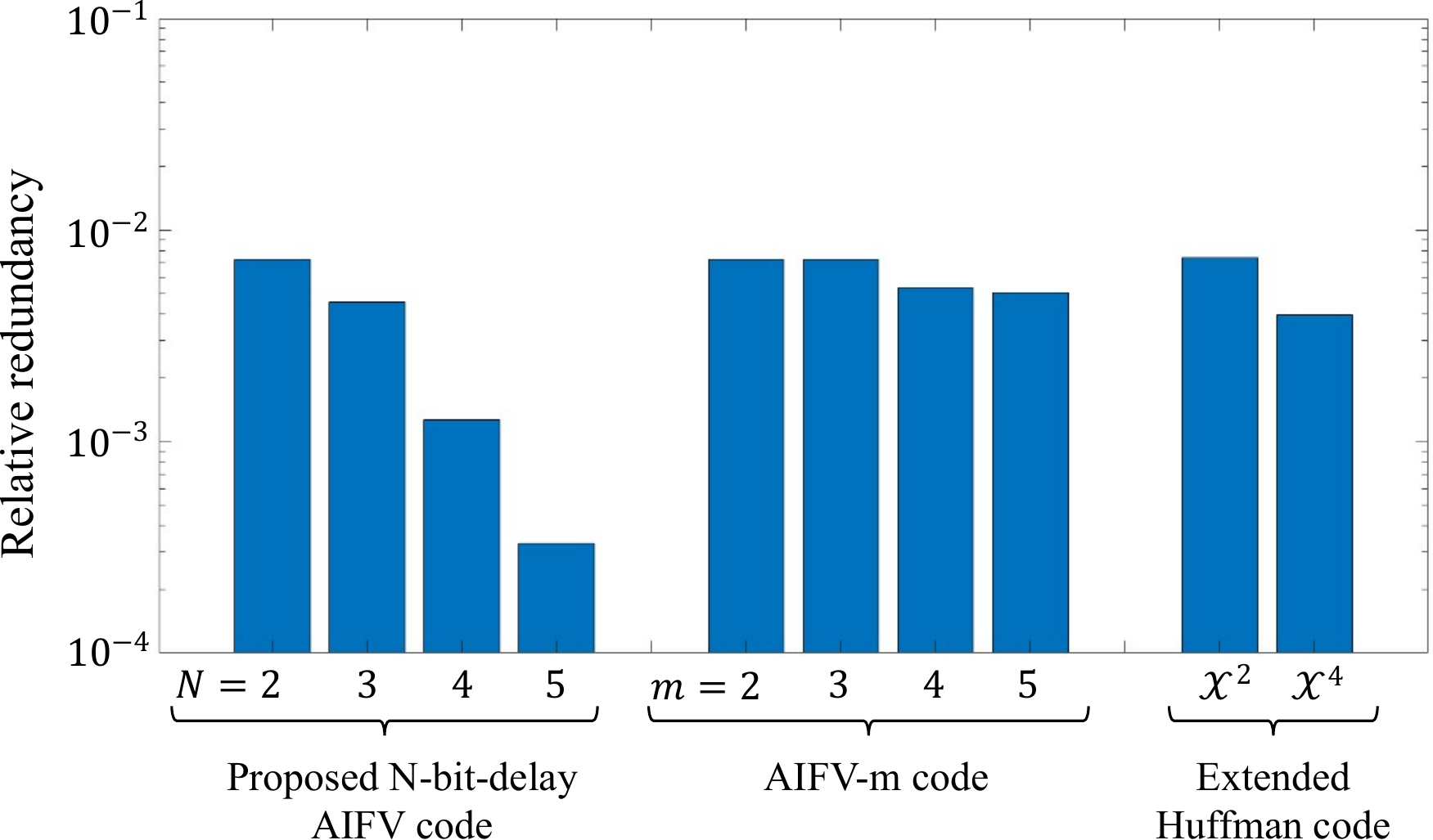}}
	\end{center}
	\caption{Theoretical relative redundancy of the constructed codebooks in non-binary-source cases.}
	\label{fig:comp_dict_simple}
\end{figure}

For the non-binary cases, we compared the codes using three source distributions introduced in the previous work \cite{ref:aifv1}:
\begin{eqnarray}
	P_\mathcal{X}^{(0)}(a_m) &=& \frac{1}{|\mathcal{X}|}\nonumber\\
	P_\mathcal{X}^{(1)}(a_m) &=& \frac{(m+1)}{\sum_{m'=0}^{|\mathcal{X}|-1}(m'+1)}\nonumber\\
	P_\mathcal{X}^{(2)}(a_m) &=& \frac{(m+1)^2}{\sum_{m'=0}^{|\mathcal{X}|-1}(m'+1)^2}.
\end{eqnarray}
In this comparison, we set as $\mathcal{X}=\mathbb{A}_5$, where the theoretical limits of code length for $P_\mathcal{X}^{(0)}$, $P_\mathcal{X}^{(1)}$, and $P_\mathcal{X}^{(2)}$ were 2.3219, 2.1493, and 1.8427 bit/sample, respectively.

Fig.~\ref{fig:comp_dict_simple} depicts the results.
The codebook sizes, {as defined above}, of the proposed codes for $P_\mathcal{X}^{(0)}$, $P_\mathcal{X}^{(1)}$, and $P_\mathcal{X}^{(2)}$ were at most $(25 \times 5)$, $(65 \times 5)$, and $(67 \times 5)$, respectively.
Therefore, we compared the extended Huffman codes having codebook sizes at most $5^4$.
The proposed codes of $N=4,5$ outperformed the conventional AIFV-$m$ and extended Huffman codes in all cases.

\subsubsection{Optimality check of the proposed codes}
\begin{table}[t]
	\caption{Optimality of the constructed $N$-bit-delay AIFV codebooks.}
	\begin{center}
		\begin{tabular}{c|c|c|}
			\cline{2-3}
			& Binary source                                                                        & Non-binary source \\ \hline
			\multicolumn{1}{|c|}{\emph{E-optimality}} & \multicolumn{2}{c|}{Guaranteed in theory}                                                                \\ \hline
			\multicolumn{1}{|c|}{\emph{F-optimality}} & \multicolumn{2}{c|}{Empirically checked using {\it Theorem \ref{thm:optimum_check}}}                     \\ \hline
			\multicolumn{1}{|c|}{\emph{G-optimality}} & \begin{tabular}{c}Partially checked \\$(N=2,3)$ \end{tabular}                        & --                \\ \hline
		\end{tabular}
	\end{center}
	\label{tbl:optimality_check}
\end{table}
Next, we checked the optimality of the proposed $N$-bit-delay AIFV codes used above.
Table \ref{tbl:optimality_check} summarizes the results.
Due to the numerical precision derived from the double precision floating point numbers, we here regarded the costs or expected code lengths as equal when their absolute difference was smaller than $10^{-14}$ (bit/sample).
\emph{E-optimality} was guaranteed by {\it Theorem \ref{thm:optimum}}, so we checked the updated costs to use {\it Theorem \ref{thm:optimum_check}}.
\emph{F-optimality} empirically held for every codebook of the proposed code, with the costs being invariant by the update.

For binary sources with small $N$s, we can optimize each tree in the code forest by a brute-force search, trying every possible code tree available under the given mode.
It can be formulated in a simple way in the binary-input cases, which we leave to Appendix \ref{app:bruteforce}.
Replacing the code-tree-wise optimization in Fig.~\ref{fig:alg_outline} with the brute-force search for each code tree, we can construct code forests without limiting the modes.
In this case, we can get a \emph{G-optimal} codebook if the algorithm achieves \emph{F-optimality}.

Using this approach, we made the proposed $N$-bit-delay AIFV codebooks for the binary sources shown above with $N=2,3$.
In every case, every cost became invariant by the update, and we were able to get a \emph{G-optimal} codebook for every $(p_{\rm src}(a_0), p_{\rm src}(a_1))$, owing to {\it Theorems \ref{thm:optimum_check}} and {\it \ref{thm:Gopt_if_Fopt}}.
The expected code lengths did not differ from the ones by the ILP approach.
Therefore, the binary-input codebooks used in the previous comparison were \emph{G-optimal}, at least for $N=2,3$.
Since the conventional AIFV-2 codes showed the same compression efficiency as the proposed $2$-bit-delay AIFV codes, they were also \emph{G-optimal}, which does not conflict with the theoretical result in the previous work \cite{ref:aifv2_opt_proof}.

\subsection{Average code length for finite sequence}
\begin{figure}[!tb]
	\begin{center}
		\subfigure[Comparison for sequence size $512$.]{
			\includegraphics[width=8.5cm,  bb=0 0 824 519]{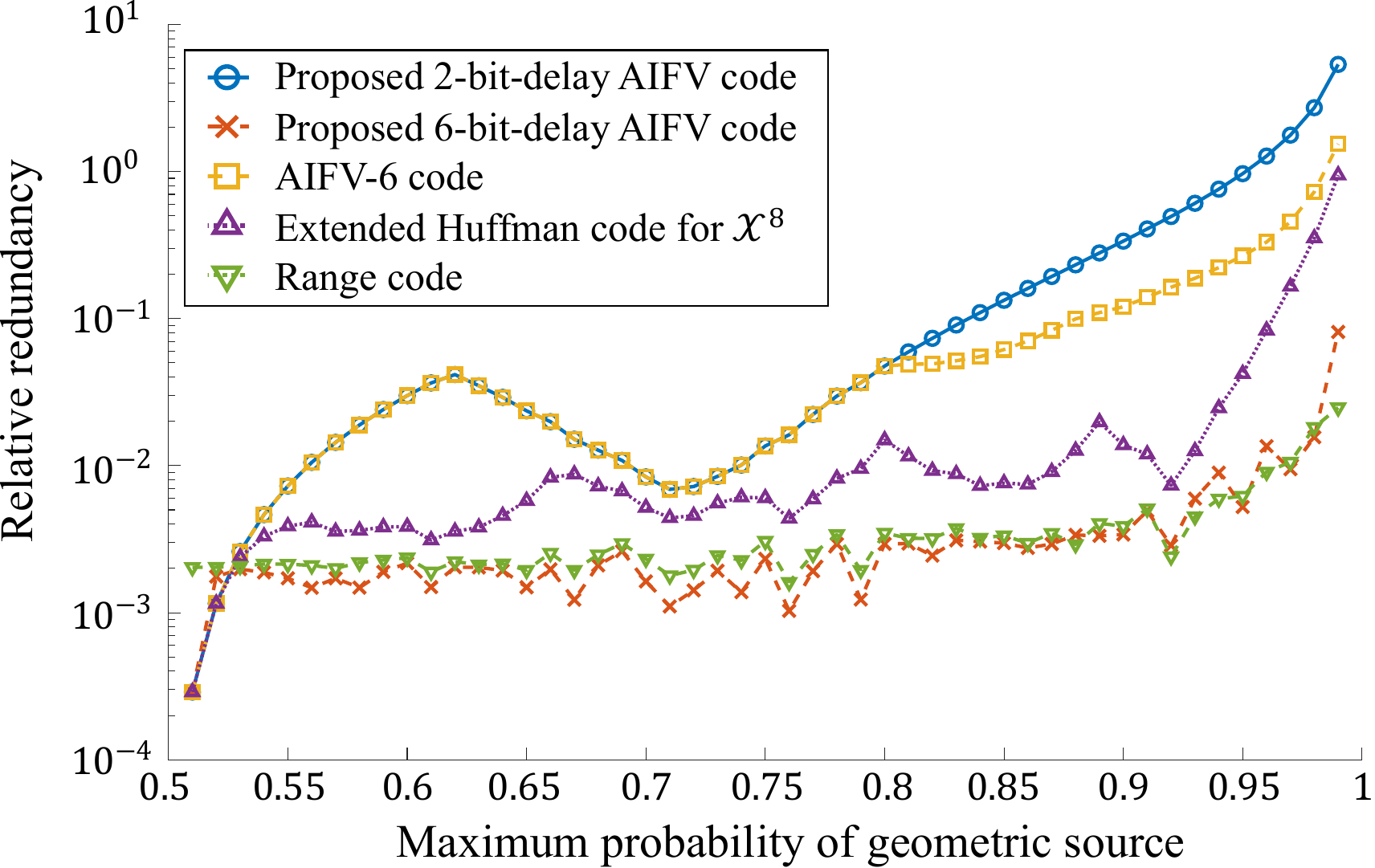}}
		\subfigure[Comparison for sequence size $1024$.]{
			\includegraphics[width=8.5cm,  bb=0 0 824 519]{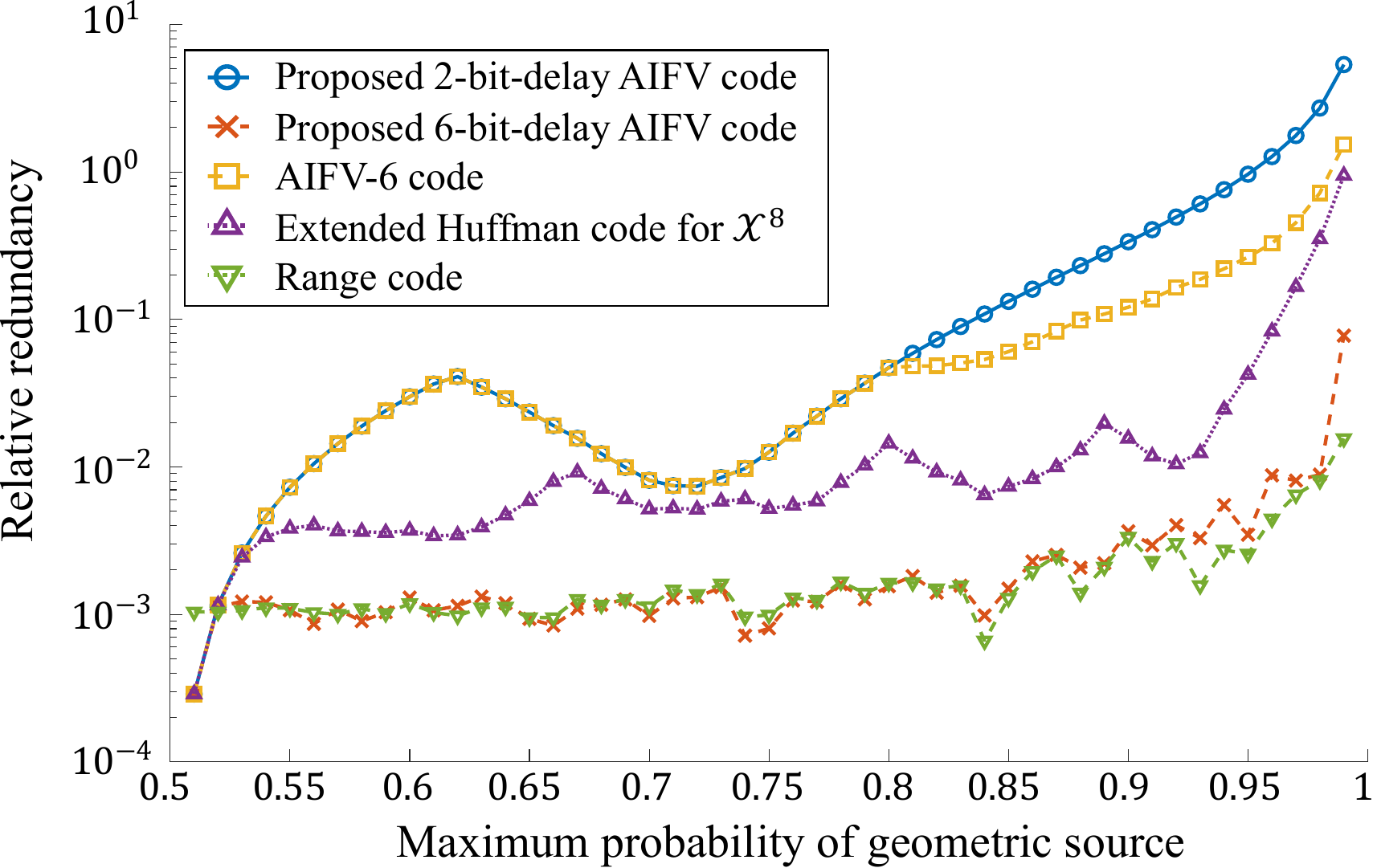}}
		\subfigure[Comparison for sequence size $2048$.]{
			\includegraphics[width=8.5cm,  bb=0 0 824 519]{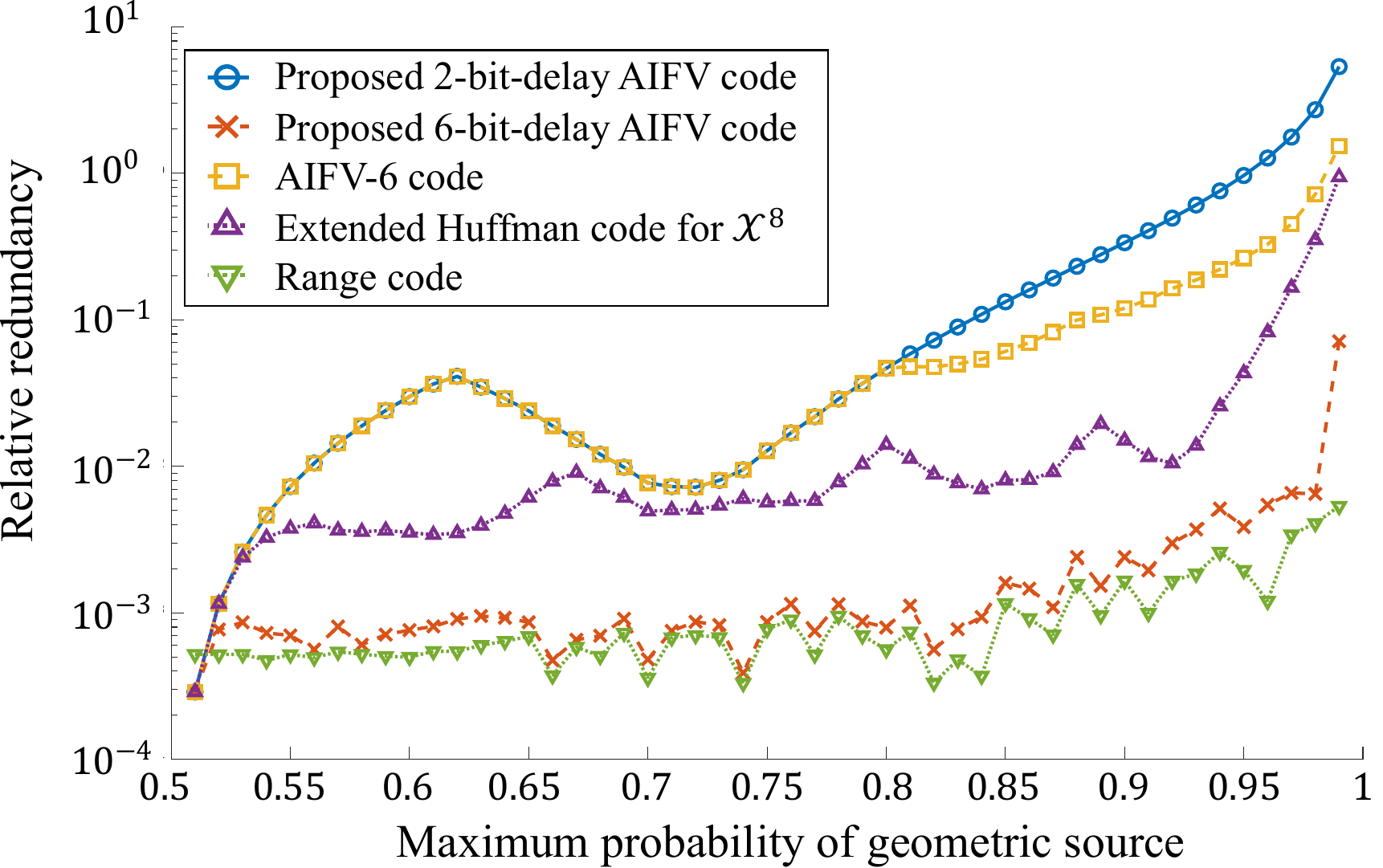}}
	\end{center}
	\caption{Averages of relative redundancy among $10000$ trials for each source in binary-source cases.}
	\label{fig:comp_enc_geo}
\end{figure}

Theoretically, the arithmetic codes achieve the expected code length identical to the entropy \cite{ref:hbdc}.
However, in practice, we have to compress data within a finite length of source symbol sequences,
and they cannot achieve entropy in such cases.
Therefore, there are some chances for the proposed codes to show higher efficiency than the arithmetic codes.

To investigate the actual performance, we compressed random numbers generated by the inversion method \cite{ref:rand}
using the same sources and codebooks as in the above experiment.
We compared with the range codes \cite{ref:hbdc}, a practical realization of the arithmetic codes, using 32-bit precision ranges.
The source distributions were given as known values for the range codes.
Different sizes of source symbol sequences were used for the comparison to see the influence of the size.
Note that the code length here does not include codes for representing code tables or source distributions since they are expected to be shared between the encoder and decoder in advance. 

Fig.~\ref{fig:comp_enc_geo} shows the average relative redundancy among the trials for binary sources of sizes $512$, $1024$, and $2048$.
The range codes were designed for $\mathcal{X}=\mathbb{A}_2$.
This result is identical to the relative redundancy of compressing non-binary exponential sources.
The proposed codes of $N=6$ performed the most efficiently in almost all cases for $512$-length sequences.

Multiplying $(1-p_{\rm src}(a_0))$ to the sequence size will give us the approximate size of source symbol sequences in the case of non-binary exponential sources:
If we interpret the input sequence as unary, the size of the non-binary source given by unary decoding it will be equivalent to the amount of $a_1$.
For example, when $p_{\rm src}(a_0)=0.75$, the size $512$ corresponds to about $128$ for non-binary sequence size, which is reasonable enough for practical use.

\begin{figure}[!tb]
	\begin{center}
		\subfigure[Comparison for $P_\mathcal{X}^{(0)}$.]{
			\includegraphics[width=8.5cm,  bb=0 0 844 524]{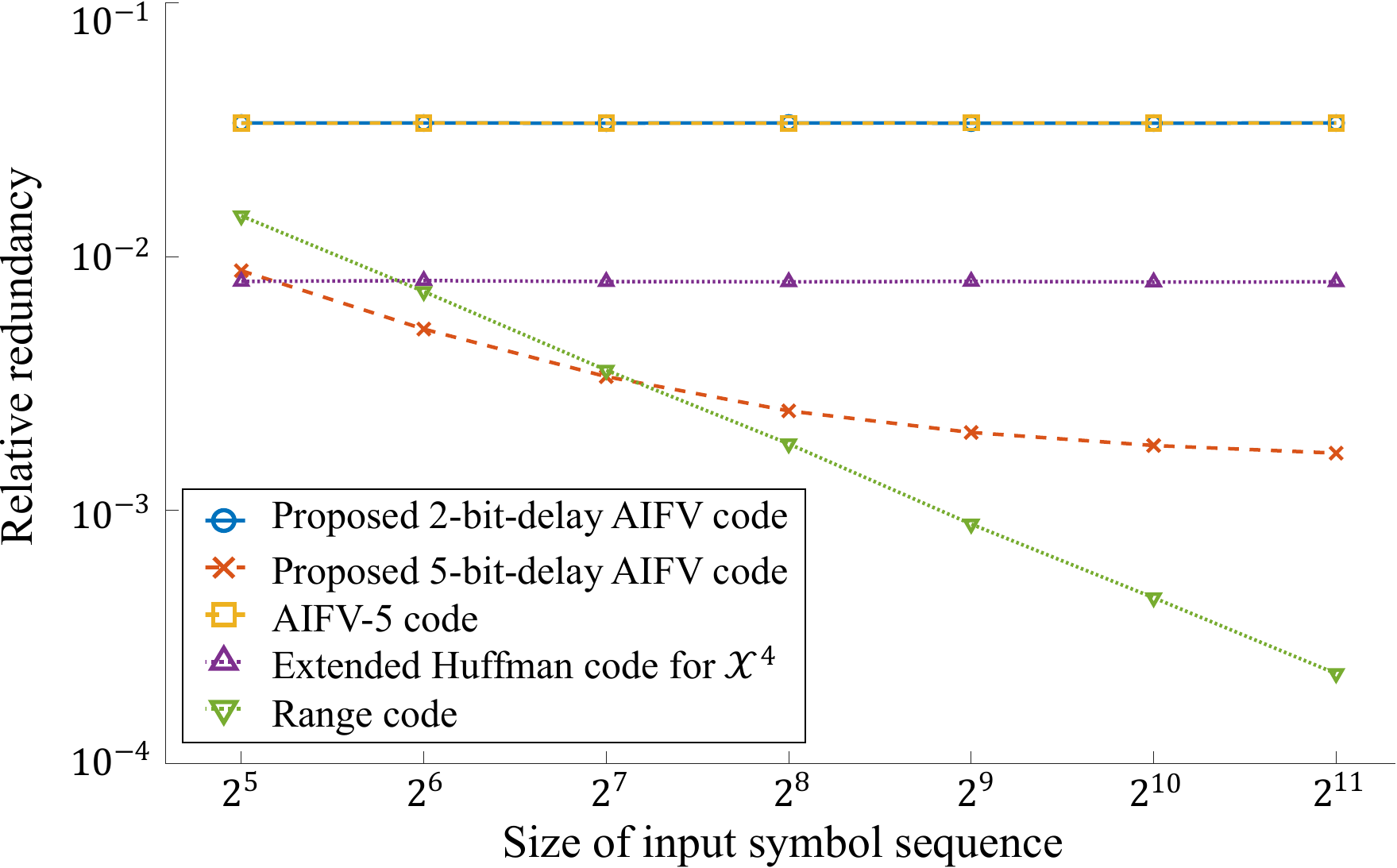}}
		\subfigure[Comparison for $P_\mathcal{X}^{(1)}$.]{
			\includegraphics[width=8.5cm,  bb=0 0 844 524]{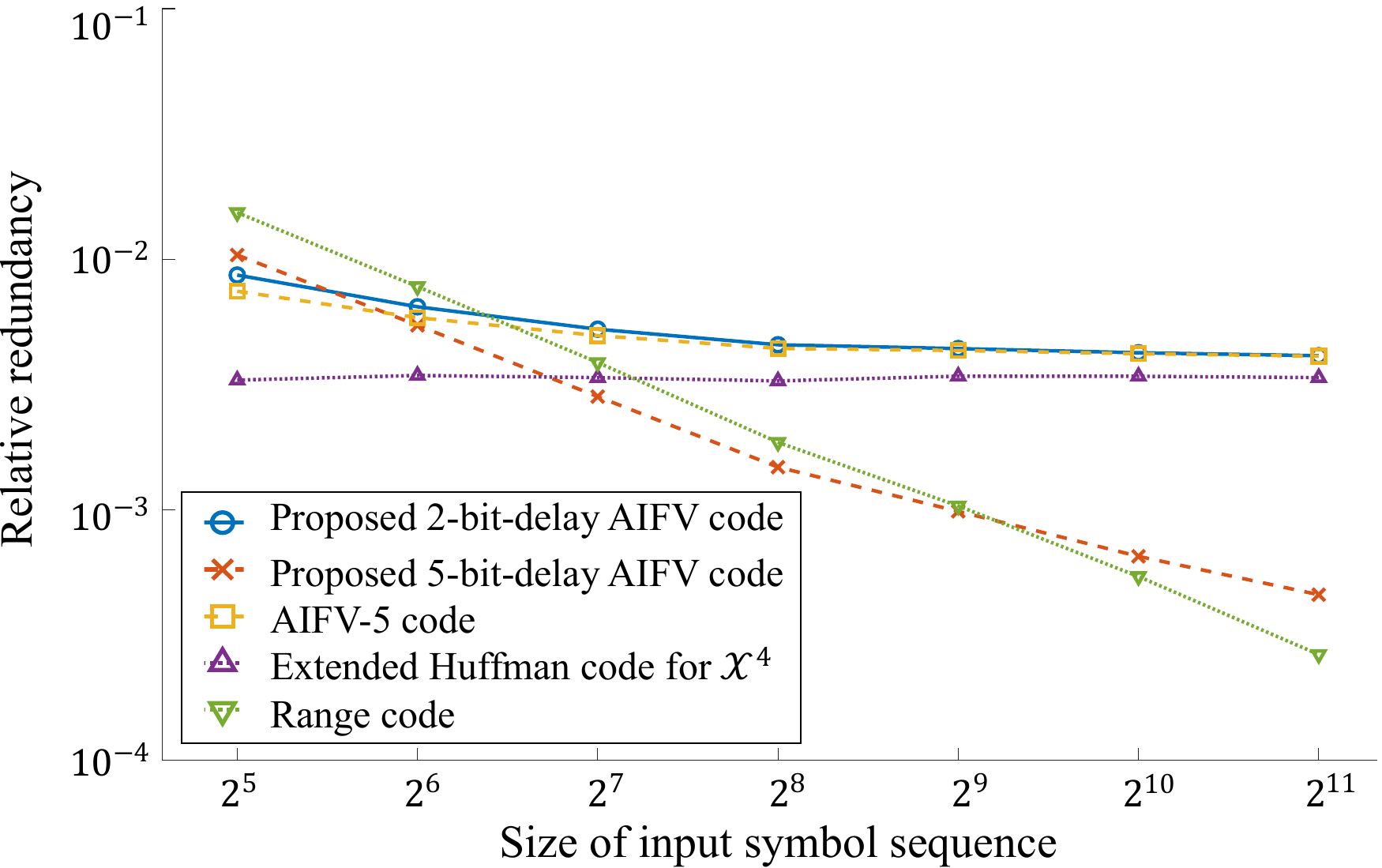}}
		\subfigure[Comparison for $P_\mathcal{X}^{(2)}$.]{
			\includegraphics[width=8.5cm,  bb=0 0 844 524]{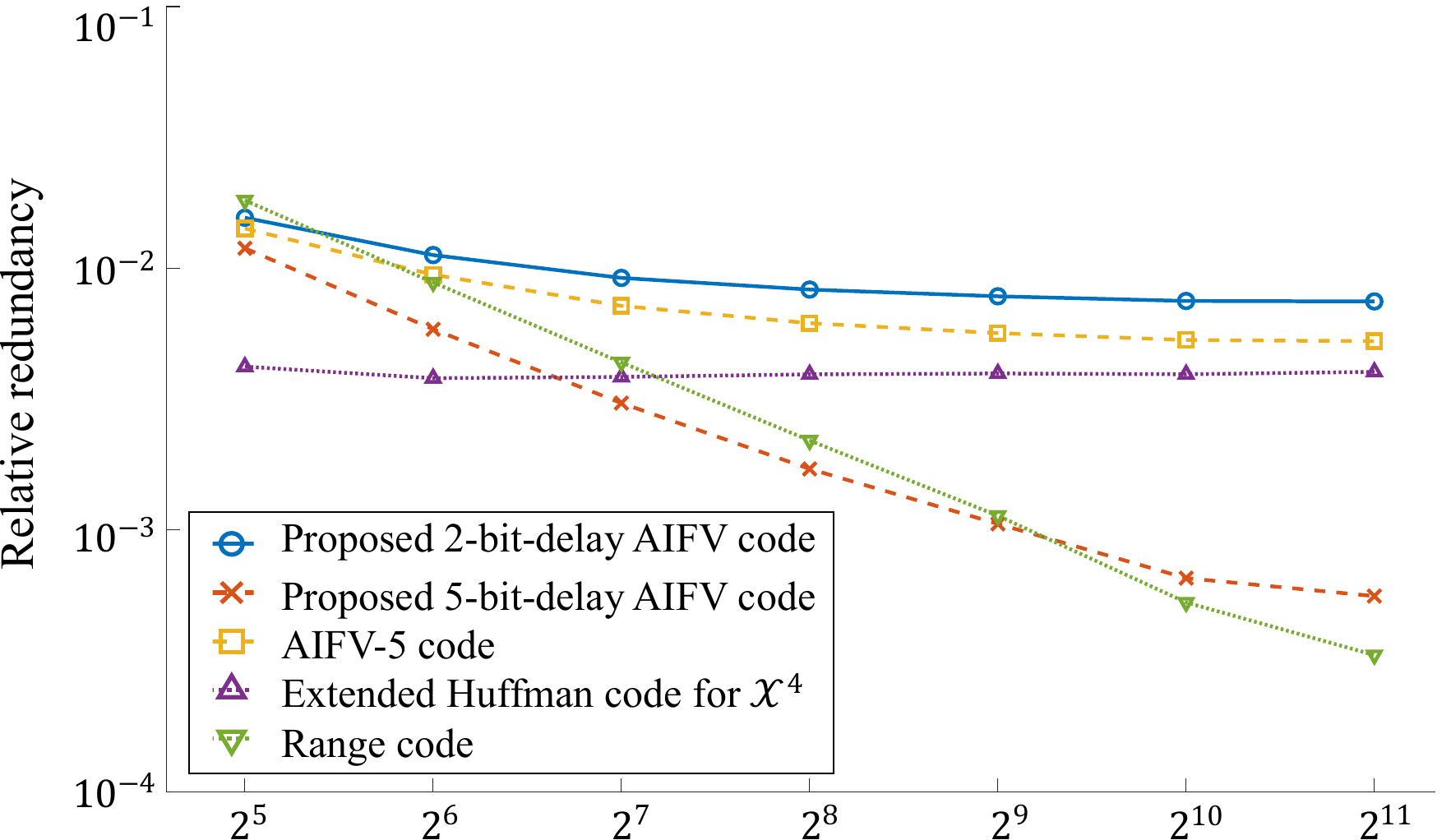}}
	\end{center}
	\caption{Averages of relative redundancy among $100000$ trials for each sequence size in non-binary-source cases.}
	\label{fig:comp_enc_simple}
\end{figure}

Fig.~\ref{fig:comp_enc_simple} compares for $P_\mathcal{X}^{(0)}$, $P_\mathcal{X}^{(1)}$, and $P_\mathcal{X}^{(2)}$ of sizes from $32$ to $2048$.
The range codes were designed for $\mathcal{X}=\mathbb{A}_5$ in this case.
As in Fig.~\ref{fig:comp_enc_simple}, there were some cases for every source where the proposed codes of $N=5$ showed higher efficiency than the other codes.

{We cannot ignore the termination codewords when the input sequence is short. Their lengths depend on the decoding delay of the code, and the range codes also use them since they can be interpreted as AIFV codes with many trees. On the other hand, codes permitting longer delay can achieve shorter expected code lengths. These fact makes the gap in performance dependent on the input size.} 
Therefore, simpler methods, such as AIFV codes with fewer trees (including Huffman codes), converged their performance at shorter inputs but had limited compression efficiency. More complex methods, such as range codes and AIFV codes with many trees, needed longer inputs to achieve closer to their theoretical performance while achieving much higher efficiency.

\section{Conclusions}
\label{sec:conclusions}
We discussed the construction of $N$-bit-delay AIFV codes, finding optimal codes for given input sources.
$N$-bit-delay AIFV codes are made by sets of code trees, namely linked code forests, and can represent every code we can make when permitting decoding delay up to $N$ bits.
The code construction problem was formulated into three important stages, and we defined some optimality for achieving each stage:
\emph{G-optimality}, where the code achieves the global optimum among all the codes decodable within $N$ bits of decoding delay;
\emph{F-optimality}, where the code achieves the optimum among the codes using a fixed set of modes;
\emph{E-optimality}, where some set of modes exists{, which we do not know in advance}, and the code achieves the optimum among the codes using them.
We presented that the construction of a code forest can be decomposed into some code-tree-wise independent problems when we focus on \emph{E-optimality}.
Additionally, we derived an empirical way to check the \emph{F-optimality} of the codes.

We then showed the theoretical properties of the optimal codes.
We detected some set of modes that make \emph{F-optimality} sufficient for \emph{G-optimality}.
We also revealed that, without loss of generality, we can let the code trees be symmetric to each other if their modes are so.

The code-tree construction method was proposed based on these ideas.
By solving code-tree-wise ILP problems iteratively, we can guarantee to get \emph{E-optimal} code forests.

In the experiments, we empirically checked that every constructed code was actually \emph{F-optimal}.
Furthermore, in cases of binary inputs with $N=2,3$, we could check every constructed code was \emph{G-optimal}.
The constructed $N$-bit-delay AIFV codes showed higher compression efficiency when $N\ge 3$ than the conventional AIFV-$m$ and extended Huffman codes.
Moreover, in the random numbers simulation, they performed better than the 32-bit-precision range codes under reasonable conditions.

Indeed, the proposed code-tree construction is still very complex, even though we introduced some methods of reducing complexity. 
However, the ideas shown here must be essential to further develop AIFV-code techniques.

\appendix
\section{Generalized code-tree optimization}
\label{app:gen_opt}
The proposed algorithm in {\it Procedure \ref{prc:iterative}} requires some code tree in the solution to be reachable from every other one.
However, this assumption does not generally hold, and several independent code forests may be given in the iteration.
In such cases, we cannot calculate the inverse of $( P^{(i)}_{\rm trans\{0\}\{0\}} - I_{K-1})$ in updating costs, and the algorithm fails to work.
Here, we extend the updating rule to work for general cases.

The transition matrix $P_{\rm trans}^{(i)}$ can always be transformed, by some permutation, into a block triangular matrix having irreducible matrices for its diagonal blocks: It is obvious from the definition of the irreducible matrix \cite{ref:probmat}.
{Since the diagonal blocks correspond to strongly-connected components \cite{ref:scc_tar,ref:intro_algo_scc} in the context of bidirectional graphs, the appropriate permutation can be found by the depth-first search regarding the transition matrix $P_{\rm trans}^{(i)}$ as an adjacency matrix, grouping the mutually reachable code trees.}

At least one block {cannot} reach any other blocks, and we here call it an absorption block.
Since absorption blocks are not connected to any other blocks, we can modify the permutation to make them block diagonal.
Therefore, we set a permutation matrix $Q^{(i)}$ to transform as
\begin{equation}
	P_{\rm trans}^{(i)}=Q^{(i){\rm t}}\left (
	\begin{array}{cccccccc}
		\hat{P}^{(i)}_{0,0}                   &                                       &        &                                                         &                                                     &                                                         &        &                                     \\
		& \hat{P}^{(i)}_{1,1}                   &        &                                                         &                                                     &                                                         &        &                                     \\
		0                                     &                                       & \ddots &                                                         &                                                     & 0                                                       &        &                                     \\
		&                                       &        & \hat{P}^{(i)}_{J_{\rm abs}^{(i)}-1,J_{\rm abs}^{(i)}-1} &                                                     &                                                         &        &                                     \\
		\hat{P}^{(i)}_{J_{\rm abs}^{(i)},0}   & \hat{P}^{(i)}_{J_{\rm abs}^{(i)},1}   & \cdots & \cdots                                                  & \hat{P}^{(i)}_{J_{\rm abs}^{(i)},J_{\rm abs}^{(i)}} &                                                         &        &                                     \\
		\hat{P}^{(i)}_{J_{\rm abs}^{(i)}+1,0} & \hat{P}^{(i)}_{J_{\rm abs}^{(i)}+1,1} & \cdots & \cdots                                                  & \cdots                                              & \hat{P}^{(i)}_{J_{\rm abs}^{(i)}+1,J_{\rm abs}^{(i)}+1} &        &                                     \\
		\vdots                                & \vdots                                & \cdots & \cdots                                                  & \cdots                                              & \cdots                                                  & \ddots &                                     \\
		\hat{P}^{(i)}_{J^{(i)}-1,0}           & \hat{P}^{(i)}_{J^{(i)}-1,1}           & \cdots & \cdots                                                  & \cdots                                              & \cdots                                                  & \cdots & \hat{P}^{(i)}_{J^{(i)}-1,J^{(i)}-1}
	\end{array}
	\right )Q^{(i)}
\end{equation}
and write the other variables as
\begin{equation}
	{\bm L}^{(i)} = Q^{(i){\rm t}}\left (
	\begin{array}{c}
		\hat{\bm L}^{(i)}_0 \\\hat{\bm L}^{(i)}_1\\ \vdots \\\hat{\bm L}^{(i)}_{J^{(i)}-1}
	\end{array}
	\right ),
	{\bm C}^{(i)} =Q^{(i){\rm t}} \left (
	\begin{array}{c}
		\hat{\bm C}^{(i)}_0 \\\hat{\bm C}^{(i)}_1\\ \vdots \\\hat{\bm C}^{(i)}_{J^{(i)}-1}
	\end{array}
	\right ),
	{\bm \Pi}^{(i)}_j = Q^{(i){\rm t}}\left (
	\begin{array}{c}
		{\bm 0}_{\sum_{j'=0}^{j-1} K_{j'}^{(i)}} \\ \\ \hat{\bm \Pi}^{(i)}_{j} \\ \\ {\bm 0}_{\sum_{j'=j+1}^{J^{(i)}-1} K_{j'}^{(i)}}
	\end{array}
	\right ),
	Q^{(i)} = \left (
	\begin{array}{c}
		\hat{Q}^{(i)}_0 \\\hat{Q}^{(i)}_1\\ \vdots \\\hat{Q}^{(i)}_{J^{(i)}-1}
	\end{array}
	\right ):
\end{equation}
$J^{(i)}$ and $J_{\rm abs}^{(i)}$ are natural numbers;
$\hat{P}_{j,j}^{(i)}$ is a $K^{(i)}_j$-by-$K^{(i)}_j$ irreducible matrix, where $K^{(i)}_j > 0$ and $\sum_j K^{(i)}_j=K$;
$\hat{Q}^{(i)}_j$ is a $K^{(i)}_j$-by-$K$ permutation matrix;
$\hat{\bm L}^{(i)}_j$, $\hat{\bm C}^{(i)}_j$, and $\hat{\bm \Pi}^{(i)}_j$ are $K^{(i)}_j$-th-order column vectors.
Every matrix $(\hat{P}^{(i)}_{j,0}, \hat{P}^{(i)}_{j,1}, \cdots, \hat{P}^{(i)}_{j,j-1})$ for $j=J_{\rm abs}^{(i)},\cdots, J^{(i)}-1$ has at least one non-zero element.
Since $\hat{P}_{j,j}^{(i)}$ is irreducible, we can always find a unique block-wise stationary distribution $\hat{\bm \Pi}^{(i)}_j$ comprising only positive values and satisfying $\hat{\bm \Pi}^{(i){\rm t}}_j\hat{P}_{j,j}^{(i)}=\hat{\bm \Pi}^{(i){\rm t}}_j$ for $j\in \mathbb{Z}^+_{<J_{\rm abs}^{(i)}}$.
Every weighted average of $\{\hat{\bm \Pi}^{(i)}_j\mid j\in \mathbb{Z}^+_{<J_{\rm abs}^{(i)}}\}$ becomes a stationary distribution of the transition matrix $P_{\rm trans}^{(i)}$.
Let us write the expected code lengths as $\bar{L}_j^{(i)}=\hat{\bm \Pi}^{(i){\rm t}}_j\hat{\bm L}^{(i)}_j$.

{Based on the above representation, each absorption block becomes an independent code forest when we set the initial code tree as one in the block. In other words, in cases where $( P^{(i)}_{\rm trans\{0\}\{0\}} - I_{K-1})$ does not have an inverse, the set of code trees contains plural independent code forests, which achieve the expected code lengths $\bar{L}_j^{(i)}$. In such cases, it is hard to control the performance of all the independent code forests. However, by generalizing the algorithm in {\it Procedure \ref{prc:iterative}}, we can ensure the upper bound of the expected code lengths to be decreased by iteration: }
\mytheory{prc}{Generalized iterative code tree construction}{
	\label{prc:iterative_gen}
	Follow the steps below with a given natural number $K$ and modes $\{\tilde{\textsc{Mode}}_k\mid k\in\mathbb{Z}^+_{<K}\}$.
	\begin{enumerate}
		\item Set an initial value ${\bm C}^{(0)}$ and $i=1$.
		\item Get code trees $\{T^{(i)}_k\}=\{(\textrm{Cword}^{(i)}_k, \textrm{Link}^{(i)}_k, \tilde{\textsc{Mode}}_k)\mid k\in\mathbb{Z}^+_{<K}\}$ by solving {\it Minimization problem \ref{mip:codetreewise}} for each tree $T^{(i)}_k$ with given $K$, $\tilde{\textsc{Mode}}_k$, and ${\bm C}^{(i-1)}$.
		\item Get the tree-wise expected code length ${\bm L}^{(i)}$ and the transition matrix $P_{\rm trans}^{(i)}$ respectively by
		\begin{eqnarray}
			L^{(i)}_k &=& \sum_m \left\|\textrm{Cword}^{(i)}_k(a_m)\right\|_{\rm len}\cdot p_{\rm src}(a_m) \nonumber\\
			P^{(i)}_{k, k'} &=& \sum_{m : \textrm{Link}^{(i)}_k(a_m)=k'} p_{\rm src}(a_m).\nonumber
		\end{eqnarray}
		\item Find a permutation matrix $Q^{(i)}$ which transforms $P_{\rm trans}^{(i)}$ into a block triangular matrix with $J_{\rm abs}^{(i)}$ absorption blocks sorted in the upper block diagonal part.
		\item Find stationary distributions $\{{\bm \Pi}_j^{(i)}\}$ satisfying Eqs.~(\ref{eq:stabledist1}) and (\ref{eq:stabledist2}) and calculate the respective expected code length $\bar{L}_j^{(i)}={\bm \Pi}_j^{(i){\rm t}}{\bm L}^{(i)}$. Say $j^{(i)}_*=\arg\max_j \bar{L}_j^{(i)}$.
		\item Update the costs as
		\begin{equation}
			\label{eq:cost_gen}
			\hat{\bm C}^{(i)}_j = \left \{
			\begin{array}{cc}
				0                                                                                                                                                                                                            & \text{if $K_j^{(i)}=1$ for $j<J_{\rm abs}^{(i)}$} \\
				\left(\begin{array}{c}
					0 \\
					\displaystyle \left(\hat{P}_{j,j\{0\}\{0\}}^{(i)}-I_{K^{(i)}_j-1}\right)^{-1}\left({\bm 1}_{K^{(i)}_j-1}\bar{L}_j^{(i)}-\hat{\bm L}^{(i)}_{j\{0\}\emptyset}\right)
				\end{array}
				\right)                                                                                                                                                                                                      & \text{if $K_j^{(i)}>1$ for $j<J_{\rm abs}^{(i)}$} \\
				\\
				\displaystyle \left(\hat{P}_{j,j}^{(i)}-I_{K^{(i)}_j}\right)^{-1}\left({\bm 1}_{K^{(i)}_j}\bar{L}_{j^{(i)}_*}^{(i)}-\hat{\bm L}^{(i)}_{j}-\sum_{j'=0}^{j-1}\hat{P}_{j,j'}^{(i)}\hat{\bm C}^{(i)}_{j'}\right) & \text{for $j\ge J_{\rm abs}^{(i)}$}
			\end{array}
			\right. .
		\end{equation}
		\item If $\hat{\bm C}_{j^{(i)}_*}^{(i)}\neq \hat{Q}^{(i)}_{j^{(i)}_*}{\bm C}^{(i-1)} - {\bm 1}_{K_{j^{(i)}_*}^{(i)}} \hat{Q}^{(i)}_{j^{(i)}_*\{j\neq 0\}\emptyset}{\bm C}^{(i-1)}$, increment $i$ and return to step b.
		\item Output $\{T_k\}=\{T^{(i)}_k\mid k\in A^{(i)}\}$ where $A^{(i)}=\{k\mid k\text{-th column of }\hat{Q}_{j^{(i)}_*}^{(i)}\text{ is not zero}\}$.
	\end{enumerate}
}
{The $j^{(i)}_*$-th block corresponds to the worst-performing code forest, and its expected code length $\max_j \bar{L}_j^{(i)}$ decreases by iteration, as we will prove in the following {\it Theorem \ref{thm:optimum_gen}}. 
	The set $\{j\neq 0\}$ refers to the set of all non-zero integers so that $\hat{Q}^{(i)}_{j^{(i)}_*\{j\neq 0\}\emptyset}$ represents the $0$-th row vector in the permutation matrix $\hat{Q}^{(i)}_{j^{(i)}_*}$.
	The termination condition of the algorithm depends on the cost corresponding to the $j^{(i)}_*$-th block, where the term $- {\bm 1}_{K_{j^{(i)}_*}^{(i)}} \hat{Q}^{(i)}_{j^{(i)}_*\{j\neq 0\}\emptyset}{\bm C}^{(i-1)}$ in step g is responsible for forcing the 0-th cost in the $j^{(i)}_*$-th block to be zero. }

We have designed {\it Procedure \ref{prc:iterative_gen}} to output the worst-performing code forest to discuss its optimality as in the following theorem.
However, in practice, it is reasonable to output the best-performing one, corresponding to the absorption block giving the minimum $\bar{L}^{(i)}_j$ in step h after the convergence.
In such a case, it is preferable to repeat {\it Procedure \ref{prc:iterative_gen}} using only the best-performing absorption block because it is not guaranteed to converge when the worst-performing one converges. 
\mytheory{thm}{E-optimality with no assumption}{
	\label{thm:optimum_gen}
	{\it Procedure \ref{prc:iterative_gen}} gives an \emph{E-optimal} code forest $\{T_k\}$ for the given $K\in\mathbb{N}$ and $\{\tilde{\textsc{Mode}}_k\mid k\in\mathbb{Z}^+_{<K}\}$ at a finite iteration.
}
\underline{\it Proof}: We take the following steps to prove the theorem.
\begin{enumerate}
	\item $\max_j \bar{L}_j^{(i-1)} \ge \max_j \bar{L}_j^{(i)}$ holds for all $i$.
	\item $\max_j \bar{L}_j^{(i-1)} > \max_j \bar{L}_j^{(i)}$ if $\hat{\bm C}_{j^{(i)}_*}^{(i)}\neq \hat{Q}^{(i)}_{j^{(i)}_*}{\bm C}^{(i-1)} - {\bm 1}_{K_{j^{(i)}_*}^{(i)}} \hat{Q}^{(i)}_{j^{(i)}_*\{j\neq 0\}\emptyset}{\bm C}^{(i-1)}$.
	\item When $\hat{\bm C}_{j^{(i)}_*}^{(i)}= \hat{Q}^{(i)}_{j^{(i)}_*}{\bm C}^{(i-1)} - {\bm 1}_{K_{j^{(i)}_*}^{(i)}} \hat{Q}^{(i)}_{j^{(i)}_*\{j\neq 0\}\emptyset}{\bm C}^{(i-1)}$, $\max_j \bar{L}_j^{(i)}$ is the minimum value of {\it Minimization problem \ref{mip:modefixedsub}} for $A=A^{(i)}$.
\end{enumerate}
\vspace{5mm}

{\bf a}. Since $\hat{P}_{j,j}^{(i)}$ is irreducible for $0\le j<J_{\rm abs}^{(i)}$, the 0-th state in $\hat{P}^{(i)}_{j,j}$ is naturally reachable from every other state in the same block, and thus {\it Theorem \ref{thm:detnzero}} guarantees $( \hat{P}_{j,j\{0\}\{0\}}^{(i)} - I_{K^{(i)}_j-1})$ to have an inverse.
In the case of $j\ge J_{\rm abs}^{(i)}$, every state in $\hat{P}_{j,j}^{(i)}$ {can} reach some state in {$\hat{P}_{j',j'}^{(i)}$} of $j'< J_{\rm abs}^{(i)}$.
Therefore, the set of states in $\hat{P}_{j,j}^{(i)}$ forms an open set, and from {\it Theorem \ref{thm:detnzero}}, we can say that $( \hat{P}_{j,j}^{(i)} - I_{K^{(i)}_j})$ for $j\ge J_{\rm abs}^{(i)}$ also has an inverse.

{Here, we use ${\bm f}^{(i)}_{\rm obj}$ defined in Eq.~(\ref{eq:obj_vec}). }
${\bm \Pi}^{(i)}_{j^{(i)}_*}$ comprises only positive numbers, and each element in ${\bm f}^{(i)}_{\rm obj}({\bm L}^{(i-1)}, P_{\rm trans}^{(i-1)})$ is not smaller than the corresponding element in ${\bm f}^{(i)}_{\rm obj}({\bm L}^{(i)}, P_{\rm trans}^{(i)})$.
Therefore, the inner products hold
\begin{equation}
	\label{eq:inp_gen}
	{\bm \Pi}^{(i){\rm t}}_{j^{(i)}_*}{\bm f}^{(i)}_{\rm obj}({\bm L}^{(i-1)}, P_{\rm trans}^{(i-1)}) \ge {\bm \Pi}^{(i){\rm t}}_{j^{(i)}_*}{\bm f}^{(i)}_{\rm obj}({\bm L}^{(i)}, P_{\rm trans}^{(i)}) .
\end{equation}
The right-hand side of Eq.~(\ref{eq:inp_gen}) is written as
\begin{equation}
	\label{eq:Lb_i_gen}
	{\bm \Pi}^{(i){\rm t}}_{j^{(i)}_*}{\bm f}^{(i)}_{\rm obj}({\bm L}^{(i)}, P_{\rm trans}^{(i)})
	={\bm \Pi}^{(i){\rm t}}_{j^{(i)}_*} {\bm L}^{(i)}  + {\bm \Pi}^{(i){\rm t}}_{j^{(i)}_*}\left ( P_{\rm trans}^{(i)} - I_K \right ){\bm C}^{(i-1)}  = \bar{L}_{j^{(i)}_*}^{(i)}.
\end{equation}

On the other hand, we have
\begin{equation}
	Q^{(i-1)}{\bm f}^{(i)}_{\rm obj}({\bm L}^{(i-1)}, P_{\rm trans}^{(i-1)}) = Q^{(i-1)}{\bm L}^{(i-1)} + Q^{(i-1)}\left( P_{\rm trans}^{(i-1)} - I_K \right )Q^{(i-1){\rm t}}Q^{(i-1)}{\bm C}^{(i-1)}
\end{equation}
and
\begin{equation}
	\label{eq:Lb_i-1_gen_perm}
	\hat{Q}_j^{(i-1)}{\bm f}^{(i)}_{\rm obj}({\bm L}^{(i-1)}, P_{\rm trans}^{(i-1)}) = \left \{
	\begin{array}{cc}
		\displaystyle {\bm 1}_{K_j^{(i-1)}}\bar{L}_j^{(i-1)}              & \text{for $j<J_{\rm abs}^{(i)}$}    \\
		\\
		\displaystyle  {\bm 1}_{K_j^{(i-1)}}\bar{L}_{j^{(i-1)}_*}^{(i-1)} & \text{for $j\ge J_{\rm abs}^{(i)}$}
	\end{array}
	\right. .
\end{equation}
{The above result for $j<J_{\rm abs}^{(i)}$ is given by a similar way to step a in {\it Proof of Theorem \ref{thm:optimum}} for each absorption block. The one for $j\ge J_{\rm abs}^{(i)}$ is given by simply substituting Eq.~(\ref{eq:cost_gen}) into ${\bm f}^{(i)}_{\rm obj}({\bm L}^{(i-1)}, P_{\rm trans}^{(i-1)})$}.
Note that the above holds even if $K^{(i-1)}_j = 1$ for $j<J_{\rm abs}^{(i-1)}$ because in that case, $\hat{\bm \Pi}^{(i-1)}_j=1$ and $\hat{\bm L}^{(i-1)}_j=\bar{L}_{j}^{(i-1)}$.
Since $j^{(i-1)}_*=\arg\max \bar{L}_j^{(i-1)}$, we can get the upper bound of the left-hand side of Eq.~(\ref{eq:inp_gen}) as
\begin{equation}
	\label{eq:Lb_i-1_gen}
	{\bm \Pi}^{(i){\rm t}}_{j^{(i)}_*}{\bm f}^{(i)}_{\rm obj}({\bm L}^{(i-1)}, P_{\rm trans}^{(i-1)})
	\le {\bm \Pi}^{(i){\rm t}}_{j^{(i)}_*}{\bm 1}_{K}\bar{L}_{j^{(i-1)}_*}^{(i-1)},
\end{equation}
resulting in
\begin{equation}
	\bar{L}_{j^{(i-1)}_*}^{(i-1)} \ge \bar{L}_{j^{(i)}_*}^{(i)}.
\end{equation}
\\

{\bf b}. Let us assume $\hat{\bm C}_{j^{(i)}_*}^{(i)}\neq \hat{Q}^{(i)}_{j^{(i)}_*}{\bm C}^{(i-1)} - {\bm 1}_{K_{j^{(i)}_*}^{(i)}} \hat{Q}^{(i)}_{j^{(i)}_*\{j\neq 0\}\emptyset}{\bm C}^{(i-1)}$ when $\bar{L}_{j^{(i-1)}_*}^{(i-1)} = \bar{L}_{j^{(i)}_*}^{(i)}$.
In the case of $\hat{Q}^{(i)}_{j^{(i)}_*}{\bm f}_{\rm obj}^{(i)}({\bm L}^{(i-1)}, P_{\rm trans}^{(i-1)})\neq {\bm 1}_{K^{(i)}_{j^{(i)}_*}}\bar{L}_{j^{(i-1)}_*}^{(i-1)}$, we have from Eq.~(\ref{eq:Lb_i-1_gen})
\begin{equation}
	{\bm \Pi}^{(i){\rm t}}_{j^{(i)}_*}{\bm f}^{(i)}_{\rm obj}({\bm L}^{(i-1)}, P_{\rm trans}^{(i-1)})
	< {\bm \Pi}^{(i){\rm t}}_{j^{(i)}_*}{\bm 1}_{K}\bar{L}_{j^{(i-1)}_*}^{(i-1)},
\end{equation}
and thus $\bar{L}_{j^{(i-1)}_*}^{(i-1)} > \bar{L}_{j^{(i)}_*}^{(i)}$, which would conflict with the assumption.

In the case of $\hat{Q}^{(i)}_{j^{(i)}_*}{\bm f}_{\rm obj}^{(i)}({\bm L}^{(i-1)}, P_{\rm trans}^{(i-1)})= {\bm 1}_{K^{(i)}_{j^{(i)}_*}}\bar{L}_{j^{(i-1)}_*}^{(i-1)}$,
from Eq.~(\ref{eq:Lb_i-1_gen_perm}), we have
\begin{equation}
	\label{eq:if_Li_eq_Li-1_gen}
	\hat{Q}^{(i)}_{j^{(i)}_*}{\bm f}_{\rm obj}^{(i)}({\bm L}^{(i-1)}, P_{\rm trans}^{(i-1)}) = {\bm 1}_{K^{(i)}_{j^{(i)}_*}} \bar{L}_{j^{(i-1)}_*}^{(i-1)} = {\bm 1}_{K^{(i)}_{j^{(i)}_*}} \bar{L}_{j^{(i)}_*}^{(i)}= \hat{Q}^{(i)}_{j^{(i)}_*}{\bm f}^{(i+1)}_{\rm obj}({\bm L}^{(i)}, P_{\rm trans}^{(i)}).
\end{equation}
Here,
\begin{eqnarray}
	\hat{Q}^{(i)}_{j^{(i)}_*}{\bm f}^{(i+1)}_{\rm obj}({\bm L}^{(i)}, P_{\rm trans}^{(i)}) &=&
	\hat{\bm L}_{j^{(i)}_*}^{(i)} + \left(\hat{P}_{j^{(i)}_*,j^{(i)}_*\{0\}\{0\}}^{(i)}-I_{{K^{(i)}_{j^{(i)}_*}}-1}\right) \hat{Q}^{(i)}_{j^{(i)}_*}{\bm C}^{(i)}
\end{eqnarray}
and
\begin{eqnarray}
	&&\hat{Q}^{(i)}_{j^{(i)}_*}{\bm f}^{(i)}_{\rm obj}({\bm L}^{(i)}, P_{\rm trans}^{(i)}) \nonumber\\
	&=&	\hat{\bm L}_{j^{(i)}_*}^{(i)} + \left(\hat{P}_{j^{(i)}_*,j^{(i)}_*\{0\}\{0\}}^{(i)}-I_{{K^{(i)}_{j^{(i)}_*}}-1}\right) \hat{Q}^{(i)}_{j^{(i)}_*}{\bm C}^{(i-1)}\nonumber\\
	&=&\hat{\bm L}_{j^{(i)}_*}^{(i)} + \left(\hat{P}_{j^{(i)}_*,j^{(i)}_*\{0\}\{0\}}^{(i)}-I_{{K^{(i)}_{j^{(i)}_*}}-1}\right) \left(\hat{Q}^{(i)}_{j^{(i)}_*}{\bm C}^{(i-1)} - {\bm 1}_{K_{j^{(i)}_*}^{(i)}} \hat{Q}^{(i)}_{j^{(i)}_*\{j\neq 0\}\emptyset}{\bm C}^{(i-1)}\right)
\end{eqnarray}
because $(\hat{P}_{j^{(i)}_*,j^{(i)}_*\{0\}\{0\}}^{(i)}-I_{{K^{(i)}_{j^{(i)}_*}}-1}){\bm 1}_{K_{j^{(i)}_*}^{(i)}}={\bm 0}_{K_{j^{(i)}_*}^{(i)}}$.
Since the 0-th elements of $\hat{\bm C}_{j^{(i)}_*}^{(i)}$ and $\hat{Q}^{(i)}_{j^{(i)}_*}{\bm C}^{(i-1)} - {\bm 1}_{K_{j^{(i)}_*}^{(i)}} \hat{Q}^{(i)}_{j^{(i)}_*\{j\neq 0\}\emptyset}{\bm C}^{(i-1)}$ are 0,
we can get $\hat{Q}^{(i)}_{j^{(i)}_*}{\bm f}^{(i+1)}_{\rm obj}({\bm L}^{(i)}, P_{\rm trans}^{(i)})\neq \hat{Q}^{(i)}_{j^{(i)}_*}{\bm f}^{(i)}_{\rm obj}({\bm L}^{(i)}, P_{\rm trans}^{(i)})$ from the assumption.
According to Eq.~(\ref{eq:if_Li_eq_Li-1_gen}), it must be
\begin{equation}
	\hat{Q}^{(i)}_{j^{(i)}_*}{\bm f}^{(i)}_{\rm obj}({\bm L}^{(i-1)}, P_{\rm trans}^{(i-1)}) \neq \hat{Q}^{(i)}_{j^{(i)}_*}{\bm f}^{(i)}_{\rm obj}({\bm L}^{(i)}, P_{\rm trans}^{(i)}),
\end{equation}
which means at least one element in $\hat{Q}^{(i)}_{j^{(i)}_*}{\bm f}^{(i)}_{\rm obj}({\bm L}^{(i)}, P_{\rm trans}^{(i)})$ is smaller than the counterpart of $\hat{Q}^{(i)}_{j^{(i)}_*}{\bm f}^{(i)}_{\rm obj}({\bm L}^{(i-1)}, P_{\rm trans}^{(i-1)})$.
The other elements in $\hat{Q}^{(i)}_{j^{(i)}_*}{\bm f}^{(i)}_{\rm obj}({\bm L}^{(i)}, P_{\rm trans}^{(i)})$ are smaller than or equal to the counterparts of $\hat{Q}^{(i)}_{j^{(i)}_*}{\bm f}^{(i)}_{\rm obj}({\bm L}^{(i-1)}, P_{\rm trans}^{(i-1)})$,
and all of the elements in ${\bm \Pi}^{(i)}_{j^{(i)}_*}$ are larger than 0.
Thus, taking inner products gives
\begin{eqnarray}
	{\bm \Pi}^{(i){\rm t}}_{j^{(i)}_*}{\bm f}^{(i)}_{\rm obj}({\bm L}^{(i-1)}, P_{\rm trans}^{(i-1)}) &>& {\bm \Pi}^{(i){\rm t}}_{j^{(i)}_*}{\bm f}^{(i)}_{\rm obj}({\bm L}^{(i)}, P_{\rm trans}^{(i)})\nonumber\\
	\iff \bar{L}_{j^{(i-1)}_*}^{(i-1)} &>& \bar{L}_{j^{(i)}_*}^{(i)},
\end{eqnarray}
which would conflict with the assumption.
Combining the above fact with step a of this proof, we have
\begin{equation}
	\hat{\bm C}_{j^{(i)}_*}^{(i)}\neq \hat{Q}^{(i)}_{j^{(i)}_*}{\bm C}^{(i-1)} - {\bm 1}_{K_{j^{(i)}_*}^{(i)}} \hat{Q}^{(i)}_{j^{(i)}_*\{j\neq 0\}\emptyset}{\bm C}^{(i-1)}\Longrightarrow \bar{L}_{j^{(i-1)}_*}^{(i-1)} > \bar{L}_{j^{(i)}_*}^{(i)}.
\end{equation}
\\

{\bf c}. Let us assume $\bar{L}^{(i)}_{j_*^{(i)}}>\mathcal{L}_{\rm subset}(N, p_{\rm src}, K,\{\tilde{\textsc{Mode}}_k\}, A^{(i)})$ when $\hat{\bm C}_{j^{(i)}_*}^{(i)}= \hat{Q}^{(i)}_{j^{(i)}_*}{\bm C}^{(i-1)} - {\bm 1}_{K_{j^{(i)}_*}^{(i)}} \hat{Q}^{(i)}_{j^{(i)}_*\{j\neq 0\}\emptyset}{\bm C}^{(i-1)}$.
From this assumption, we can set some feasible code forest $\{T^*_k\}=\{(\textrm{Cword}^*_k, \textrm{Link}^*_k, \tilde{\textsc{Mode}}_k)\mid k\in\mathbb{Z}^+_{<K}\}$ whose subset $\{T^*_k\mid k\in A^{(i)}\}$ belongs to $\mathbb{LF}_{M,N}$ and achieves the minimum expected code length $\bar{L}^{\ast}=\mathcal{L}_{\rm subset}(N, p_{\rm src}, K,\{\tilde{\textsc{Mode}}_k\}, A^{(i)})$ ($<\bar{L}^{(i)}_{j_*^{(i)}}$).

We can define the code-tree-wise expected code lengths, transition matrix, and stable distribution of $\{T^*_k\}$ respectively as ${\bm L}^{\ast}$, $P_{\rm trans}^{\ast}$, and ${\bm \Pi}^{\ast}$.
Since $\{T^*_k\}$ is defined to make its subset $\{T^*_k\mid k\in A^{(i)}\}$ achieve the minimum, we can assume without loss of generality that $\hat{Q}^{(i)}_j{\bm \Pi}^{\ast}$ is a zero vector when $j\neq j^{(i)}_*$ and $\hat{Q}^{(i)}_{j^{(i)}_*}{\bm \Pi}^{\ast}$ adds up to 1.

According to steps a and b of this proof, we can formulate as
\begin{eqnarray}
	{\bm 1}_{K^{(i)}_{j^{(i)}_*}} \bar{L}^{(i)}_{j^{(i)}_*}&=&\hat{\bm L}_{j^{(i)}_*}^{(i)} + \left(\hat{P}_{j^{(i)}_*,j^{(i)}_*\{0\}\{0\}}^{(i)}-I_{{K^{(i)}_{j^{(i)}_*}}-1}\right) \hat{Q}^{(i)}_{j^{(i)}_*}{\bm C}^{(i)}\nonumber\\
	&=&\hat{\bm L}_{j^{(i)}_*}^{(i)} + \left(\hat{P}_{j^{(i)}_*,j^{(i)}_*\{0\}\{0\}}^{(i)}-I_{{K^{(i)}_{j^{(i)}_*}}-1}\right) \left(\hat{Q}^{(i)}_{j^{(i)}_*}{\bm C}^{(i-1)} - {\bm 1}_{K_{j^{(i)}_*}^{(i)}} \hat{Q}^{(i)}_{j^{(i)}_*\{j\neq 0\}\emptyset}{\bm C}^{(i-1)}\right)\nonumber\\
	&=&\hat{Q}^{(i)}_{j^{(i)}_*}{\bm f}^{(i)}_{\rm obj}({\bm L}^{(i)}, P_{\rm trans}^{(i)}).
	\label{eq:obj_vec_l4_gen}
\end{eqnarray}
Since solving {\it Minimization problem \ref{mip:codetreewise}} in the $i$-th iteration minimizes Eq.~(\ref{eq:obj_vec_l4_gen}), each element does not become smaller when replacing ${\bm L}^{(i)}$ and $P_{\rm trans}^{(i)}$ with ${\bm L}^{\ast}$ and $P_{\rm trans}^{\ast}$.
Therefore, taking an inner product with $\hat{Q}^{(i)}_{j^{(i)}_*}{\bm \Pi}^{\ast}$, having only non-negative values and adding up to 1, leads to an inequality
\begin{eqnarray}
	\bar{L}^{(i)}_{j^{(i)}_*} &\le& \bar{L}^{\ast},
\end{eqnarray}
which conflicts with the assumption.
Thus, $\bar{L}^{(i)}_{j^{(i)}_*}=\mathcal{L}_{\rm subset}(N, p_{\rm src}, K,\{\tilde{\textsc{Mode}}_k\}, A^{(i)})$ when $\hat{\bm C}_{j^{(i)}_*}^{(i)}= \hat{Q}^{(i)}_{j^{(i)}_*}{\bm C}^{(i-1)} - {\bm 1}_{K_{j^{(i)}_*}^{(i)}} \hat{Q}^{(i)}_{j^{(i)}_*\{j\neq 0\}\emptyset}{\bm C}^{(i-1)}$. \\

From step b, the cost ${\bm C}$ does not oscillate unless $\hat{\bm C}_{j^{(i)}_*}^{(i)}= \hat{Q}^{(i)}_{j^{(i)}_*}{\bm C}^{(i-1)} - {\bm 1}_{K_{j^{(i)}_*}^{(i)}} \hat{Q}^{(i)}_{j^{(i)}_*\{j\neq 0\}\emptyset}{\bm C}^{(i-1)}$.
For the same reason as {\it Theorem \ref{thm:optimum}}, the algorithm must converge within a finite iteration.
Therefore, {\it Procedure \ref{prc:iterative_gen}} always gives an \emph{E-optimal} $\{T_k\}$.
$\qquad \blacksquare$\\\\

As the above theorem suggests, we can guarantee \emph{E-optimality} even if we use the worst-performing code forest given by the iteration. If we use the best-performing one after the convergence, its optimality can be checked just as {\it Procedure \ref{prc:iterative}}:
\mytheory{thm}{F-optimality check}{
	\label{thm:optimum_check_gen}
	If we have ${\bm C}^{(i)}={\bm C}^{(i-1)}$ in {\it Procedure \ref{prc:iterative_gen}},
	$\{\tilde{T}_k\}=\{T^{(i)}_k\mid k\in A\}$, where $j_{**}=\arg\min_j \bar{L}_j^{(i)}$ and $A=\{k\mid k\text{-th column of }\hat{Q}_{j_{**}}^{(i)}\text{ is not zero}\}$, is \emph{F-optimal} for the given $K$ and $\{\tilde{\textsc{Mode}}_k\mid k\in\mathbb{Z}^+_{<K}\}$.
}
\underline{\it Proof}: If ${\bm C}^{(i)}={\bm C}^{(i-1)}$, for any code forest $\{T^*_k\}=\{(\textrm{Cword}^*_k, \textrm{Link}^*_k, \tilde{\textsc{Mode}}_k)\mid k\in\mathbb{Z}^+_{<K}\}$ with the tree-wise expected code length ${\bm L}^{\ast}$, transition matrix $P_{\rm trans}^{\ast}$, stationary distribution ${\bm \Pi}^{\ast}$, and expected code length $\bar{L}^{\ast}$, we have
\begin{equation}
	\min_j\bar{L}^{(i)}_j\le {\bm \Pi}^{\ast{\rm t}}{\bm f}^{(i+1)}_{\rm obj}({\bm L}^{(i)}, P_{\rm trans}^{(i)})
	= {\bm \Pi}^{\ast{\rm t}}{\bm f}^{(i)}_{\rm obj}({\bm L}^{(i)}, P_{\rm trans}^{(i)})\nonumber\\
	\le {\bm \Pi}^{\ast{\rm t}}{\bm f}^{(i)}_{\rm obj}({\bm L}^{\ast}, P_{\rm trans}^{\ast})=\bar{L}^{\ast}.
\end{equation}
Therefore, $\min_j\bar{L}^{(i)}_j=\mathcal{L}_{\rm fixed}(N, p_{\rm src}, K,\{\tilde{\textsc{Mode}}_k\})$.
$\qquad \blacksquare$

Furthermore, we can also guarantee that the constructed code forests achieve the expected code length not worse than Huffman codes.
If we make the initial costs $C^{(0)}_k$ in {\it Procedure \ref{prc:iterative_gen}} for $k\neq 0$ be sufficiently large values,
$\textrm{Link}_k^{(1)}(a)=0$ becomes the optimal choice for every link. In this case, $T^{(1)}_0$ becomes a Huffman tree, and $\{T^{(1)}_0\}$ will be the only absorption block in the code forest because every link points to $T^{(1)}_0$. Therefore, the worst-performing code forest after the first iteration becomes equivalent to the Huffman code, and its expected code length will be the upper bound of the later iterations.

\section{Generality of AIFV-2 codes}
\label{app:gen_aifv2}
\begin{figure}[!tb]
	\begin{center}
		\subfigure[Structures of $T^*_2$, $T^*_3$, and the trees made by copying the subtrees.]{
			\includegraphics[width=4cm,  bb=0 0 331 522]{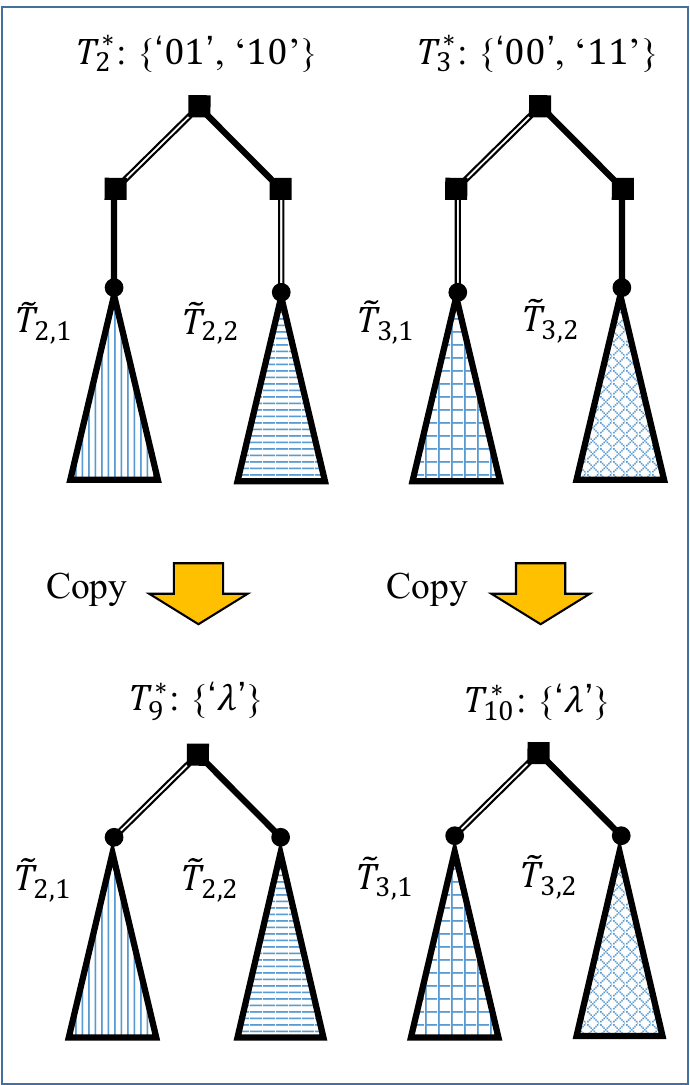}}
		\subfigure[All the cases where there are some nodes pointing to either $T^*_2$ or $T^*_3$ (upper half) and their respective reassignments (lower half). ]{
			\includegraphics[width=10.6cm,  bb=0 0 879 523]{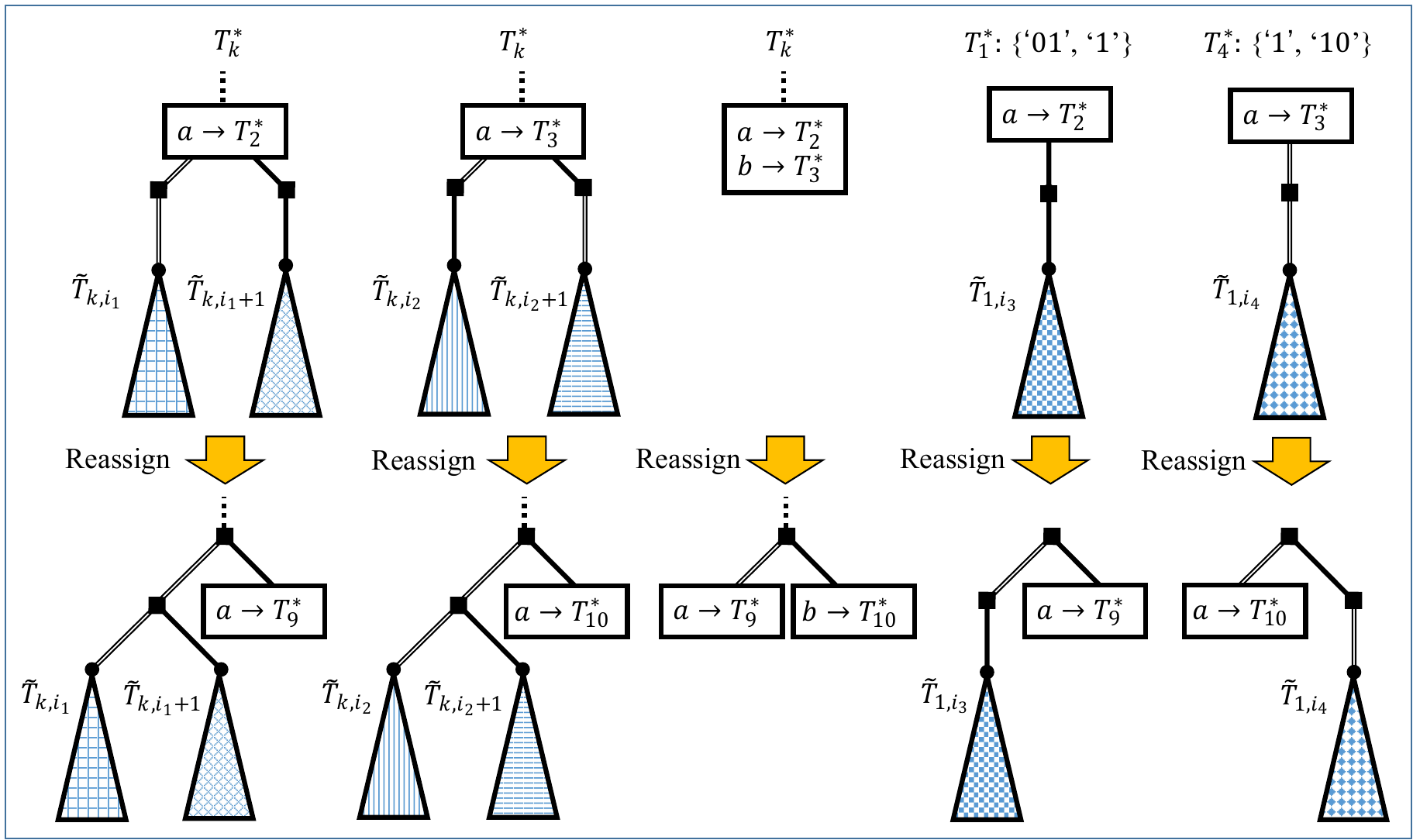}}
	\end{center}
	\caption{Tree structures related to the modes $\{$`01', `10'$\}$ and $\{$`00', `11'$\}$. Black squares indicate the nodes with no symbol assigned. }
	\label{fig:aifv2_reassignment1}
\end{figure}
It is recently reported that the conventional AIFV-2 codes can achieve the minimum expected code length among all the codes decodable with 2 bits of decoding delay \cite{ref:aifv2_opt_proof}. We here explain this fact from the perspective of $N$-bit-delay AIFV codes. 
According to the results in our previous work \cite{ref:n_aifv},
2-bit-delay AIFV codes can represent VV codes decodable within 2 bits of decoding delay without loss of generality.
Moreover, due to {\it Theorem \ref{thm:mos}}, we need at most 9 code trees, assigned different modes described in Fig.~\ref{fig:2bit_mode_patt}, to construct an optimal code forest for a given source.

Say $\{T^*_k\mid k\in\mathbb{Z}^+_{<9}\}$ is \emph{G-optimal} where $T^*_k=(\textrm{Cword}^*_k, \textrm{Link}^*_k, \textsc{Mode}^*_k)$, $\textsc{Mode}^*_0=\{$`$\lambda$'$\}$, and $\textsc{Mode}^*_1=\{$`01', `1'$\}$.
We first claim that we do not necessarily need the modes $\{$`01', `10'$\}$, $\{$`00', `11'$\}$, $\{$`00', `10'$\}$ and $\{$`01', `11'$\}$ to construct a \emph{G-optimal} code.
Let us write as $\textsc{Mode}^*_2=\{$`01', `10'$\}$, and $\textsc{Mode}^*_3=\{$`00', `11'$\}$ without loss of generality.
As illustrated in the upper half of Fig.~\ref{fig:aifv2_reassignment1} (a), $T^*_2$ and $T^*_3$ never have a symbol assigned to the nodes of `$\lambda$' (root), `0', or `1':
If they do, they would infringe {\it Rule \ref{rle:decodable}} \ref{srle:prefix_in_mode}.
Therefore, we can always make some additional code trees $T^*_9$ and $T^*_{10}$ where $\textsc{Mode}^*_9=\textsc{Mode}^*_{10}=\{$`$\lambda$'$\}$ using the subtrees of $T^*_2$ and $T^*_3$, as in the lower half of Fig.~\ref{fig:aifv2_reassignment1} (a).

From {\it Rule \ref{rle:decodable}}, there are 5 patterns available for the structure below the node pointing to $T^*_2$ or $T^*_3$,
as in the upper half of Fig.~\ref{fig:aifv2_reassignment1} (b).
Among all the code trees in $\{T^*_k\mid k\in\mathbb{Z}^+_{<11}\}$, reassigning the nodes and subtrees for each pattern as in the lower half of Fig.~\ref{fig:aifv2_reassignment1} (b) does not change the expected code length of the code forest:
This reassignment never infringes {\it Rule \ref{rle:decodable}};
although the nodes linked to $T^*_2$ or $T^*_3$ get 1 bit deeper, the next code tree $T^*_9$ or $T^*_{10}$ gives an exactly 1-bit shorter codeword for any symbol.
Note that the reassignment changes the length of the termination codeword, but it does not affect the expected code length because it appears only once for any length of the source symbol sequence.

The reassigned code forest never links to $T^*_2$ or $T^*_3$ but achieves the minimum expected code length.
So, after the reassignment, the code forest $\{T^*_k\mid k\in\mathbb{Z}^+_{<11}\setminus \{2,3\}\}$ constructs an optimal code where $\textsc{Mode}^*_0=\textsc{Mode}^*_9=\textsc{Mode}^*_{10}=\{$`$\lambda$'$\}$.
Owing to {\it Theorem \ref{thm:mos}}, we can make a \emph{G-optimal} code with a single tree for each mode,
and thus some code forest $\{T_k\mid k\in\mathbb{Z}^+_{<9}\setminus \{2,3\}\}$, where $T_k=(\textrm{Cword}_k, \textrm{Link}_k, \textsc{Mode}^*_k)$,
can achieve the optimum.
Similarly, we do not have to use the modes $\{$`00', `10'$\}$ and $\{$`01', `11'$\}$ to construct an optimal code.

Secondly, we claim that the modes $\{$`0', `10'$\}$, $\{$`0', `11'$\}$ and $\{$`00', `1'$\}$ are not necessary to construct an optimal code.
Since we have shown that 4 out of 9 modes are unnecessary for constructing an optimum,
we write an optimal code forest as $\{T^{**}_k\mid k\in\mathbb{Z}^+_{<5}\}$ where $T^{**}_k=(\textrm{Cword}^{**}_k, \textrm{Link}^{**}_k, \textsc{Mode}^{**}_k)$, $\textsc{Mode}^{**}_0=\{$`$\lambda$'$\}$, and $\textsc{Mode}^{**}_1=\{$`01', `1'$\}$.

\begin{figure}[!tb]
	\begin{center}
		\subfigure[Structures of $T^{**}_2$ and the tree made by copying the subtrees.]{
			\includegraphics[width=4.8cm,  bb=0 0 360 270]{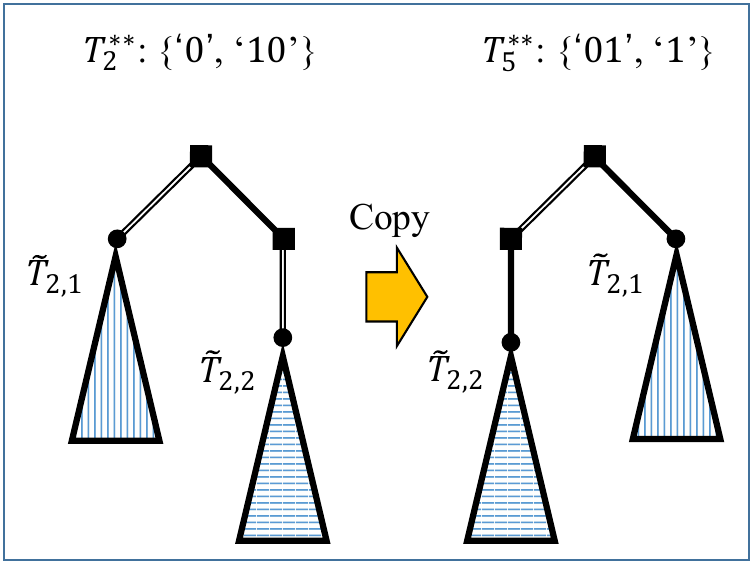}}
		\subfigure[Node pointing to $T^{**}_2$ and its reassignment. ]{
			\includegraphics[width=3.4cm,  bb=0 0 257 270]{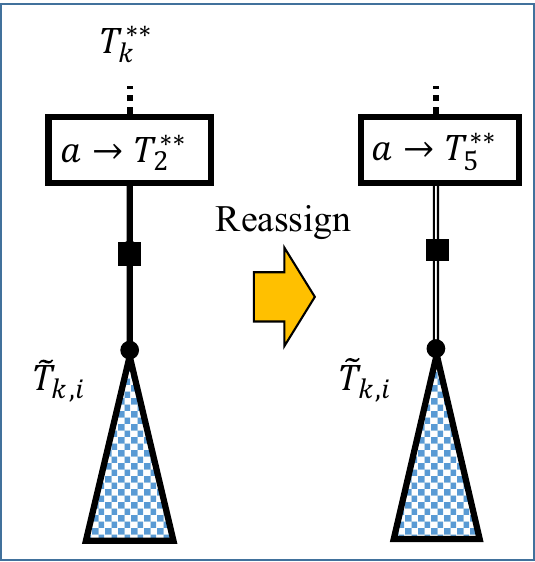}}
	\end{center}
	\caption{Tree structures related to the mode $\{$`0', `11'$\}$. Black squares indicate the nodes with no symbol assigned. }
	\label{fig:aifv2_reassignment2}
\end{figure}

Let us assume $\textsc{Mode}^{**}_2=\{$`0', `10'$\}$ without loss of generality.
As illustrated in Fig.~\ref{fig:aifv2_reassignment2} (a), $T^{**}_2$ never has a symbol assigned to the nodes of `$\lambda$' or `1',
and therefore we can always make a code tree $T^{**}_5$ with $\textsc{Mode}^{**}_5=\{$`01', `1'$\}$ using the subtrees of $T^{**}_2$.
Obeying {\it Rule \ref{rle:decodable}}, there is one pattern available for the structure below the node pointing to $T^{**}_2$,
as in the left-hand side of Fig.~\ref{fig:aifv2_reassignment2} (b).
Similar to the above discussion, we can reassign the nodes and subtrees as in the right-hand side of Fig.~\ref{fig:aifv2_reassignment2} (b) without changing the expected code length.
So, after the reassignment, the code forest $\{T^{**}_k\mid k\in\mathbb{Z}^+_{<6}\setminus \{2\}\}$ constructs an optimal code where $\textsc{Mode}^{**}_1=\textsc{Mode}^{**}_5=\{$`01', `1'$\}$.
Owing to {\it Theorem \ref{thm:mos}}, some code forest $\{T_k\mid k\in\mathbb{Z}^+_{<5}\setminus \{2\}\}$, where $T_k=(\textrm{Cword}_k, \textrm{Link}_k, \textsc{Mode}^{**}_k)$, can be \emph{G-optimal}.
The same can be said for $\{$`0', `11'$\}$ and $\{$`00', `1'$\}$.

As a result, two code trees with modes $\{$`$\lambda$'$\}$ and $\{$`01', `1'$\}$ are enough to construct a \emph{G-optimal} code.
This means that the conventional AIFV-2 can achieve the minimum expected code length among all the codes decodable within a 2-bit decoding delay.

\section{Brute-force search for code-tree-wise optimization}
\label{app:bruteforce}
If we focus on binary inputs ($M=2$), it is possible to try every available code tree with a given mode.
For small $N$s, we can solve {\it Minimization problem \ref{mip:codetreewise}} using every mode in $\mathbb{BM}_N$.
Under the full code forest condition, every code tree can be represented as a simple partition of some codeword set:
\mytheory{thm}{Code-tree representation in binary-input cases}{
	\label{thm:bintree}
	Say $T_k=(\textrm{Cword}_k, \textrm{Link}_k, \textrm{Mode}_k)$ is an arbitrary code tree in full code forests of $N$-bit-delay AIFV codes for $\mathbb{A}_2$ under MoS condition.
	When $\textrm{Mode}_k=F_{\rm red}((\text{`0'}\oplus \textsc{Lb})\cup (\text{`1'}\oplus \textsc{Ub})) \in \mathbb{BM}_N$ where $\textsc{Lb},\textsc{Ub}\subset \mathbb{W}_{N-1}$, $T_k$ can be represented by some non-empty $W,\bar{W}\subset \mathbb{W}_{N}$ satisfying $W\cup\bar{W}=(\text{`0'}\oplus \textsc{Lb})\cup (\text{`1'}\oplus \textsc{Ub})$ as follows.
	\begin{eqnarray}
		\textrm{Cword}_k(a_0) &=& f_{\rm cmn}(W)\\
		\textrm{Mode}_{\textrm{Link}_k(a_0)} &=& F_{\rm red}(f_{\rm cmn}(W)\oslash W)\\
		\textrm{Cword}_k(a_1) &=& f_{\rm cmn}(\bar{W})\\
		\textrm{Mode}_{\textrm{Link}_k(a_1)} &=& F_{\rm red}(f_{\rm cmn}(\bar{W})\oslash \bar{W})
	\end{eqnarray}
}
The above uses an additional notation:
\begin{itemize}
	\item $f_{\rm cmn}$: $\mathbb{M}\to \mathbb{W}$. $f_{\rm cmn}(\textsc{Words})$ outputs the maximum-length common prefix of $\textsc{Words} \in \mathbb{M}$.
\end{itemize}
According to this theorem, we can make every code tree of the mode $F_{\rm red}((\text{`0'}\oplus \textsc{Lb})\cup (\text{`1'}\oplus \textsc{Ub}))$ only by thinking of the partitions of $(\text{`0'}\oplus \textsc{Lb})\cup (\text{`1'}\oplus \textsc{Ub})$.
Note that $\textrm{Link}_k(a)$ can be uniquely determined from the value of $\textrm{Mode}_{\textrm{Link}_k(a)}$ due to MoS condition.

For example, in making a code tree $T_k$ with $\textrm{Mode}_k=\{$`001', `01', `1'$\}$ for $N=3$, the mode can be written as $\textrm{Mode}_k=F_{\rm red}(\{$`001', `010', `011', `100', `101', `110', `111'$\})$.
There are 126 partitions which makes $\{$`001', `010', `011', `100', `101', `110', `111'$\}$ into $W$ and $\bar{W}$, counting the permutations.
We can try every partition to find the best tree for {\it Minimization problem \ref{mip:codetreewise}}.
If we try $W=\{$`001', `010'$\}$ and $\bar{W}=\{$`011', `100', `101', `110', `111'$\}$, for instance,
we have $\textrm{Cword}_k(a_0)=$ `0', $\textrm{Mode}_{\textrm{Link}_k(a_0)}=\{$`01', `10'$\}$, $\textrm{Cword}_k(a_1)=$ `$\lambda$', and  $\textrm{Mode}_{\textrm{Link}_k(a_1)}=\{$`011', `1'$\}$.

To prove the above theorem, we introduce a lemma about the expanded codewords.
\mytheory{lmm}{Expanded codewords in binary-input cases}{
	\label{lmm:expcd_bin}
	In full code forests of $N$-bit-delay AIFV codes for $\mathbb{A}_2$, every expanded codeword is at most $N$-bit length.
}
\underline{{\it Proof of Lemma \ref{lmm:expcd_bin}}}:
For an arbitrary code tree $T_k$, we can write one of the expanded codewords corresponding to a source symbol $a_0$ as
\begin{equation}
	\textrm{Expcw}=\textrm{Cword}_k(a_0)\oplus \textrm{Query}\in \textsc{Expand}_k(a_0)
\end{equation}
where $\textrm{Query}\in \textsc{Mode}_{\textrm{Link}_k(a_0)}$ $(\in \mathbb{BM}_N)$.
Let us assume $\|\textrm{Expcw}\|_{\rm len}>N$.
In this case, $\textrm{Cword}_k(a_0)$ is at least 1-bit length because $\textrm{Query}\le N$.
Since $\|\textrm{Expcw}\|_{\rm len}>N \ge 2$, we can write it as
\begin{equation}
	\textrm{Expcw}=(y_{\rm head}\oplus w_{\rm middle}) \oplus y_{\rm tail}
\end{equation}
using some $y_{\rm head}, y_{\rm tail}\in \mathbb{W}_1$ and $w_{\rm middle}\in\mathbb{W}$.

Under this assumption, the following facts must hold.
\begin{enumerate}
	\item $(y_{\rm head}\oplus w_{\rm middle}) \oplus (\neg y_{\rm tail})\notin \textsc{Expand}_k(a_0)$
	\item $(y_{\rm head}\oplus w_{\rm middle}) \oplus (\neg y_{\rm tail})\in \textsc{Expand}_k(a_1)$
	\item $\exists \textrm{Suffix}\in \mathbb{W}: (\neg y_{\rm head})\oplus \textrm{Suffix}\in \textsc{Expand}_k(a_1)$\\
\end{enumerate}

[Reason for a]
It obviously holds when $\textrm{Query}=$ `$\lambda$' because $\textsc{Expand}_k(a_0)=\textrm{Cword}_k(a_0)\oplus \{\text{`$\lambda$'}\}=\{(y_{\rm head}\oplus w_{\rm middle}) \oplus y_{\rm tail}\}$.
In cases of $\textrm{Query}\neq $ `$\lambda$', if we assume $(y_{\rm head}\oplus w_{\rm middle}) \oplus (\neg y_{\rm tail})\in \textsc{Expand}_k(a_0)$,
both $(\textrm{Cword}_k(a_0)\oslash(y_{\rm head}\oplus w_{\rm middle})) \oplus y_{\rm tail}$ and $(\textrm{Cword}_k(a_0)\oslash(y_{\rm head}\oplus w_{\rm middle})) \oplus (\neg y_{\rm tail})$ should belong to $\textsc{Mode}_{\textrm{Link}_k(a_0)}$.
This fact conflicts with $\textsc{Mode}_{\textrm{Link}_k(a_0)}\in \mathbb{BM}_N$ because $\mathbb{BM}_N$ requires the members not to have any full partial tree.

[Reason for b]
If $(y_{\rm head}\oplus w_{\rm middle}) \oplus (\neg y_{\rm tail})\notin \textsc{Expand}_k(a_1)$, $(y_{\rm head}\oplus w_{\rm middle}) \oplus (\neg y_{\rm tail})$ cannot be a member of $\textsc{Expands}_k$.
Therefore, with the proposition a, $\textrm{Expcw}$ becomes a member of $F_{\rm red}(\textsc{Expands}_k)$ because it cannot make any full partial tree.
Combining this fact with the full code forest condition, $\textsc{Mode}_k$ must contain $\textrm{Expcw}$.
However, it conflicts with $\textsc{Mode}_k\in\mathbb{BM}_N$ because $\|\textrm{Expcw}\|_{\rm len}>N$.

[Reason for c]
Since $y_{\rm head}\preceq\textrm{Cword}_k(a_0)$, $\forall \textrm{Suffix}\in \mathbb{W}: (\neg y_{\rm head})\oplus \textrm{Suffix}\notin \textsc{Expand}_k(a_0)$.
If $\textsc{Expand}_k(a_1)$ also does not include any codeword beginning with $\neg y_{\rm head}$, $\textsc{Expands}_k$ must have a common prefix $y_{\rm head}$.
In this case, from the full code forest condition, $\textsc{Mode}_k$ must also have a common prefix $y_{\rm head}$, which conflicts with $\textsc{Mode}_k\in\mathbb{BM}_N$.
\\

From the propositions b and c, $\textrm{Cword}_k(a_1)$ must be `$\lambda$'.
Therefore, from the proposition b, $\textsc{Mode}_{\textrm{Link}_k(a_1)}$ must contain $(y_{\rm head}\oplus w_{\rm middle}) \oplus (\neg y_{\rm tail})$.
However, it is longer than $N$ bits, which conflicts with $\textsc{Mode}_{\textrm{Link}_k(a_1)}\in\mathbb{BM}_N$.
So, every expanded codeword is at most $N$-bit length. $\qquad \blacksquare$\\\\
\underline{{\it Proof of Theorem \ref{thm:bintree}}}:
For arbitrary fixed-length binary-string sets $\textsc{Words}_N, \textsc{Words}'_N\subset\mathbb{W}_N$,
\begin{equation}
	\label{eq:reduce_uniqueness}
	\textsc{Words}_N=\textsc{Words}'_N\iff F_{\rm red}(\textsc{Words}_N)=F_{\rm red}(\textsc{Words}'_N)
\end{equation}
holds:
When we interpret $\textsc{Words}_N$ and $\textsc{Words}'_N$ as trees, the way of cutting off all of their full partial trees is obviously unique;
when we interpret $F_{\rm red}(\textsc{Words}_N)$ and $F_{\rm red}(\textsc{Words}'_N)$ as trees, there is only one way to append some full partial trees to their leaves when we make every leaf be $N$-bit depth.

According to {\it Lemma \ref{lmm:expcd_bin}}, every expanded codeword is at most $N$-bit length.
Therefore,
\begin{equation}
	F_{\rm red}(\textsc{Expands}_k)=F_{\rm red}\left(\bigcup_{\textrm{Expcw}\in\textsc{Expands}_k}\textrm{Expcw}\oplus \mathbb{W}_{N-\|\textrm{Expcw}\|_{\rm len}}\right).
\end{equation}
Owing to the full code forest condition, $\textrm{Mode}_k=F_{\rm red}((\text{`0'}\oplus \textsc{Lb})\cup (\text{`1'}\oplus \textsc{Ub}))$, and Eq.~(\ref{eq:reduce_uniqueness}),
\begin{equation}
	\bigcup_{\textrm{Expcw}\in\textsc{Expands}_k}\textrm{Expcw}\oplus \mathbb{W}_{N-\|\textrm{Expcw}\|_{\rm len}} = (\text{`0'}\oplus \textsc{Lb})\cup (\text{`1'}\oplus \textsc{Ub}).
\end{equation}
Here, from {\it Rule \ref{rle:decodable}} \ref{srle:prefix_free_expansion}, there is no overlap between $\textsc{Expand}_k(a_0)$ and $\textsc{Expand}_k(a_1)$, and thus
we can define a partition $(W, \bar{W})$ of $(\text{`0'}\oplus \textsc{Lb})\cup (\text{`1'}\oplus \textsc{Ub})$ for an arbitrary code tree $T_k$ that makes
\begin{eqnarray}
	\label{eq:extend_expcw}
	\bigcup_{\textrm{Expcw}\in\textsc{Expand}_k(a_0)}\textrm{Expcw}\oplus \mathbb{W}_{N-\|\textrm{Expcw}\|_{\rm len}} &=& W,\\
	\bigcup_{\textrm{Expcw}'\in\textsc{Expand}_k(a_1)}\textrm{Expcw}'\oplus \mathbb{W}_{N-\|\textrm{Expcw}'\|_{\rm len}} &=& \bar{W}.
\end{eqnarray}

From Eq.~(\ref{eq:extend_expcw}), we have
\begin{equation}
	\label{eq:extend_mode}
	\textrm{Cword}_k(a_0)\oplus \left(\bigcup_{\textrm{Query}\in\textsc{Mode}_{\textrm{Link}_k(a_0)}}\textrm{Query}\oplus \mathbb{W}_{N-\|\textrm{Cword}_k(a_0)\oplus\textrm{Query}\|_{\rm len}}\right) = W.
\end{equation}
Since $\textsc{Mode}_{\textrm{Link}_k(a_0)}\in \mathbb{BM}_N$, $\textsc{Mode}_{\textrm{Link}_k(a_0)}$ has no common prefix, and thus
\begin{equation}
	\textrm{Cword}_k(a_0) = f_{\rm cmn}(W).
\end{equation}
Therefore, from Eq.~(\ref{eq:extend_mode}),
\begin{equation}
	F_{\rm red}\left(\bigcup_{\textrm{Query}\in\textsc{Mode}_{\textrm{Link}_k(a_0)}}\textrm{Query}\oplus \mathbb{W}_{N-\|\textrm{Cword}_k(a_0)\oplus\textrm{Query}\|_{\rm len}}\right) = F_{\rm red}(\textsc{Mode}_{\textrm{Link}_k(a_0)})= F_{\rm red}(f_{\rm cmn}(W)\oslash W).
\end{equation}
Due to $\textsc{Mode}_{\textrm{Link}_k(a_0)}\in \mathbb{BM}_N$, $\textsc{Mode}_{\textrm{Link}_k(a_0)}$ has no full partial tree, so it is invariant by $F_{\rm red}$.
As a result, we have
\begin{equation}
	\textsc{Mode}_{\textrm{Link}_k(a_0)}= F_{\rm red}(f_{\rm cmn}(W)\oslash W).
\end{equation}
The same can be said for $a_1$ with $\bar{W}$. $\qquad \blacksquare$\\\\

Note that in cases of non-binary inputs, we cannot try every code tree as easily as stated above.
Even in the binary-input cases, the brute-force search becomes impractical quickly as $N$ increases:
For $N=3$, we only have to think of $254$ partitions for $T_0$, making two labeled non-empty subsets from a set of size 8,
but it grows up to $65534$ patterns when $N=4$.
Therefore, in general, it is reasonable to use ILP approach as in Section \ref{sec:design}.

\end{document}